\documentclass{aa}  
\usepackage{natbib}
\usepackage{graphicx}
\usepackage{lscape}

\usepackage{txfonts}
\begin{document}

   \title{Stellar parameters and chemical abundances of 223 evolved stars with and without planets\thanks{Based on spectral data retrieved from the ELODIE and SOPHIE archives at Observatoire de Haute-Provence \citep{Moultaka2004}.}\fnmsep\thanks{Based on data obtained from the ESO Science Archive Facility collected at the La Silla Paranal Observatory, ESO (Chile) with the HARPS and FEROS spectrographs.}}

   \author{E. Jofr\'{e} 
           \inst{1,4}\fnmsep\thanks{Visiting Astronomer, Complejo Astron\'omico El Leoncito operated under agreement between the Consejo Nacional de Investigaciones Cient\'{i}ficas y T\'{e}cnicas de la            Rep\'{u}blica Argentina and the National Universities of La Plata, C\'{o}rdoba and San Juan.}
           , R. Petrucci\inst{2,4}, C. Saffe\inst{3,4}, L. Saker\inst{1}, 
           E. Artur de la Villarmois\inst{1}, C. Chavero\inst{1,4}, 
           M. G\'{o}mez\inst{1,4} \and P. J. D. Mauas\inst{2,4}  
          }

   \institute{Observatorio Astron\'{o}mico de C\'{o}rdoba (OAC), Laprida 854, X5000BGR, C\'ordoba, Argentina \\
              \email{emiliano@oac.uncor.edu}
         \and
             Instituto de Astronom\'{i}a y F\'{i}sica del Espacio (IAFE), CC.67, suc. 28, 1428, Buenos Aires, Argentina 
             \and
             Instituto de Ciencias Astron\'omicas, de la Tierra y del Espacio (ICATE), CC.467, 5400, San Juan, Argentina 
            \and
            Consejo Nacional de Investigaciones Cient\'{i}ficas y T\'{e}cnicas (CONICET), Argentina  
             }

   \date{Received 25 June 2014 / accepted 14 October 2014}

 
  \abstract
   {}
   {We present fundamental stellar parameters, chemical abundances, and rotational velocities for a sample of 86 evolved stars with planets (56 giants; 30 subgiants), and for a control sample of 137 stars (101 giants; 36 subgiants) without planets. The analysis was based on both high signal-to-noise and resolution echelle spectra. The main goals of this work are i) to investigate chemical differences between evolved stars that host planets and those of the control sample without planets; ii) to explore potential differences between the properties of the planets around giants and subgiants; and iii) to search for possible correlations between these properties and the chemical abundances of their host stars. Implications for the scenarios of planet formation and evolution are also discussed.}
   {The fundamental stellar parameters ($T_{\mathrm{eff}}$, $\log g$, [Fe/H], $\xi_{t}$) were computed homogeneously using the FUNDPAR code. The chemical abundances of 14 elements (Na, Mg, Al, Si, Ca, Sc, Ti, V, Cr, Mn, Co, Ni, Zn, and Ba) were obtained using the MOOG code. Rotational velocities were derived from the full width at half maximum of iron isolated lines.}
   {In agreement with previous studies, we find that subgiants with planets are, on average, more metal-rich than subgiants without planets by $\sim$ 0.16 dex. The [Fe/H] distribution of giants with planets is centered at slightly subsolar metallicities and there is no metallicity enhancement relative to the [Fe/H] distribution of giants without planets. Furthermore, contrary to recent results, we do not find any clear difference between the metallicity distributions of stars with and without planets for giants with $M_{\mathrm{\star}}$ > 1.5 $M_{\odot}$. With regard to the other chemical elements, the analysis of the [X/Fe] distributions shows differences between giants with and without planets for some elements, particularly V, Co, and Ba. Subgiants with and without planets exhibit similar behavior for most of the elements. On the other hand, we find no evidence of rapid rotation among the giants with planets or among the giants without planets. Finally, analyzing the planet properties, some interesting trends might be emerging: i) multi-planet systems around evolved stars show a slight metallicity enhancement compared with single-planet systems; ii) planets with $a$ $\lesssim$ 0.5 AU orbit subgiants with [Fe/H] > 0 and giants hosting planets with $a$ $\lesssim$ 1 AU have [Fe/H] < 0; iii) higher-mass planets tend to orbit more metal-poor giants with $M_{\mathrm{\star}}$ $\leq$ 1.5 $M_{\odot}$, whereas planets around subgiants seem to follow the planet-mass metallicity trend observed on dwarf hosts; iv) [X/Fe] ratios for Na, Si, and Al seem to increase with the mass of planets around giants; v) planets orbiting giants show lower orbital eccentricities than those orbiting subgiants and dwarfs, suggesting a more efficient tidal circularization or the result of the engulfment of close-in planets with larger eccentricities.}
   {}

\authorrunning{Jofr\'{e} et al.}
\titlerunning{Stellar parameters and chemical abundances of 223 evolved stars with and without planets}

   \maketitle
%

\section{Introduction}
Nearly two decades ago, \cite{Gonzalez1997} showed the first evidence of a correlation between the metallicities of the host stars and the presence of planets, suggesting that stars with giant planets tend to be more metal-rich in comparison with nearby average dwarfs. This initial trend has been confirmed by several uniform studies on large samples \citep[e.g.,][]{Fischer2005, Santos2004, Ghezzi10a} and now it is widely accepted that FGKM-type dwarfs hosting gas giant planets\footnote{i.e. $M_{p} \gtrsim 1 M_{Jup}$, where $M_{p}$ is the planetary mass and $M_{Jup}$ is the mass of Jupiter.} are, on average, more metal-rich than stars without detected planets. Furthermore, it has been shown that the frequency of stars having giant planets is a strong rising function of the stellar metallicity \citep{Santos2004, Fischer2005}.   
  
The causes of this high metallicity in dwarf stars hosting giant planets have been largely debated in the literature, without reaching a complete consensus to date. One of the scenarios most favored in the literature states that stars hosting giant planets have been formed from intrinsic high metallicity clouds of gas and dust. In this scenario, which is usually called \textit{primordial hypothesis}, the star should be metal-rich throughout its entire radius. This hypothesis is also supported by the core accretion theory for planet formation, where the high metal content allows the planetary cores to grow rapidly enough and to start to accrete gas before it photo-evaporates \citep{Pollack1996, Ida2004, Johnson2012}. However, observational surveys have not succeeded in finding young metal-rich T-Tauri stars in nearby star forming regions \citep{Padgett1996, James2006, Santos2008}. On the other hand, the so-called \textit{pollution hypothesis} states that the high metallicity is a by-product of planet formation. Here, metal-rich rocky material (e.g., planetesimals, asteroids, debris) is accreted onto the star during this process as a result of interactions between young planets with the surrounding disk. Hence, the star would be metal-rich only at the surface convective envelope. However, stellar theorical simulations of  \cite{Theado2012} state that just a fraction of the metal-rich accreted material can remain on the stellar outer regions and would be too small to be observable. For a more complete description of the evidence for both hypotheses, the reader is referred to the review of \cite{Gonzalez2006} and references therein. 

The frequency of planets and how this relates to the properties of their host stars, has been studied for dwarf stars with masses below 1.2 $M_{\odot}$. More massive stars (spectral type earlier than F7) have an insufficient number of spectral lines and high rotation velocities which reduce the precision of the Doppler technique to detect planets \citep[see][]{Galland2005, Galland2006}. In order to overcome this difficulty, several radial velocity surveys to search planets around evolved stars (subgiants and giants) have begun in the last years \citep[e.g.,][]{Frink2002, Hatzes2003, Setiawan2003, Setiawan2004, Sato2005, Johnson2007a, Niedzielski2007}, making it possible to investigate the occurrence of planets around stars in the mass range 1.5-4 $M_{\odot}$. To date, these surveys have resulted in the discovery of $\sim$ 123 evolved stars with planets ($\sim$ 81 giant hosts and $\sim$ 42 subgiants).  

The combination of the results of planet-search surveys at the higher mass range with those at the lower end of the mass scale \citep[red dwarfs,][]{Johnson2010, Haghighipour2010, Bonfils2013} is providing evidence that the stellar mass might also have a role in the planet formation process. Observational studies indicate that planetary frequency increases with stellar mass (Lovis \& Mayor 2007; Johnson et al. 2007a, 2010). Further support for the increase in the planet-formation efficiency as a function of the stellar mass is provided by theoretical models \citep[e.g.,][]{Ida2005, Kennedy2008, Mordasini2009, Villaver2009}.  

In addition, the growing number of evolved stars with planets has enabled, in the last decade, to analyze the link between metallicity and the presence of giant planets in this type of stars. A few studies, based on relatively small samples ($N \lesssim  16$) of subgiant hosts, agree that the same metallicity trend found for dwarfs remains \citep{Fischer2005, Ghezzi10a, Maldonado2013}. \citet{Johnson2010}, with a higher number of subgiants with planets (N = 36), confirmed this finding. The results for giant hosts, on the other hand, have been more controversial in the last years \citep{Hekker2007, Pasquini2007, Takeda2008, Santos2009}. The first studies, although based on small or inhomogeneous samples, showed that giant hosts were metal-poor (Sadakane et al. 2005; Schuler et al. 2005). Hekker \& Melendez (2007) suggested that giant hosts follow the same trend than dwarf hosts. Recent studies found the opposite trend rather than supporting this result \citep[e.g.,][]{Mortier2013, Maldonado2013}. However, Maldonado et al. suggested that massive giants with planets have a metallicity excess compared with the control sample without planets.

Correlations between planet properties (mass, period, eccentricity) and the metallicities of the host stars can provide additional contraints on the planetary formation models. So far, the exploration of such relations has been carried out mainly for dwarf stars \citep[e.g.,][]{Fischer2005, Kang2011}. For example, a recent study suggests that planets orbiting metal-poor stars are in wider orbits than those around metal-rich stars \citep[e.g.,][]{Adibekyan2013}. Furthermore, one of the most interesting results obtained so far, suggests that the planet-metallicity correlation is weaker for Neptunian and lower mass planets \citep[e.g.,][]{Udry2007, Sousa2008, Sousa2011, Bouchy2009, Johnson2009, Ghezzi10a, Buchhave2012, Neves2013}. Interestingly, \citet{Maldonado2013}, analyzing a sample of evolved stars, found a decreasing trend between the stellar metallicity and the mass of the most massive planets. Another remarkable finding is the complete lack of inner planets with semimajor axes below 0.5 UA orbiting giant stars \citep[e.g.,][]{Johnson2007b, Sato2008, Sato2010}. Several studies have suggested that short-period planets might be engulfed as the star evolves off the main-sequence \citep[e.g.,][]{Siess1999, Johnson2007a, Massarotti2008b, Villaver2009, Kunitomo2011}. 

In view of the lack of consensus regarding the metallicity of evolved stars with planets, including the intriguing results obtained recently by Maldonado et al., in this paper we present a homogeneous spectrocopic analysis of 223 evolved stars (56 giants and 30 subgiants with planets), which constitutes one of the largest sample of evolved stars  with planets analyzed uniformly, so far. The size of the sample allows us to search for correlations between chemical abundances and the ocurrence of planets and planet properties. We also analyze the abundances of chemical elements other than Fe, something which has been extensively carried out for dwarf stars with planets, but only ocassionally on evolved stars with planets. The search of such trends can provide important constraints for the models of planet formation and evolution around more massive stars.

 The paper is organized as follows: In Section 2 we describe the samples analyzed along with the spectroscopic observations and data reduction. The determination method of fundamental parameters and chemical abundances and the comparison with the literature are presented in Section 3. This section also includes the calculation of evolutionary parameters, space-velocity components, Galaxy population membership, and projected stellar rotational velocities. The metallicity distributions are analyzed in Section 4, whereas the results for elements other than iron are presented in Section 5. The properties of planets around evolved stars and the study of correlations with chemical abundances are presented in Section 6. In this section we also discuss some implications for the models of planet formation and evolution. Finally, the summary and conclusions are presented in Section 7. 

\section{Observations}

   \begin{figure}
   \centering
   \includegraphics[width=.50\textwidth]{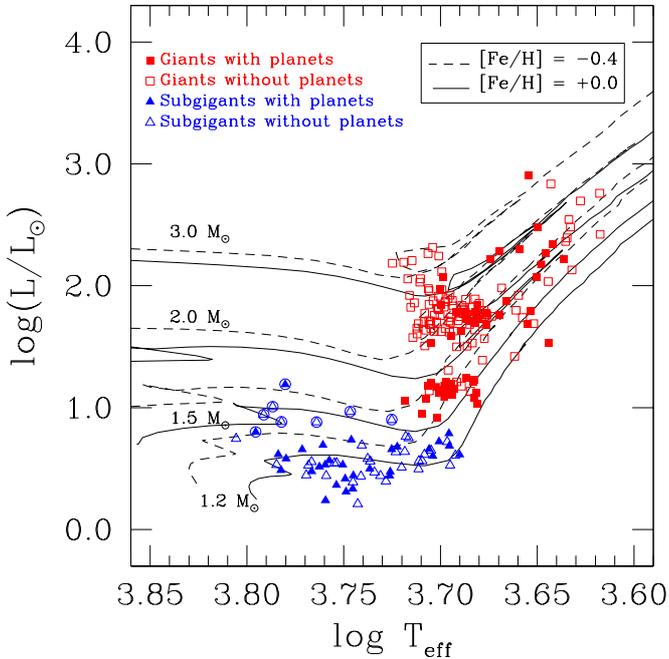}

   \caption{Location of the 223 evolved stars in an H-R diagram. Giant stars are indicated with squares and subgiants with triangles. Filled symbols correspond to stars with planets and empty symbols depict stars without planets. Evolutionary tracks for masses between 1.2 and 3 $M_{\mathrm{\odot}}$ from \citet{Girardi2000} are overplotted. Tracks for [Fe/H]=-0.4 dex are marked with dashed lines and those corresponding to [Fe/H]=+0.0 dex are indicated with solid lines.}
              \label{FigGam}%
    \end{figure}

   \begin{table*}
   \tiny
      \caption[]{Observation log.}
         \label{table:1}
     \centering
         \begin{tabular}{c c c c c}
            \hline\hline
              Instrument/Telescope & Observatory & Spectral resolution & Spectral range ({\AA}) & N  \\
            \hline
HARPS/3.6 m ESO  & La Silla (Chile) &120000 & 3780 - 6910 & 84     \\
ELODIE/1.93 m OHP & OHP (France) &40000 & 3850 - 6800 & 69     \\
FEROS/2.20 m MPG/ESO  & La Silla (Chile) &48000 & 3500 - 9200 & 43     \\
SOPHIE/1.93 m OHP &  OHP (France) &75000 & 3872 - 6943 & 20     \\
EBASIM/2.15 m CASLEO & CASLEO (Argentina) &30000 & 5000 - 7000 & 7     \\
            \hline
         \end{tabular}
        \end{table*}


\subsection{Sample}
The complete sample contains 223 evolved stars (giants and subgiants) including 86 stars with planets. The stars with planets were compiled from the catalog of planets detected by the radial velocity (RV) technique at the Extrasolar Planets Encyclopaedia database\footnote{http://exoplanet.eu/catalog/}. We selected those stars with high S/N spectra available in public databases (see below) and/or are observable from the southern hemisphere. 

In addition to the sample of stars with planets, we built a comparison sample of 137 stars that belong to RV surveys for planets around evolved stars, but for which no planet has been reported so far. We selected 67 stars from the Okayama planet search program \citep{Sato2005}, 34 stars from the ESO FEROS planet search \citep{Setiawan2003} and 36 stars from the retired A stars program \citep{Fischer2005, Johnson2007a}. As in the case of stars with planets, we chose stars with spectra publicly accessible. The control stars from the retired A stars program were obtained from a list of 850 stars for which there are enough RV measurements (N > 10) spanning more than 4 years, to securely detect the presence of planets with orbital periods shorter than 4 years and velocity amplitudes K > 30 m $s^{-1}$ \citep{Fischer2005}. In the case of the stars from the Okayama and the FEROS program we conservatively selected stars with N > 20 spanning more than 4 years of observations. Therefore, the probabilities that these stars harbor planets with similar characteristics to those found so far are low\footnote{In this work when we refer to stars without planets we mean stars that do not harbor planetary companions with similar properties to those reported so far. In this sense, these stars might have planets with other characteristics (e.g., very low mass and/or long period planets) that are harder to detect with the ongoing RV surveys.}.

Figure 1 shows the Hertzsprung-Russell diagram for our sample. In this work, as in \citet{Ghezzi10b} and in \citet{Maldonado2013}, stars are classified as giants (red squares) or subgiants (blue triangles) according to their  bolometric magnitudes, $M_{bol}$. Stars with $M_{bol}$ < 2.82 are classified as giants, whereas those with $M_{bol}$ > 2.82 are marked as subgiants. However, 8 stars (\object{HD 16175}, \object{HD 60532}, \object{HD 75782}, \object{HD 164507}, \object{HD 57006}, \object{HD 67767}, \object{HD 121370}, and \object{HD 198802}, indicated with empty circles in Figure 1) are considered as subgiants although their magnitudes are above the cut-off value. These stars have not evolved to the red giant branch (RGB) yet, and have surface gravities ($\log g \sim$ 3.9) more consistent with the subgiant class. According to this classification, the final sample includes 56 giants with planets (GWP), 101 giants without planets (GWOP), 30 subgiants with planets (SGWP) and 36 subgiants without planets (SGWOP). All the GWOP turned out to be from the Okayama and FEROS surveys, whereas the 36 SGWOP are part of the retired A stars program.

\subsection{Observations and data reduction}
For most of the objects in our sample we used publicly available high signal-to-noise (S/N $\gtrsim$ 150) and high resolution spectra gathered with four different instruments: HARPS (3.6 m ESO telescope, La Silla, Chile), FEROS (2.2 m ESO/MPI telescope, La Silla, Chile), ELODIE (1.93 m telescope, OHP, France), and SOPHIE (1.93 m telescope, OHP, France). In addition we obtained high resolution and high signal-to-noise (S/N $\gtrsim$ 150) spectra with the EBASIM spectrograph \citep{Pintado2003} at the \textit{Jorge Sahade} 2.15 m telescope (CASLEO, San Juan, Argentina) for 7 stars in our sample. Table 1 provides the spectral resolution achieved and spectral range covered with these instruments and the number of stars observed. 

Our EBASIM observations were taken between June 2012 and August 2013. The spectra were manually reduced following standard procedures, employing the tasks within the echelle package in IRAF\footnote{IRAF is distributed by the National Optical Astronomy Observatories, which are operated by the Association of Universities for Research in Astronomy, Inc., under cooperative agreement with the National Science Foundation.}. The procedure included typical corrections such as bias level, flat-fielding, scattered light contribution and blaze-shape removal, order extraction, wavelength calibration, merge of individual orders, continuum normalization and cosmic rays removal. Reduced spectra were corrected for radial velocity shifts with the \textit{dopcor} task. Radial velocities were measured cross-correlating our program stars with standard stars using the \textit{fxcor} task.  

Archival data are already reduced with pipelines designed for each instrument. In general, it was only necessary to normalize the spectra and applied corrections for cosmic rays and for radial velocity to these data. Multiple spectra, for a given star and instrument, were combined using the \textit{scombine} task.

\section{Data analysis}
\subsection{Fundamental parameters}
Fundamental stellar parameters such as effective temperature ($T_{\mathrm{eff}}$), surface gravity ($\log g$), metallicity ([Fe/H]) and microturbulent velocity ($\xi_{t}$) were derived homogeneously using the FUNDPAR program\footnote{Available at http://icate-conicet.gob.ar/saffe/fundpar/} \citep{Saffe2011}. This code uses ATLAS9 \citep{Kurucz1993}, LTE plane-parallel model atmosphere with the NEWODF opacities \citep{Castelli2003} and solar-scaled abundances from \citet{Grevesse1998} and the 2009 version of the MOOG program \citep{Sneden1973}. 

Basically, atmospheric parameters are calculated from the equivalent widths (EWs) of iron lines (\ion{Fe}{I} and \ion{Fe}{II}) by requiring excitation and ionization equilibrium and the
independence between abundances and EWs. The iron line list (72 of \ion{Fe}{I} and 12 of \ion{Fe}{II}) as well as the atomic parameters (excitation potential and $\log gf$) were compiled from \citet{Silva2011}. Lines giving abundances departing $\pm 3\sigma$ from the average were removed, and the fundamental parameters were re-calculated. Atmospheric models were computed including convection and overshooting. 

 \begin{figure}
   \centering
   \includegraphics[width=.47\textwidth]{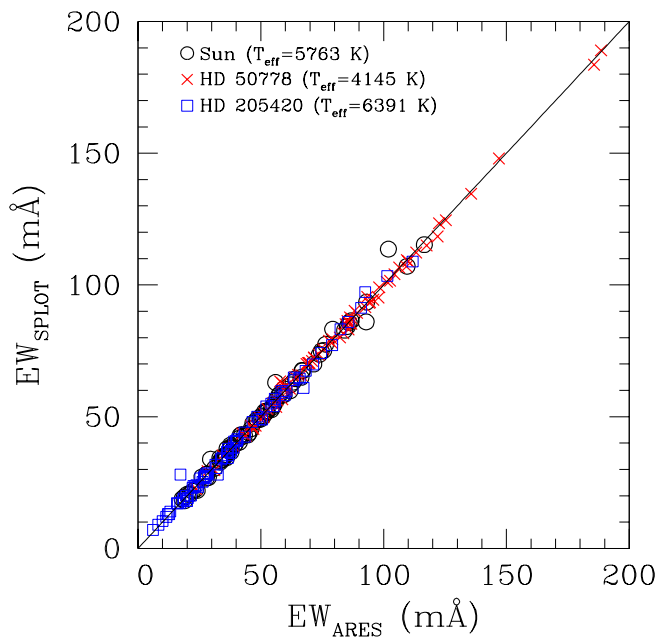}
   \caption{Comparison between EWs measured by ARES and with the IRAF-splot task. Red crosses, blue squares, and black circles correspond to the \object{HD 50778}, \object{HD 205420}, and the Sun, respectively. The continuos black-line represents the identity function.}
              \label{FigGam}%
    \end{figure}

The EWs were automatically measured using the ARES code \citep{Sousa2007}, choosing the appropriate \textit{rejt} parameter depending on the S/N of each individual spectrum \citep{Sousa2008}. In order to check for possible differences between the EWs from ARES and those manually obtained with the IRAF-\textit{splot} task, we compared these measurements (see Figure 2) for 3 stars: HD 50778 (a cool star, $T_{\mathrm{eff}}$ = 4145 K), HD 205420 (a hot star, $T_{\mathrm{eff}}$ = 6391 K) and the Sun ($T_{\mathrm{eff}}$ = 5763 K). The mean differences between the two sets of EWs are $\langle EW_{ARES} - EW_{SPLOT} \rangle$ = 0.12 $\pm$ 1.54 m{\AA}, -0.06 $\pm$ 1.95 m{\AA}, and 0.01 $\pm$ 2.04 m{\AA} for HD 50778, HD 205420, and the Sun, respectively. Thus, ARES EWs show no significant differences with those measured manually. 

Final fundamental parameters, along with their statistical uncertainties, are listed in Table 2. Intrinsic uncertainties are based on the scatter of the individual iron abundances from each individual line, the standard deviations in the slopes of the least-squares fits of iron abundances with reduced EW, excitation, and ionization potential \citep{Gonzalez1998}. A discussion of systematic errors in the fundamental parameters and their influence in the abundance determinations are presented in the next sections. 

In order to check for possible systematic offsets in the determination of the atmospheric parameters introduced by the use of different instruments and reduction pipelines, we calculated fundamental parameters for the Sun using spectra gathered with each of the five spectrographs listed in Table 1. In addition, in Table 3 we show the parameters calculated for other stars observed with more than one instrument. No significant differences between the fundamental parameters obtained from different spectrographs are found \citep[see also][]{Santos2004}. Considering that the spectra have good S/N ($\gtrsim$ 150), for objects with observations from more than one instrument, we chose the parameters derived from the highest resolution spectrum. 

Regarding differences between plane-parallel (Kurucz-ATLAS9) and spherical models (MARCS), \citet{Carlberg2012} compared stellar parameters for 27 stars, calculated from Kurucz-ATLAS9 and MARCS models, finding, in general, a good agreement, with only a small difference in microturbulence of 0.04 km/s (spherical with respect to plane-parallel). Hence, we do not expect differences with spherical models. On the other hand, several studies show that non-LTE effects are significant for very metal-poor evolved stars with $T_{\mathrm{eff}}$ > 6000 K \citep{Mashonkina2010, Gehren2001, Lind2012}. Most of the stars in our sample have cooler temperatures and thus non-LTE effect should not compromise our results. 

\begin{longtab}
\begin{longtable}{l c c c c c c c c}
            \caption{\label{kstars} Fundamental stellar parameters and rotation velocities.}\\                      
\hline \hline  
Star 	&	$T_{\mathrm{eff}}$			&	$\log g$			&	$[Fe/H]$			&	$\xi_{t}$			&	N\ion{Fe}{I}	&	N\ion{Fe}{II}	&	Instrument	&	v$\sin i$	\\
HD/other	&	(K)			&	(cm $s^{-2}$)			&	(dex)			&	(km $s^{-1}$)			&		&		&		&	(km $s^{-1}$)	\\
\hline
\endfirsthead
\caption{continued.}\\
\hline \hline  
Star 	&	$T_{\mathrm{eff}}$			&	$\log g$			&	$[Fe/H]$			&	$\xi_{t}$			&	N\ion{Fe}{I}	&	N\ion{Fe}{II}	&	Instrument	&	v$\sin i$	\\
HD/other	&	(K)			&	(cm $s^{-2}$)			&	(dex)			&	(km $s^{-1}$)			&		&		&		&	(km $s^{-1}$)	\\
\hline
\endhead
\hline
\endfoot
\multicolumn{9}{c}{Giant stars with planets}\\ 
\hline
1502	&	5011	$\pm$	27	&	3.30	$\pm$	0.10	&	-0.01	$\pm$	0.06	&	1.18	$\pm$	0.04	&	60	&	11	&	FEROS	&	1.57	$\pm$	0.40	\\
1690	&	4406	$\pm$	32	&	2.01	$\pm$	0.07	&	-0.31	$\pm$	0.03	&	1.69	$\pm$	0.05	&	50	&	10	&	HARPS	&	3.02	$\pm$	0.45	\\
4313	&	5005	$\pm$	10	&	3.45	$\pm$	0.08	&	0.11	$\pm$	0.07	&	1.13	$\pm$	0.08	&	51	&	11	&	FEROS	&	1.91	$\pm$	0.25	\\
4732*	&	4989	$\pm$	15	&	3.46	$\pm$	0.04	&	0.15	$\pm$	0.02	&	0.85	$\pm$	0.06	&	50	&	9	&	EBASIM	&	1.67	$\pm$	1.36	\\
5608	&	4929	$\pm$	32	&   3.32	$\pm$	0.05	&	0.14	$\pm$ 0.04	    &	1.11	$\pm$	0.04	&	53	&	9	&	FEROS	&	2.09	$\pm$ 0.23		\\
5891	&	4816	$\pm$	20	&	2.67	$\pm$	0.09	&	-0.37	$\pm$	0.04	&	1.63	$\pm$	0.07	&	41	&	10	&	FEROS	&	0.63	$\pm$	0.45	\\
11977	&	5008	$\pm$	21	&	2.97	$\pm$	0.05	&	-0.19	$\pm$	0.04	&	1.41	$\pm$	0.02	&	45	&	10	&	HARPS	&	0.74	$\pm$	0.45	\\
\hline
    \end{longtable}
    \tablefoot{* Stars hosting multi-planet systems.}
     \end{longtab}

 \begin{table}
  \tiny
      \caption[]{Comparison between fundamental parameters of the same stars observed with different spectrographs.}
         \label{table:1}
     \centering
         \begin{tabular}{c c c c c}
            \hline\hline
              Star & Instrument & $T_{\mathrm{eff}}$ & $\log g$ & [Fe/H]   \\
               & &  (K) & (cm $s^{-2}$) &    \\
            \hline
Sun & HARPS  & 5763 $\pm$ 24 & 4.42 $\pm$ 0.03 & 0.01 $\pm$ 0.05     \\
 & SOPHIE & 5764 $\pm$ 49 & 4.44 $\pm$ 0.05 & -0.03 $\pm$ 0.04    \\
 & FEROS & 5739 $\pm$ 32 & 4.43 $\pm$ 0.03 & -0.03 $\pm$ 0.04     \\
 & ELODIE & 5748 $\pm$ 37 & 4.36 $\pm$ 0.05 & -0.03 $\pm$ 0.03     \\
 & EBASIM & 5735 $\pm$ 50 & 4.35 $\pm$ 0.08 & -0.08 $\pm$ 0.04 \\
\hline
\object{HD 16400} & ELODIE & 4814 $\pm$ 28 & 2.68 $\pm$ 0.10 & -0.05 $\pm$ 0.04 \\
 & HARPS & 4867 $\pm$ 33 & 2.73 $\pm$ 0.08 & 0.00 $\pm$ 0.04 \\
\hline
\object{HD 28305} & ELODIE & 4948 $\pm$ 28 & 2.93 $\pm$ 0.05 & 0.01 $\pm$ 0.03 \\
 & HARPS & 5016 $\pm$ 33 & 2.97 $\pm$ 0.04 & 0.01 $\pm$ 0.02 \\
 & FEROS & 4995 $\pm$ 33 & 2.83 $\pm$ 0.05 & 0.01 $\pm$ 0.03 \\
\hline
\object{HD 62509} & ELODIE & 4895 $\pm$ 24 & 3.08 $\pm$ 0.06 & 0.03 $\pm$ 0.06 \\
 & HARPS & 4946 $\pm$ 18 & 3.07 $\pm$ 0.05 & 0.07 $\pm$ 0.05 \\
 \hline
\object{HD 168723} & ELODIE & 4942 $\pm$ 35 & 3.24 $\pm$ 0.11 & -0.18 $\pm$ 0.05 \\
 & HARPS & 4966 $\pm$ 40 & 3.28 $\pm$ 0.07 & -0.20 $\pm$ 0.07 \\
            \hline
         \end{tabular}
        \end{table}   
        
 \begin{figure*}
   \centering
   \includegraphics[width=.33\textwidth]{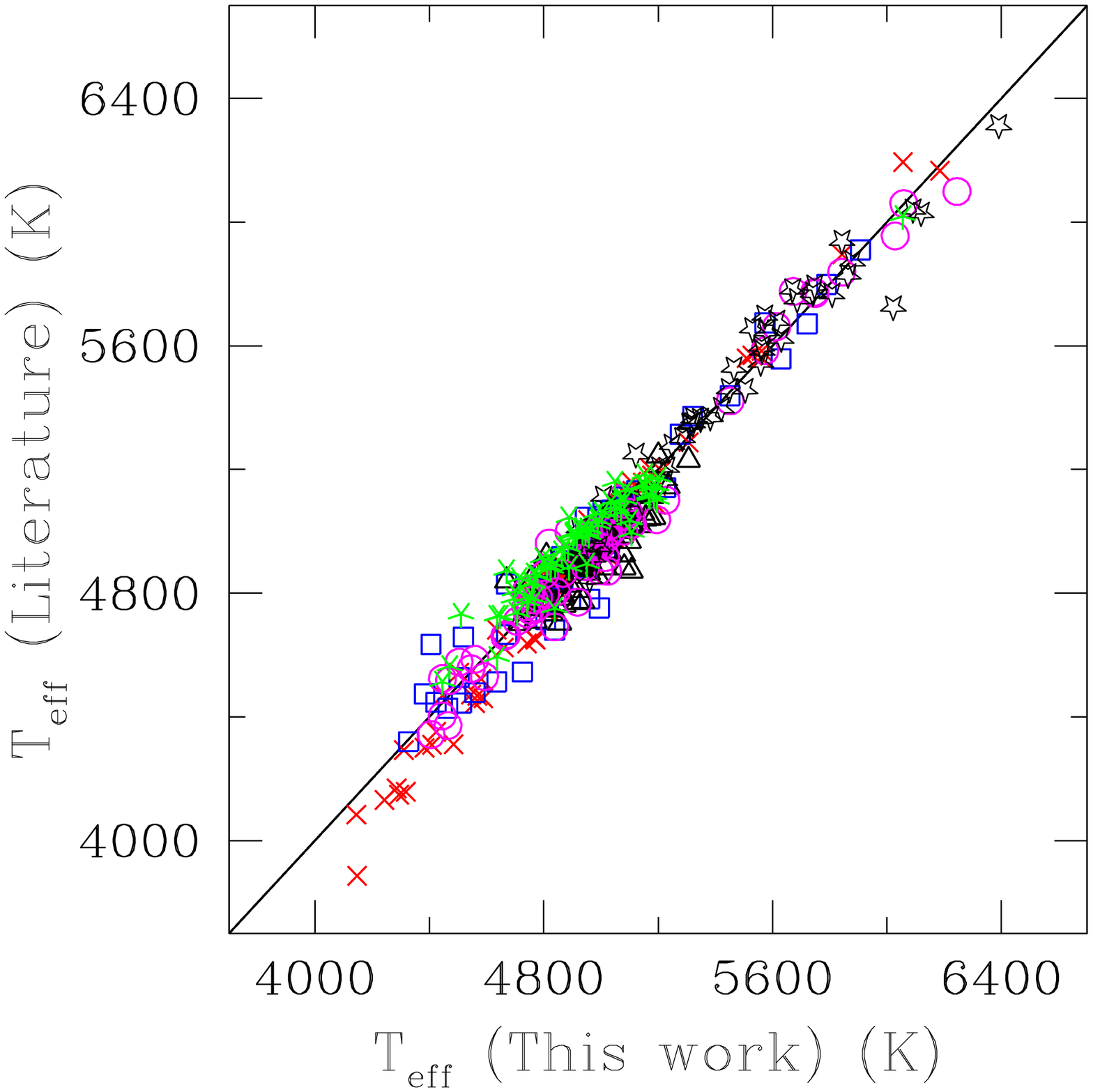}
    \includegraphics[width=.33\textwidth]{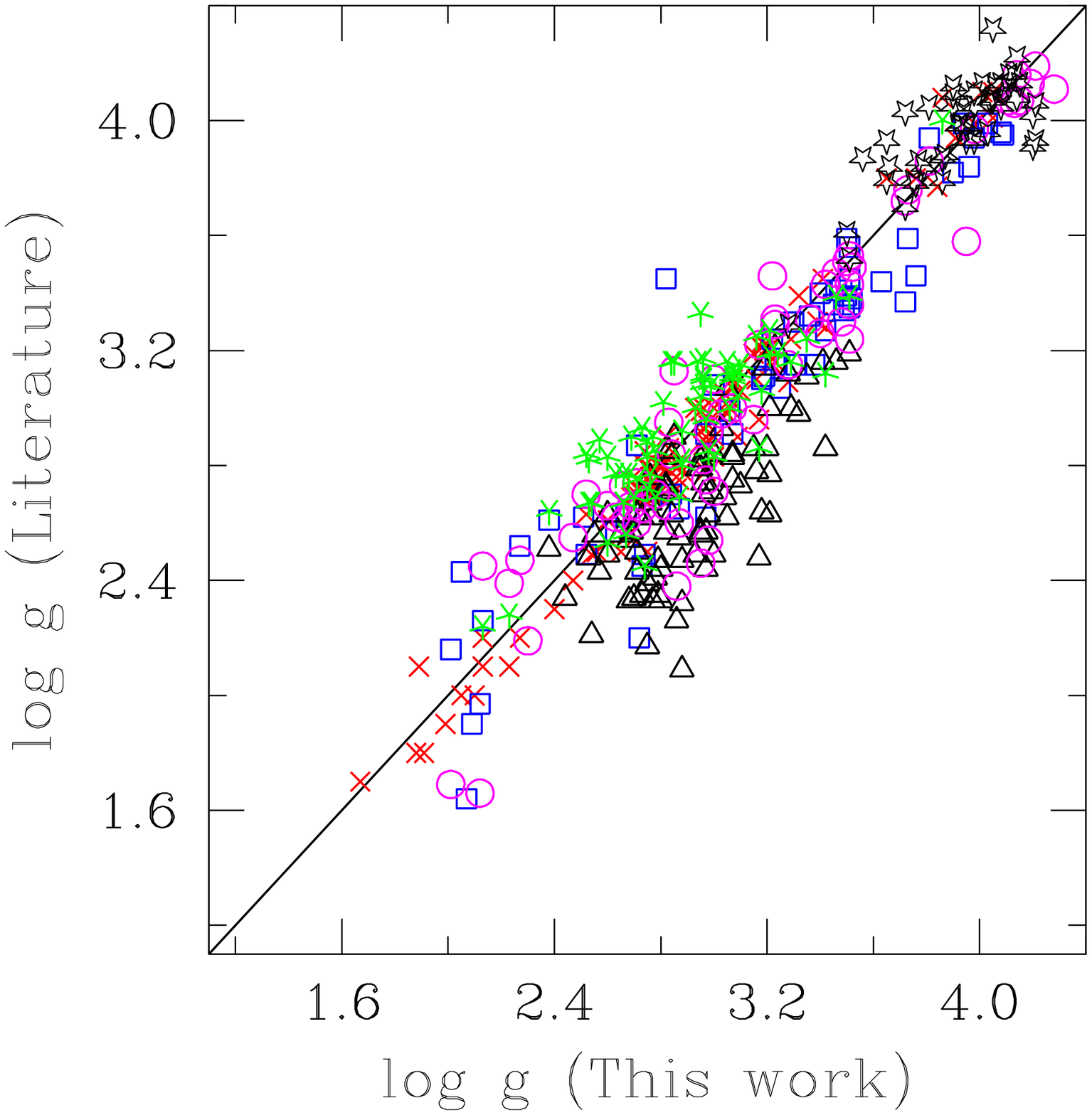}
\includegraphics[width=.33\textwidth]{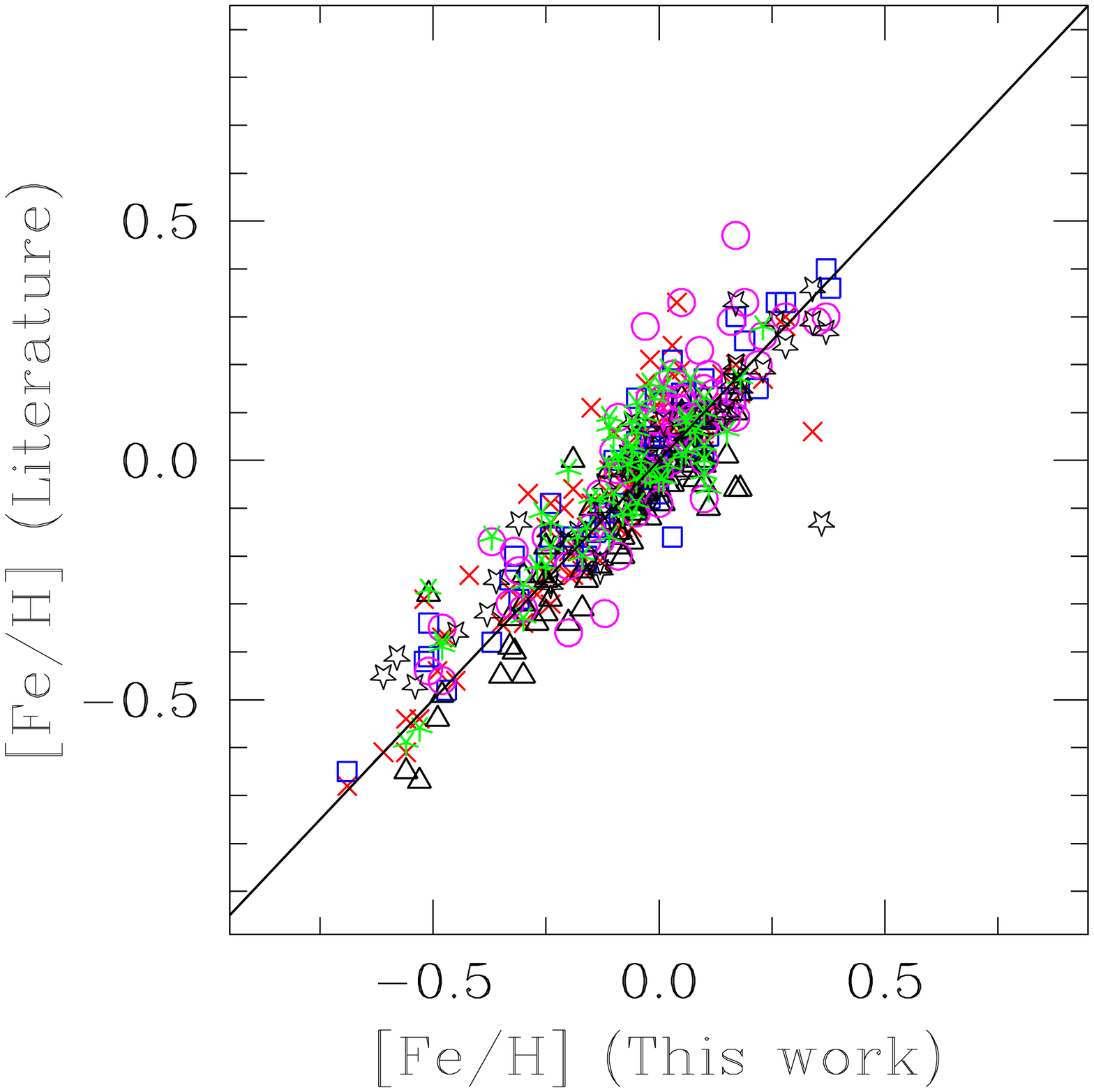}
   \caption{Comparison between the fundamental parameters obtained in this work and those calculated by other authors: effective temperature (\textit{left panel}), surface gravity (\textit{middle panel}), and metallicity (\textit{right panel}). Different symbols represent comparisons with different references: \citet{Maldonado2013} (magenta circles), \citet{Mortier2013} (blue squares), \citet{Silva2006, Silva2011} (red crosses), \citet{Takeda2008} (black triangles), \citet{Luck2007} (green asterisks), and \citet{Valenti2005} (black stars).} 
              \label{FigGam}%
    \end{figure*}

\subsubsection{Comparison with other studies} 
In order to check the consistency of our results, we compared stellar parameters obtained here with those recently published by other authors based on homogeneous spectroscopic analysis. Figure 3, shows the comparison of our effective temperatures (left panel), surface gravities (middle panel) and metallicities (right panel) with those derived by \citet[][hereafter MA13, magenta circles]{Maldonado2013}, \citet[][hereafter MO13, blue squares]{Mortier2013}, \citet[][hereafter S0611, red crosses]{Silva2006, Silva2011}, \citet[][hereafter TA08, black triangles]{Takeda2008}, \citet[][hereafter LH07, green asterisks]{Luck2007}, and \citet[][hereafter VF05, black stars]{Valenti2005}. Table 4 summarizes the results, where the mean differences (this work -- literature) and the scatter around the mean differences of the fundamental parameters are represented by $\Delta$ and $\sigma$, respectively. In general, we find good agreement with previous works. However, as it has been reported by other authors \citep[e.g.,][]{Silva2011, Takeda2008}, the most noticeable discrepancy corresponds to the surface gravity values of TA08, which are systematically lower than the values obtained in this work.

       \begin{table}
       \tiny
      \caption[]{Results of the comparison between our stellar parameters and those of Maldonado et al. (2013); Mortier et al. (2013); da Silva et al. (2006, 2011); Takeda et al. (2008); Luck \& Heiter (2007), and Valenti \& Fischer (2005). All the differences are our values minus the literature values.}
         \label{table:1}
     \centering
         \begin{tabular}{c c c c c}
            \hline\hline
               & $T_{\mathrm{eff}} \pm \sigma$ (K) & $\log g \pm \sigma$ & [Fe/H] $\pm \sigma$ & $N_{common}$  \\
               &  (K) & (cm $s^{-2}$) &  &   \\

            \hline
$\Delta$[S0611] & 22.23 $\pm$ 62.95  & 0.03 $\pm$ 0.08 & -0.04 $\pm$ 0.09 & 98   \\
\hline
$\Delta$[TA08]  & 50.08 $\pm$ 67.70 &  0.27 $\pm$ 0.19 & 0.04 $\pm$ 0.24   & 83 \\
\hline
$\Delta$[MO13]  & 7.55 $\pm$ 81.04 & 0.07 $\pm$ 0.19 & -0.03 $\pm$ 0.07  & 58  \\
\hline
$\Delta$[MA13] & 14.07 $\pm$ 74.49 & 0.05 $\pm$ 0.18 & -0.04 $\pm$ 0.10  & 61  \\
\hline
$\Delta$[LH07]  & -45.11  $\pm$ 72.24 & -0.06 $\pm$ 0.16 & -0.04 $\pm$ 0.09   & 66  \\
\hline
$\Delta$[VF05]  & -6.89 $\pm$ 73.74  & -0.05 $\pm$ 0.13 & -0.01 $\pm$ 0.10   & 47  \\
            \hline
         \end{tabular}
        \end{table}

\subsection{Photometric and evolutionary parameters} 
As an independent check, we derived photometric effective temperatures using the calibrations of \citet[][hereafter GHB09]{Gonzalez2009} and the one of \citet[][hereafter A99]{Alonso1999} for the (B -- V) and (V -- K) indices. The (B -- V) colors were extracted from the Hipparcos and Tycho Catalogues \citep{Perryman1997} and were de-reddened using the visual extinction ($A_{V}$) provided by Fr\'ed\'eric Arenou's online calculator \footnote{http://wwwhip.obspm.fr/cgi-bin/afm}, which is based on the galactic coordinates and distances \citep{Arenou1992}. Color excesses E(B -- V) were obtained from the relation with $A_{V}$\footnote{$A_{V}$=3.10 $\times$ E(B -- V)}. (V -- K) colors and K magnitudes were taken from the 2MASS catalog \citep{Cutri2003} and were de-reddened using the relationship between E(B -- V) and E(V -- K)\footnote{E(V -- K)= 0.86 $\times$ E(B -- V) \citep{Luck1995, Takeda2005}. }. 

The upper and lower panels of Figure 4 show the photometric temperatures obtained from (B -- V) and (V -- K) colors, respectively, for the A99 and GH09 calibrations, as a function of our spectroscopic $T_{\mathrm{eff}}$. The agreement is quite good for both calibrations and colors. The (B -- V) temperatures seem cooler than our spectroscopic values, with an average difference (spectroscopic -- photometric) of $\langle \Delta T_{eff} \rangle \thicksim$ 118 $\pm$ 95 K for the GHB09 calibration, and $\langle \Delta T_{eff} \rangle \thicksim$ 40 $\pm$ 137 K or the A99 calibration. The photometric temperatures based on the (V -- K) colors, also agree well with the spectroscopic $T_{\mathrm{eff}}$. The mean differences are $\langle \Delta T_{eff} \rangle \thicksim$ 35 K $\pm$ 117 K and 49 $\pm$ 136 K for GHB99 and A99, respectively. The differences between photometric and spectroscopic effective temperatures provide an estimation of the accuracy error in this parameter \citep{Sousa2011b}. The sensitivity of the chemical abundances to errors of this order in $T_{\mathrm{eff}}$ are evaluated in Section 3. 

The evolutionary parameters such as the stellar luminosities, masses, radii and ages were calculated as follows: based on V magnitudes from the Hipparcos and Tycho Catalogues \citep{Perryman1997}, revised Hipparcos parallaxes $\pi$ from \citet{van2007} and the $A_{V}$ previously calculated, we computed absolute magnitudes $M_{V}$ following the classical formula\footnote{$M_{V} = V + 5 + 5 \log \pi - A_{V}$ .}. For three giant hosts there are no Hipparcos data available. For two of them, NGC 2423-3 and NGC 4349-127, we used instead the values from the Extrasolar Planet Encyclopedia. For BD+48738 we found no astrometric data from any source. Bolometric magnitudes were computed from $M_{V}$ and bolometric corrections were calculated from the empirical formula of \citet{Alonso1999}, using the atmospheric parameters from Table 2. Finally, stellar luminosities were estimated using the usual relation\footnote{$L/L_{\mathrm{\odot}} = -0.4 (M_{bol} - M_{bol \odot}$); $M_{bol \odot}$ = 4.77 (Girardi et al. 2002).}. Uncertainties were derived using error propagation. 

\begin{figure}
   \centering
   \includegraphics[width=.47\textwidth]{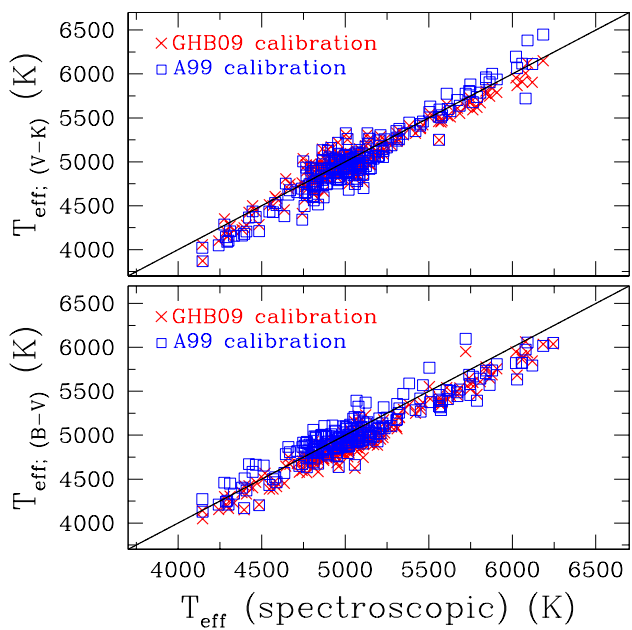}
   \caption{Comparison between the spectroscopic effective temperatures measured in this work and those calculated photometrically from the (B -- V) and (V -- K) colors.}
              \label{FigGam}%
    \end{figure}

The rest of the evolutionary parameters, such as: masses, radii, and ages were derived using L. Girardi's online code, PARAM 1.1\footnote{Version 1.1: http://stev.oapd.inaf.it/cgi-bin/param\_1.1} \citep{Silva2006}. This code requires different parameters, such as: $T_{\mathrm{eff}}$, [Fe/H], V magnitude, and  parallax. For $T_{\mathrm{eff}}$ and [Fe/H] we used the computed values given in Table 2. The other two parameters were taken from the sources indicated above. Additionally, the code calculates trigonometric gravities based on parallaxes. In Figure 5 we show a comparison between these trigonometric gravities and our spectroscopic $\log g$ values. Trigonometric $\log g$ values tend to be systematically lower than the spectroscopic determinations obtained here, by $\thicksim$ 0.16 dex (average), with a standard deviation of 0.13 dex. Several authors have reported this trend \citep[e.g.,][]{Silva1986, Silva2006, Maldonado2013} that is believed to be caused by departures from LTE of the \ion{Fe}{I} lines \citep{Bensby2003, Thevenin1999, Gratton1999}. As we shall see in the next section, differences of this order have very low impact on the determination of the chemical abundances. The resulting evolutionary and photometric parameters are listed in Table 5. Here, the luminosity, mass and radius values of BD+48738 were taken from \citet{Gettel2012}. 

 \begin{figure}
   \centering
   \includegraphics[width=.47\textwidth]{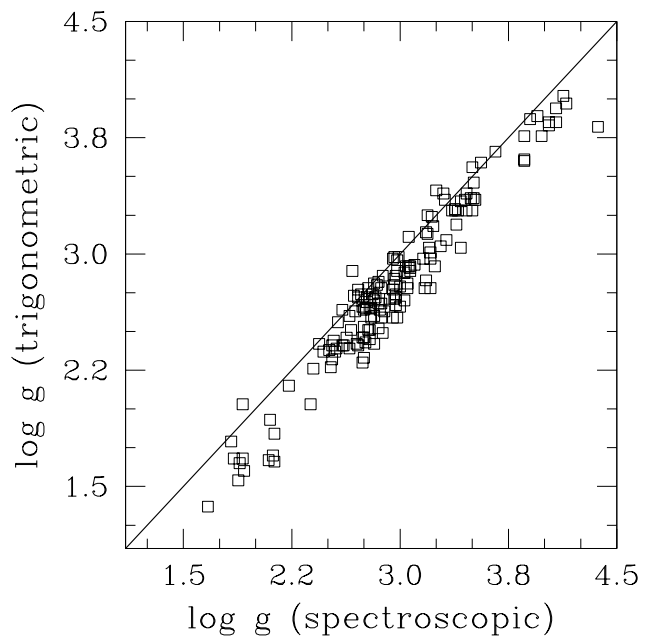}
   \caption{Comparison between the spectroscopic gravities measured in this work and the ones based on trigonometric parallaxes.}
              \label{FigGam}%
    \end{figure}
    
\begin{longtab}
\begin{longtable}{c c c c c c c}
            \caption{\label{kstars} Photometric and evolutionary parameters.}\\                      
\hline \hline  
Star &	$A_{V}$	&	$\log(L/L_{\odot})$			&	Age			&	Mass			&	Radius			&	$\log g$ (trigonometric)			\\
HD/other	&	(mag)	&				&	(Gyr)			&	($M_{\odot}$)			&	$R_{\odot}$			&	(cm $s^{-2}$)		\\
\hline
\endfirsthead
\caption{continued.}\\
\hline \hline  
Star &	$A_{V}$	&	$\log(L/L_{\odot})$			&	Age			&	Mass			&	Radius			&	$\log g$ (trigonometric)			\\
HD/other	&	(mag)	&				&	(Gyr)			&	($M_{\odot}$)			&	($R_{\odot}$)			&	(cm $s^{-2}$)			\\
\hline
\endhead
\hline
\endfoot
\multicolumn{7}{c}{Giant stars with planets}\\ 
\hline
1502	&	0.11	&	1.14	$\pm$	0.19	&	2.34	$\pm$	0.78	&	1.53	$\pm$	0.17	&	3.98	$\pm$	0.76	&	3.39	$\pm$	0.12	\\
1690	&	0.09	&	1.53	$\pm$	0.43	&	8.37	$\pm$	2.67	&	1.02	$\pm$	0.10	&	19.60	$\pm$	2.08	&	1.83	$\pm$	0.08	\\
4313	&	0.10	&	1.14	$\pm$	0.15	&	1.72	$\pm$	0.37	&	1.71	$\pm$	0.13	&	4.38	$\pm$	0.55	&	3.35	$\pm$	0.08	\\
4732	&	0.10	&	1.18	$\pm$	0.08	&	1.47	$\pm$	0.20	&	1.81	$\pm$	0.09	&	4.94	$\pm$	0.34	&	3.28	$\pm$	0.06	\\
5608	&	0.06	& 1.15	$\pm$ 0.05		&	1.72	$\pm$ 0.21		& 1.72	$\pm$ 0.07	&	4.93	$\pm$ 0.19		& 3.25		$\pm$ 0.04		\\
5891	&	0.19	&	1.23	$\pm$	0.20	&	8.30	$\pm$	2.40	&	1.03	$\pm$	0.08	&	5.80	$\pm$	0.70	&	2.89	$\pm$	0.09	\\
11977	&	0.16	&	1.84	$\pm$	0.06	&	1.00	$\pm$	0.20	&	2.13	$\pm$	0.15	&	9.94	$\pm$	0.35	&	2.74	$\pm$	0.04	\\
\hline
    \end{longtable}
    \tablefoot{* Luminosity, mass and radius were taken from Gettel et al. (2012).}
     \end{longtab}

\subsection{Chemical analysis}

In addition to Fe abundances, we have computed chemical abundances of  17 ions (Na, Mg, Al, Si, Ca, \ion{Sc}{I}, \ion{Sc}{II}, \ion{Ti}{I}, \ion{Ti}{II}, V, \ion{Cr}{I}, \ion{Cr}{II}, Mn, Co, Ni, Zn, and \ion{Ba}{II}) based on the EWs of several unblended lines, measured with ARES. The abundance computation was done using the MOOG program (abfind driver) in combination with the LTE Kurucz model atmosphere previuosly calculated with FUNDPAR. The line-list and atomic parameters for most of the elements were compiled from \citet{Neves2009} and from \citet{Chavero2010} for Zn and Ba. We emphasize that for ions such as Na, Mg, Al, \ion{Sc}{I}, \ion{Cr}{II}, Zn and \ion{Ba}{II} the line-list comprises only two or three lines and therefore the conclusions involving these species should be taken with caution. In the case of the EBASIM spectra, we did not measure Zn abundances because of the spectral coverage of this instrument. 

The calculated abundances, relative to the solar values from \citet{Anders1989}, and the dispersions around the mean are listed in Table 6 and Table 7. Similarly to the comparison we made for the fundamental parameters in the Section 3.1, we also compared the solar chemical abundances based on spectra taken with different instruments. We found no significant differences for most of the species ($\sigma$ $\lesssim$ 0.04 dex), which agree with the conclusions of \citet{Gilli2006} whom analyzed 8 stars observed with the FEROS, UVES, CORALIE, and SARG spectrographs. However, we found that Zn shows slightly larger differences ($\sigma$ $\sim$ 0.09 dex).

\begin{longtab}
 \tiny
\begin{longtable}{c r r r r r r r r r}
            \caption{\label{kstars} Derived abundances of Na, Mg, Al, Si, Ca, \ion{Sc}{I}, \ion{Sc}{II}, \ion{Ti}{I}, and \ion{Ti}{II}.}\\                      
\hline \hline  
HD/other & $[Na/H]$ &	$[Mg/H]$  &	 $[Al/H]$	&	$[Si/H]$ &	$[Ca/H]$ &	$[\ion{Sc}{I}/H]$		&	$[\ion{Sc}{II}/H]$	&	$[\ion{Ti}{I}/H]$	&	$[\ion{Ti}{II}/H]$ \\
\hline
\endfirsthead
\caption{continued.}\\
\hline \hline  
HD/other & $[Na/H]$ &	$[Mg/H]$  &	 $[Al/H]$	&	$[Si/H]$ &	$[Ca/H]$ &	$[\ion{Sc}{I}/H]$		&	$[\ion{Sc}{II}/H]$	&	$[\ion{Ti}{I}/H]$	&	$[\ion{Ti}{II}/H]$ \\
\hline
\endhead
\hline
\endfoot
\hline
\multicolumn{10}{c}{Giant stars with planets}\\ 													\hline																							
1502	&	0.08	$\pm$	0.07	&	0.06	$\pm$	0.11	&	0.03	$\pm$	0.04	&	0.06	$\pm$	0.08	&	-0.01	$\pm$	0.07	&	-0.02	$\pm$	0.14	&	-0.03	$\pm$	0.05	&	0.03	$\pm$	0.07	&	0.01	$\pm$	0.09	\\
1690	&	0.00	$\pm$	0.15	&	-0.09	$\pm$	0.07	&	0.02	$\pm$	0.08	&	-0.10	$\pm$	0.10	&	-0.21	$\pm$	0.08	&	-0.18	$\pm$	0.15	&	-0.31	$\pm$	0.14	&	-0.10	$\pm$	0.08	&	-0.14	$\pm$	0.12	\\
4313	&	0.17	$\pm$	0.06	&	0.13	$\pm$	0.10	&	0.12	$\pm$	0.03	&	0.17	$\pm$	0.08	&	0.07	$\pm$	0.07	&	0.13	$\pm$	0.16	&	0.14	$\pm$	0.06	&	0.14	$\pm$	0.07	&	0.14	$\pm$	0.11	\\
4732	&	0.17	$\pm$	0.08	&	0.12	$\pm$	0.12	&	0.10	$\pm$	0.10	&	0.10	$\pm$	0.08	&	0.03	$\pm$	0.06	&	-0.10	$\pm$	0.15	&	0.00	$\pm$	0.09	&	0.10	$\pm$	0.11	&	0.11	$\pm$	0.15	\\
5608	&	0.27	$\pm$	0.08	&	0.18	$\pm$	0.07	&	0.18	$\pm$	0.06	&	0.2	$\pm$	0.08	&	0.11	$\pm$	0.07	&	0.19	$\pm$	0.19	&	0.17	$\pm$	0.10	&	0.21	$\pm$	0.08	&	0.13	$\pm$	0.10	\\
5891	&	-0.19	$\pm$	0.10	&	-0.03	$\pm$	0.01	&	-0.03	$\pm$	0.03	&	-0.05	$\pm$	0.08	&	-0.24	$\pm$	0.07	&	-0.28	$\pm$	0.15	&	-0.20	$\pm$	0.09	&	-0.13	$\pm$	0.09	&	-0.19	$\pm$	0.12	\\
11977	&	-0.07	$\pm$	0.15	&	-0.12	$\pm$	0.11	&	-0.08	$\pm$	0.02	&	-0.06	$\pm$	0.05	&	-0.12	$\pm$	0.05	&	-0.18	$\pm$	0.11	&	-0.11	$\pm$	0.06	&	-0.12	$\pm$	0.07	&	-0.04	$\pm$	0.07	\\
\end{longtable}
     \end{longtab}  
  
 \begin{longtab}
 \tiny
\begin{longtable}{c r r r r r r r r}
            \caption{\label{kstars} Derived abundances of V, \ion{Cr}{I}, \ion{Cr}{II}, Mn, Co, Ni, Zn, and \ion{Ba}{II}.}\\                      
\hline \hline  
HD/other &	$[V/H]$ &	$[\ion{Cr}{I}/H]$ &$[\ion{Cr}{II}/H]$	&	$[Mn/H]$ &	$[Co/H]$ &	$[Ni/H]$	&	$[Zn/H]$	&	$[\ion{Ba}{II}/H]$	\\
\hline
\endfirsthead
\caption{continued.}\\
\hline \hline  
HD/other &	$[V/H]$ &	$[\ion{Cr}{I}/H]$ &$[\ion{Cr}{II}/H]$	&	$[Mn/H]$ &	$[Co/H]$ &	$[Ni/H]$	&	$[Zn/H]$	&	$[\ion{Ba}{II}/H]$	\\
\hline
\endhead
\hline
\endfoot
\hline
\multicolumn{9}{c}{Giant stars with planets}\\ 													\hline																				
1502	&	0.13	$\pm$	0.07	&	-0.03	$\pm$	0.07	&	-0.11	$\pm$	0.04	&	0.18	$\pm$	0.02	&	0.10	$\pm$	0.11	&	-0.02	$\pm$	0.07	&	0.16	$\pm$	0.02	&	0.18	$\pm$	0.08	\\
1690	&	0.22	$\pm$	0.14	&	-0.29	$\pm$	0.08	&	-0.46	$\pm$	0.13	&	-0.12	$\pm$	0.15	&	-0.12	$\pm$	0.10	&	-0.32	$\pm$	0.09	&	-0.26	$\pm$	0.18	&	-0.27	$\pm$	0.05	\\
4313	&	0.32	$\pm$	0.10	&	0.08	$\pm$	0.08	&	0.06	$\pm$	0.01	&	0.39	$\pm$	0.06	&	0.24	$\pm$	0.11	&	0.13	$\pm$	0.07	&	0.29	$\pm$	0.12	&	0.22	$\pm$	0.05	\\
4732	&	0.10	$\pm$	0.10	&	0.01	$\pm$	0.09	&	0.08	$\pm$	0.10	&	0.06	$\pm$	0.11	&	-0.02	$\pm$	0.20	&	0.05	$\pm$	0.13	&	--		&	0.41	$\pm$	0.16	\\
5608	&	0.42	$\pm$	0.11	&	0.13	$\pm$	0.10	&	0.12	$\pm$	0.03	&	0.60	$\pm$	0.08	&	0.32	$\pm$	0.11	&	0.19	$\pm$	0.08	&	0.14	$\pm$	0.13	&	0.67	$\pm$	0.07	\\
5891	&	-0.09	$\pm$	0.06	&	-0.43	$\pm$	0.08	&	-0.49	$\pm$	0.02	&	-0.39	$\pm$	0.08	&	-0.16	$\pm$	0.09	&	-0.33	$\pm$	0.06	&	0.10	$\pm$	0.15	&	-0.43	$\pm$	0.07	\\
11977	&	-0.08	$\pm$	0.05	&	-0.18	$\pm$	0.08	&	-0.22	$\pm$	0.13	&	-0.17	$\pm$	0.04	&	-0.11	$\pm$	0.07	&	-0.19	$\pm$	0.05	&	-0.30	$\pm$	0.06	&	0.22	$\pm$	0.04	\\
\end{longtable}
     \end{longtab}

\subsubsection{Abundance uncertainties and comparison with other studies}

Uncertainties in the parameters used to build the atmospheric models may introduce errors in the abundance calculations. Table 8 shows the abundance sensitivity due to variations of 100 K in $T_{\mathrm{eff}}$, 0.2 dex in $\log g$, 0.2 dex in [Fe/H], and 0.1 km$s^{-1}$ in microturbulence, for \object{HD 114613} (an average subgiant) and  \object{HD 219449} (an average giant).

For the subgiant star, species such as: \ion{Ti}{II}, \ion{Sc}{II}, and \ion{Cr}{II} are specially sensitives to changes in $\log g$ ($\Delta[X/H]\sim$ 0.08 dex) whereas other species, such as \ion{Ti}{I}, V and Sc are more sensitive to variations in $T_{\mathrm{eff}}$ ($\Delta[X/H]\sim$ 0.1 dex). The giant star follows the same trends as for the subgiant, but the sensitivity $\Delta[X/H]$ due to perturbations in $T_{\mathrm{eff}}$ are larger ($\sim$0.15 dex). In both cases (giant and subgiant), we also found a notable sensitivity in the Barium abundance due to the change in microturbulence.

Hidden blends or poor continuum location are other sources of errors when abundances are obtained from EWs measurements. However, as we usually have more than 3 lines for each element, the dispersion of the average values provides an indication of the errors introduced by the EWs measurements. The last column of Table 8 indicates the total error in the abundances of all elements. These errors were determined adding in quadrature the dispersion of each abundance from the mean value and abundance variations due to changes in the atmospheric parameters. These errors, depicted in Figure 6 for each element, are about 0.10 dex for the subgiant and of 0.15 dex for the giant. 
     
      \begin{table}
      \tiny
      \caption[]{Sensitivity of the abundances to changes of 100 K in temperature, 0.2 dex in gravity, 0.2 dex in metallicity, and 0.1 km $s^{-1}$ in microturbulence.}
         \label{table:1}
     \centering
         \begin{tabular}{c c c c c c}
         
            \hline\hline
Ion	&	$\Delta T_{eff}$	&	$\Delta \log g$ & $\Delta [Fe/H]$ &	$\Delta \xi_{t}$	& $(\Sigma \sigma^{2})^{1/2}$	\\
	&	+100 K	&	+0.2	&	+0.2	&	+0.1 km $s^{-1}$	&		\\
\hline
\multicolumn{6}{c}{\object{HD 219449} - giant star} \\
\hline
Na	&	-0.09	&	0.02	&	-0.02	&	0.02	&	0.10	\\
Mg	&	-0.06	&	0.01	&	-0.03	&	0.02	&	0.07	\\
Al	&	-0.08	&	0.00	&	-0.01	&	0.01	&	0.08	\\
Si	&	0.01	&	-0.04	&	-0.06	&	0.01	&	0.06	\\
Ca	&	-0.10	&	0.02	&	-0.03	&	0.04	&	0.12	\\
\ion{Sc}{I}	&	-0.14	&	0.00	&	0.00	&	0.02	&	0.14	\\
\ion{Sc}{II}	&	0.01	&	-0.08	&	-0.09	&	0.04	&	0.11	\\
\ion{Ti}{I}	&	-0.15	&	0.00	&	-0.01	&	0.04	&	0.16	\\
\ion{Ti}{II}	&	0.01	&	-0.08	&	0.07	&	0.04	&	0.09	\\
V	&	-0.17	&	-0.01	&	-0.02	&	0.04	&	0.18	\\
\ion{Cr}{I}	&	-0.10	&	0.01	&	0.00	&	0.03	&	0.11	\\
\ion{Cr}{II}	&	0.05	&	-0.09	&	0.06	&	0.02	&	0.08	\\
Mn	&	-0.09	&	0.01	&	-0.02	&	0.05	&	0.12	\\
Co	&	-0.06	&	-0.03	&	-0.03	&	0.03	&	0.08	\\
Ni	&	-0.04	&	-0.04	&	-0.05	&	0.02	&	0.07	\\
Zn	&	0.03	&	-0.05	&	-0.08	&	0.05	&	0.11	\\
\ion{Ba}{II}	&	-0.03	&	-0.07	&	0.06	&	0.08	&	0.13	\\
\ion{Fe}{I}	&	-0.05	&	-0.01	&	-0.03	&	0.03	&	0.07	\\
\ion{Fe}{II}	&	0.08	&	-0.04	&	-0.04	&	0.02	&	0.09	\\
 \hline	
\multicolumn{6}{c}{\object{HD 114613} - subgiant star} \\
\hline
Na	&	-0.06	&	0.03	&	-0.01	&	0.01	&	0.15	\\
Mg	&	-0.05	&	0.04	&	-0.01	&	0.02	&	0.10	\\
Al	&	-0.05	&	0.01	&	0.00	&	0.01	&	0.05	\\
Si	&	-0.02	&	0.00	&	-0.02	&	0.01	&	0.08	\\
Ca	&	-0.07	&	0.04	&	0.00	&	0.03	&	0.10	\\
\ion{Sc}{I}	&	-0.08	&	0.01	&	-0.02	&	0.01	&	0.09	\\
\ion{Sc}{II}	&	0.00	&	-0.08	&	-0.05	&	0.03	&	0.12	\\
\ion{Ti}{I}	&	-0.10	&	0.01	&	-0.01	&	0.02	&	0.11	\\
\ion{Ti}{II}	&	0.00	&	-0.08	&	-0.05	&	0.03	&	0.12	\\
V	&	-0.11	&	0.01	&	-0.02	&	0.01	&	0.11	\\
\ion{Cr}{I}	&	-0.07	&	0.02	&	-0.01	&	0.03	&	0.08	\\
\ion{Cr}{II}	&	0.03	&	-0.07	&	-0.02	&	0.04	&	0.14	\\
Mn 	&	-0.07	&	0.03	&	-0.01	&	0.04	&	0.11	\\
Co 	&	-0.07	&	0.00	&	-0.02	&	0.02	&	0.10	\\
Ni	&	-0.06	&	0.00	&	-0.02	&	0.02	&	0.07	\\
Zn	&	-0.02	&	0.00	&	-0.05	&	0.05	&	0.09	\\
\ion{Ba}{II}	&	-0.03	&	-0.04	&	-0.06	&	0.08	&	0.14	\\
\ion{Fe}{I}	&	-0.07	&	0.01	&	-0.01	&	0.02	&	0.08	\\
\ion{Fe}{II}	&	0.04	&	-0.08	&	-0.03	&	0.04	&	0.11	\\

 \hline  
        
         \end{tabular}
        \end{table}

 \begin{figure}
   \centering
   \includegraphics[width=.47\textwidth]{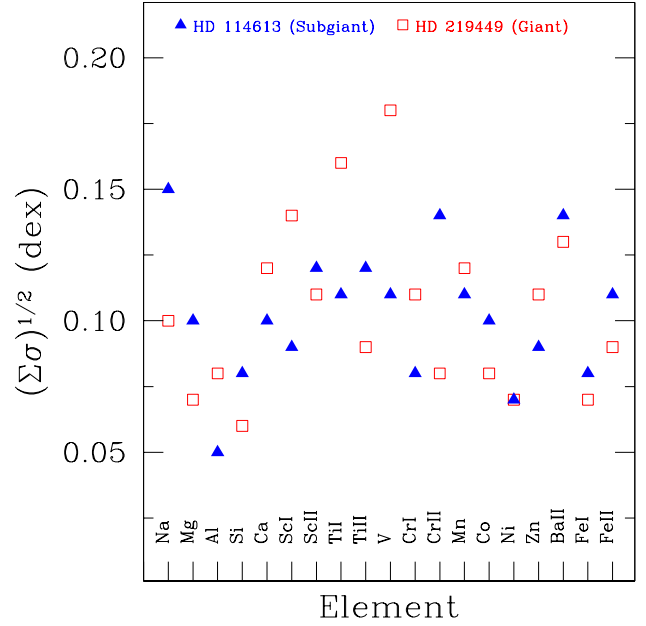}
   \caption{Total error in the abundances for each element. The empty red squares indicate the errors for a typical giant star and the blue filled triangles represent the error for a typical subgiant.}
              \label{FigGam}%
    \end{figure}

 \begin{figure}
   \centering
   \includegraphics[width=.54\textwidth]{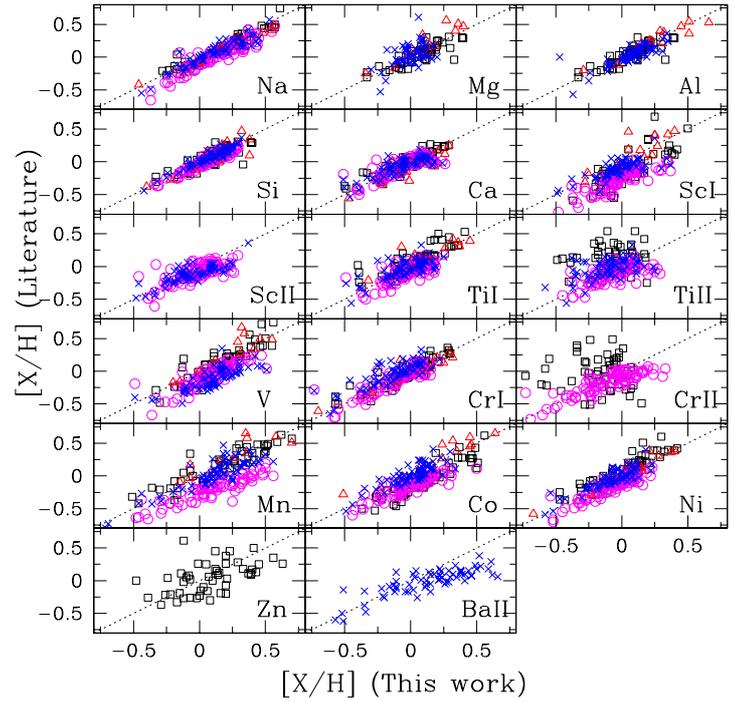}
   \caption{Comparison between the chemical abundances obtained in this work and those measured by other authors: \citet{Maldonado2013} (black squares), \citet{Gilli2006} (red triangles), \citet{Takeda2008} (magenta circles), and \citet{Luck2007} (blue crosses).}
              \label{FigGam}%
    \end{figure}

To check the reliability of our chemical abundances, in Figure 7 we made a comparison with abundances reported by MA13, TA08, LH07 and, \citet[][hereafter GI06]{Gilli2006}, using stars in common. In general, our [X/H] values agree with the abundances of other studies. For Na, Al, Si, Ca, and Ni the agreement is particularly good, being the mean differences (this work -- literature) lower than $\pm$0.08 dex. However, the Na and Ni abundances from TA08 are systematically lower. For Mg our abundances are, on average, slightly lower than those of GI06 and LH07 but agree quite well with the values of MA13. The \ion{Sc}{II} abundances obtained by LH07 and TA08 seem to be slightly shifted towards lower values. In the case of \ion{Ti}{I}, our values agree reasonably well with those of LH07, GI06, and TA08. However, the values of MA13 are higher than ours. For \ion{Cr}{I} the agreement is good with all authors except for LH07 whose abundances are marginally higher. MA13 and TA08 show systematic underabundances of Co while the values of LH07 and GI06 are slightly overabundant. For \ion{Cr}{II} our values are lower than those of MA13, but higher than those of TA08.  

The results for \ion{Sc}{I}, \ion{Ti}{II}, V, Mn, Zn, and \ion{Ba}{II} show the largest dispersions, probably due to the small number of lines used to measure these ions. For Mn, the agreement with the results of MA13 and LH07 is still relatively good, although the values from TA13 are systematically lower. The abundances of \ion{Sc}{I} obtained by TA08 and MA13 are clearly lower than our values, but our determinations agree better with the results of LH07. The V values given by TA08 and LH07 are systematically lower than our measurements, which agree better with the other studies. Despite the large scattering observed for Zn, the mean differences with the study of MA13 are essentially zero. Finally, the Ba values from LH07 seem to be smaller than those derived here, specially for larger Ba values.

 \subsection{Galaxy population membership}
 To stablish the Galaxy population membership of the stars in our sample, we calculated galactic space-velocity components (U,V,W) based on Hipparcos astrometry \citep{van2007} and radial velocities from the previous spectra cross-correlation. For the UVW computation, we developed an IRAF script following the procedure of Johnson \& Soderblom (1987). Velocities are referred to the local standard of rest (LSR), considering a solar motion of $(U,V,W)_{\odot}$=(-10.00, +5.25, +7.17) \citep{Dehnen1998}. All the velocities are listed in the last four columns of Table 9. Using these space-velocity components and following the criteria of \citet{Reddy2006}, a star can be considered as belonging to the thin disk, thick disk or halo if the probabilities $P_{\mathrm{thin}}$, $P_{\mathrm{thick}}$ or $P_{\mathrm{halo}}$  are greater than 75\%. Otherwise, if the probabilities are lower than this value, stars are considered as transition objects, either thin/thick or thick/halo stars. Figure 8 shows the Toomre diagram of our sample and the second column of Table 9 lists the population group\footnote{For the star BD+48738 no galactic space-velocity components were derived because of the lack of astrometry  and hence this star remains unclassified.}.
  
  \begin{figure}
   \centering
   \includegraphics[width=.47\textwidth]{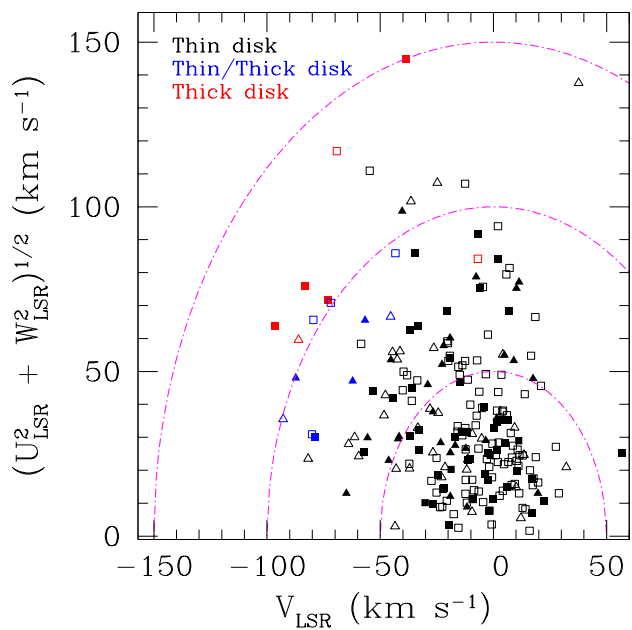}
   \caption{Toomre diagram for our complete stellar sample. Giant stars are indicated with squares and subgiants with triangles. In both cases filled symbols correspond to stars with planets. Dotted lines indicate constant total velocities, $V_{tot}=(U_{LSR}^{2}+V_{LSR}^{2}+W_{LSR}^{2})^{1/2}$, in steps of 50 $km s^{-1}$. Thin disk stars are marked in black, thick disk stars in red, and transition stars in blue.}
              \label{FigGam}%
    \end{figure}
    
 This analysis allowed us to estimate that the 93.2\% of the stars in our sample are from the thin disk, 4\% are thin/thick disk stars, and only 2\% are thick disk stars. For the 157 stars classified as giants: 147 are thin disk stars (50 GWP), 4 are from the thick disk (all GWP), and 5 transition stars (1 GWP). For the 66 stars classified as subgiants, 60 belong to the thin disk (27 SGWP), 1 is a thick disk star, and 5 are thin/thick disk objects (3 SGWP). Hence, we do not find any significant differences between the distributions of stars with planets and without planets. 
 
\begin{longtab}
\begin{longtable}{c c r r r r}
            \caption{\label{kstars} Galaxy population membership, radial velocities and galactic space-velocity components.}\\                      
\hline \hline  
Star 	&	Classification	& VR & $U_{LSR}$ & $V_{LSR}$ &	$W_{LSR}$ \\
HD/other &		&	(km $s^{-1}$)	&	(km $s^{-1}$) &	(km $s^{-1}$) &	(km $s^{-1}$)	\\
\hline
\endfirsthead
\caption{continued.}\\
\hline \hline  
Star 	&	Classification	& VR & $U_{LSR}$ & $V_{LSR}$ &	$W_{LSR}$ \\
HD/other &		&	(km $s^{-1}$)	&	(km $s^{-1}$) &	(km $s^{-1}$) &	(km $s^{-1}$)	\\
\hline
\endhead
\hline
\endfoot
\multicolumn{6}{c}{Giant stars with planets}\\ 
\hline												
1502	&	Thin	& -9.94	$\pm$	0.06	&	30.36	$\pm$	5.13	&	-36.68	$\pm$	4.28	&	-1.46	$\pm$	1.93	\\
1690	&	Thin	&	18.49	$\pm$	0.05	&	10.31	$\pm$	8.65	&	6.12	$\pm$	2.88	&	-10.72	$\pm$	0.54	\\
4313	&	Thin	&	14.40	$\pm$	0.08	&	-6.68	$\pm$	0.43	&	17.13	$\pm$	0.59	&	-2.06	$\pm$	0.33	\\
4732	&	Thin	&	26.18	$\pm$	0.15	&	-0.03	$\pm$	0.39	&	-18.67	$\pm$	0.93	&	-20.24	$\pm$	0.16	\\
5608	&	Thin	&	-23.24	$\pm$	0.15	&	-18.57	$\pm$	0.13	&	-24.65	$\pm$	0.32	&	1.83	$\pm$	0.38	\\
5891	&	Thick	&	-96.80	$\pm$	0.09	&	-66.99	$\pm$	4.74	&	-83.02	$\pm$	9.29	&	35.98	$\pm$	11.15	\\
11977	&	Thin	&	-16.20	$\pm$	0.02	&	26.48	$\pm$	0.35	&	11.20	$\pm$	0.06	&	11.81	$\pm$	0.09	\\
\end{longtable}
        \tablefoottext{a}{No galactic space-velocity components were derived due to the lack of astrometry and hence this star remains unclassified.}
     \end{longtab}   

   \begin{figure}
   \centering
   \includegraphics[width=.47\textwidth]{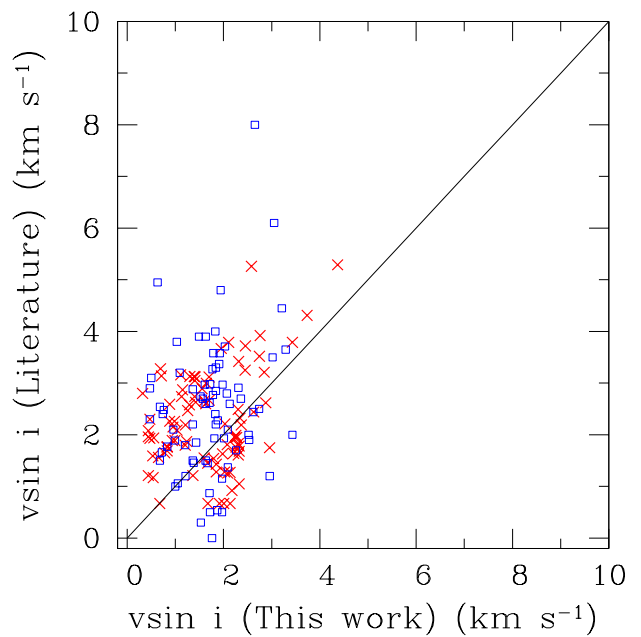}
   \caption{Comparison between the projected stellar rotational velocities derived in this work with the literature values. Crosses indicate the 89 stars in common with \citet{Takeda2008} and the empty squares represent the 74 stars with v$\sin i$ values from the exoplanets.org and exoplanet.eu databases.}
              \label{FigGam}%
    \end{figure}

  \begin{figure*}
   \centering
   \includegraphics[width=.47\textwidth]{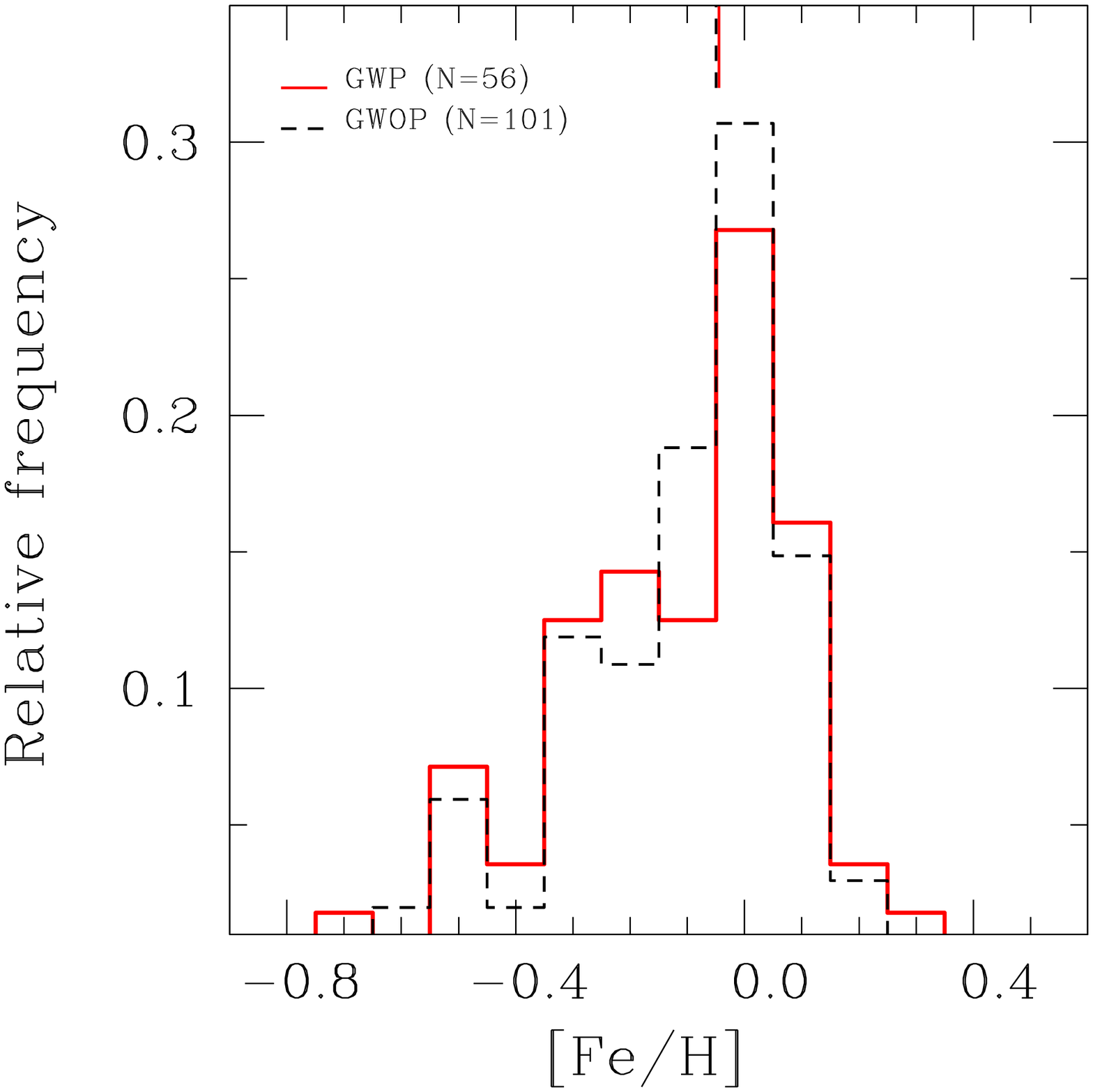}
   \includegraphics[width=.47\textwidth]{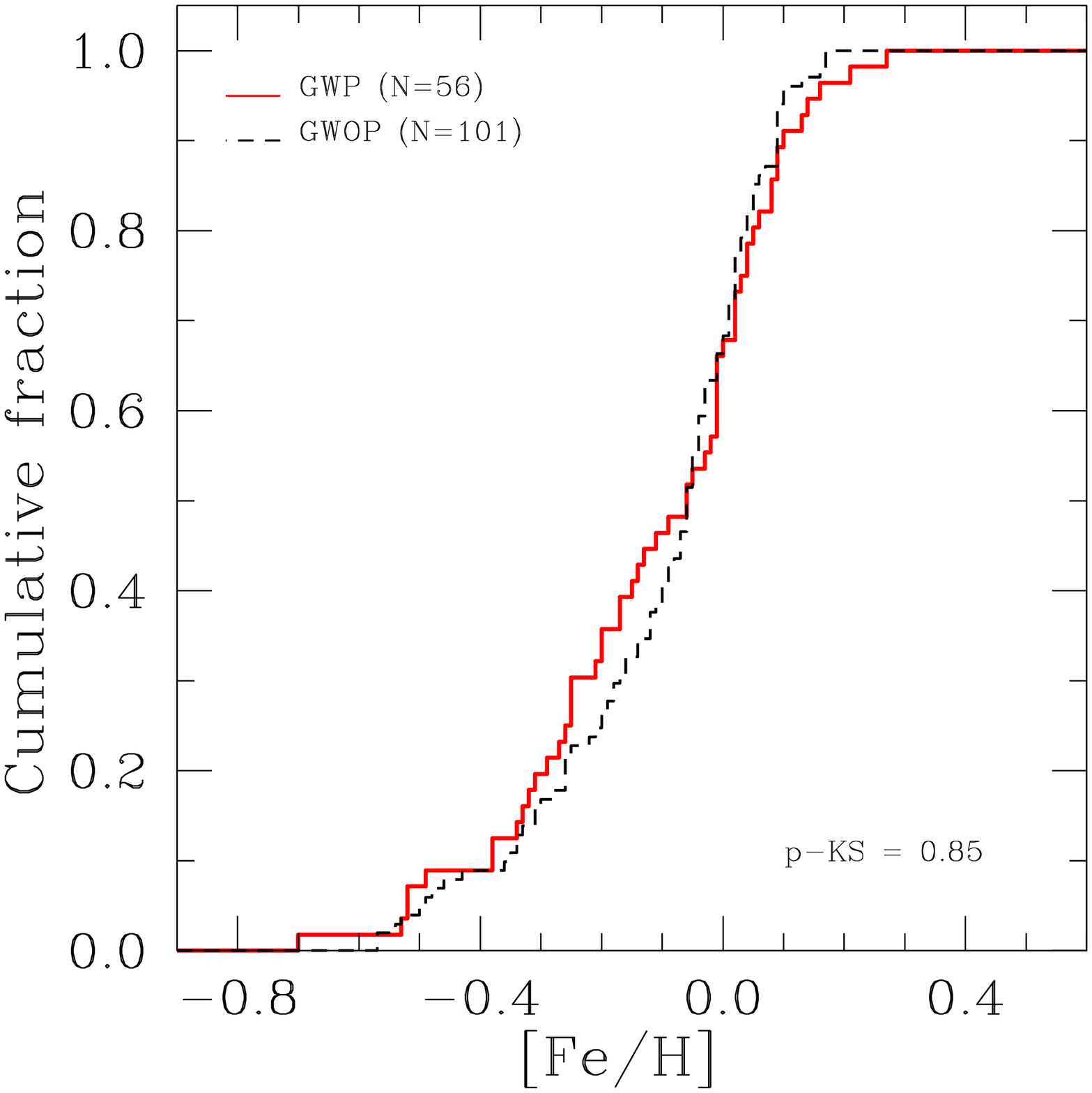}
   \caption{Normalized metallicity distribution (\textit{left panel}) and histogram of cumulative frequencies (\textit{right panel}) for the GWP sample (red continuous line) in comparison with the GWOP sample (black dashed line). Median values of the distributions are indicated with vertical lines. The Kolmogorov-Smirnov test shows that the probability that these samples derive from the same population is $\sim$ 0.85.}
              \label{FigGam}%
    \end{figure*} 
 
\subsection{Stellar rotation} 
In order to give a complete set of stellar parameters, we also measured the projected rotational velocities (v$\sin i$) of the evolved stars studied here. For this purpose, we developed an IRAF task based on the method of \citet{Fekel1997}, which automatically computes the v$\sin i$ from the widths of the spectral lines. Basically, given the stellar spectrum, as a first step, the script employs the \textit{splot} task to measure the \textit{full width at half maximum} (FWHM) of 13 relatively isolated iron lines located at 5778.45, 6027.05, 6151.62, 6173.33, 6432.68, 6452.68, 6454.99, 6455.60, 6456.38, 6469.15, 6471.66, 6733.15, and 6750.15 Å. Then, it computes the average ($FWHM_{measured}$) and the standard deviation of the measurements. We adopted the value of the dispersion as the error in the $FWHM_{measured}$. To achieve a better precision and more reliability in the derived values of v$\sin i$, our code uses twice as many spectral lines as in other determinations \citep[see, e.g.,][]{Hekker2007}. We computed the contribution to the spectral line broadening of the employed spectrograph ($FWHM_{instrument}$). To do this, we measured the FWHM of emission lines in Th-Ar calibration lamps taken with each of the spectrographs. As an independent check, we compared these values with those obtained from measurements of telluric lines. Taking this into account, then, the script calculates the intrinsic stellar broadening as: $FWHM_{intrinsic}  =  \sqrt{FWHM^{2}_{measured} - FWHM^{2}_{instrument}} $ .
As these values are in {\AA}, they are converted  to $km s^{-1}$. \citet{Gray1989}, employed a set of stars to build a calibration between the intrinsic stellar broadening in {\AA} and their total broadening ($FWHM_{total}$) in $km s^{-1}$. For each spectrograph, we selected a set of stars (usually more than 40) common to those observed by Gray and built a calibration. Knowing $T_{\mathrm{eff}}$ and $\log g$ (from Table 2), the code determines the star’s luminosity class and computes the macroturbulence velocity ($v_{macro}$) adopting the same criteria and relations applied by Hekker \& Mel\'{e}ndez (2007). Finally, combining $v_{macro}$ and the total broadening previously calculated, the program computes  v$\sin i$ as:  $\sqrt{FWHM_{total}^{2} - v^{2}_{macro}}$. 

We evaluated the  formal error in the rotational velocity adding in quadrature the error in the macroturbulence velocity and the uncertainty in the calibration derived for each spectrograph. For the error in macroturbulence, which depends on the the stellar luminosity, the script adopts the same procedure as in \citet{Hekker2007}. On the other hand, the uncertainty in the calibration is related to the total broadening ($FWHM_{total}$), for which we applied propagation of errors.

Table 2, in the last column, lists the values of  v$\sin i$ for the stars in our sample. In Figure 9, we compare our determinations with those obtained by \citet{Takeda2008} for 89 stars in common and with 74 stars taken from the exoplanets.org and the exoplanet.eu databases. We find a reasonable good agreement. The average difference is of 0.61 km$s^{-1}$ ($\sigma$ = 1.04 km$s^{-1}$) with Takeda et al. and of -0.79 (km$s^{-1}$) ($\sigma$ = 1.33 km$s^{-1}$) with the online databases.

\section{Metallicity distributions}

Several studies that analyzed the metallic content in the atmospheres of evolved stars hosting planets have been published in the last decade. The first results were based on small samples and/or the abundances were not obtained with a homogeneous technique. \citet{Schuler2005} derived the metallicity for one GWP and gathered abundances from the literature for another seven. They reported that, on average, GWP were metal-poor compared with planet hosting dwarfs. Similar results were found by \citet{Sadakane2005} analyzing 4 GWP. In 2007, Pasquini et al. studying 14 GWP (4 from the literature), concluded that in contrast to the distribution of main-sequence stars with planets, the GWP distribution does not favor high metallicity objects. Conversely, \citet{Hekker2007}, analyzing a sample of 380 GK giant stars including 20 with planets (15 from the literature), found an enhancement for GWP of 0.13 dex compared with stars without planets. \citet{Takeda2008}, with a sample of 322 giants, including 10 planet-hosts, did not find any metallicity offset. \citet{Ghezzi10b} found that the metallicity distribution of 16 GWP displays an average that is 0.17 dex more metal-poor than the sample of 117 planet-hosting dwarfs and, furthermore, that the subgiant sample is more metal-rich by 0.12 dex. \citet{Johnson2010} ruled out a flat metallicity relationship among their sample of 246 subgiants (36 with planets, including unpublished candidates). More recently, with larger samples, \citet{Mortier2013} did not find any metallicity enhancement in 71 evolved stars with planets in comparison with 733 evolved stars without planets, with metallicity values gathered from the literature. Finally, \citet{Maldonado2013} found a metallicity enhancement in 16 subgiants with planets relative to 55 without planets (50 from literature). These authors did not find evidence of a metallicity offset between giants with and without planets, analyzing  43 GWP and 67 GWOP. However, for stars with masses above 1.5 $M_{\sun}$, they reported a slight metallicity enhancement for giants with planets relative to the control sample. 

 \begin{figure*}
   \centering
   \includegraphics[width=.46\textwidth]{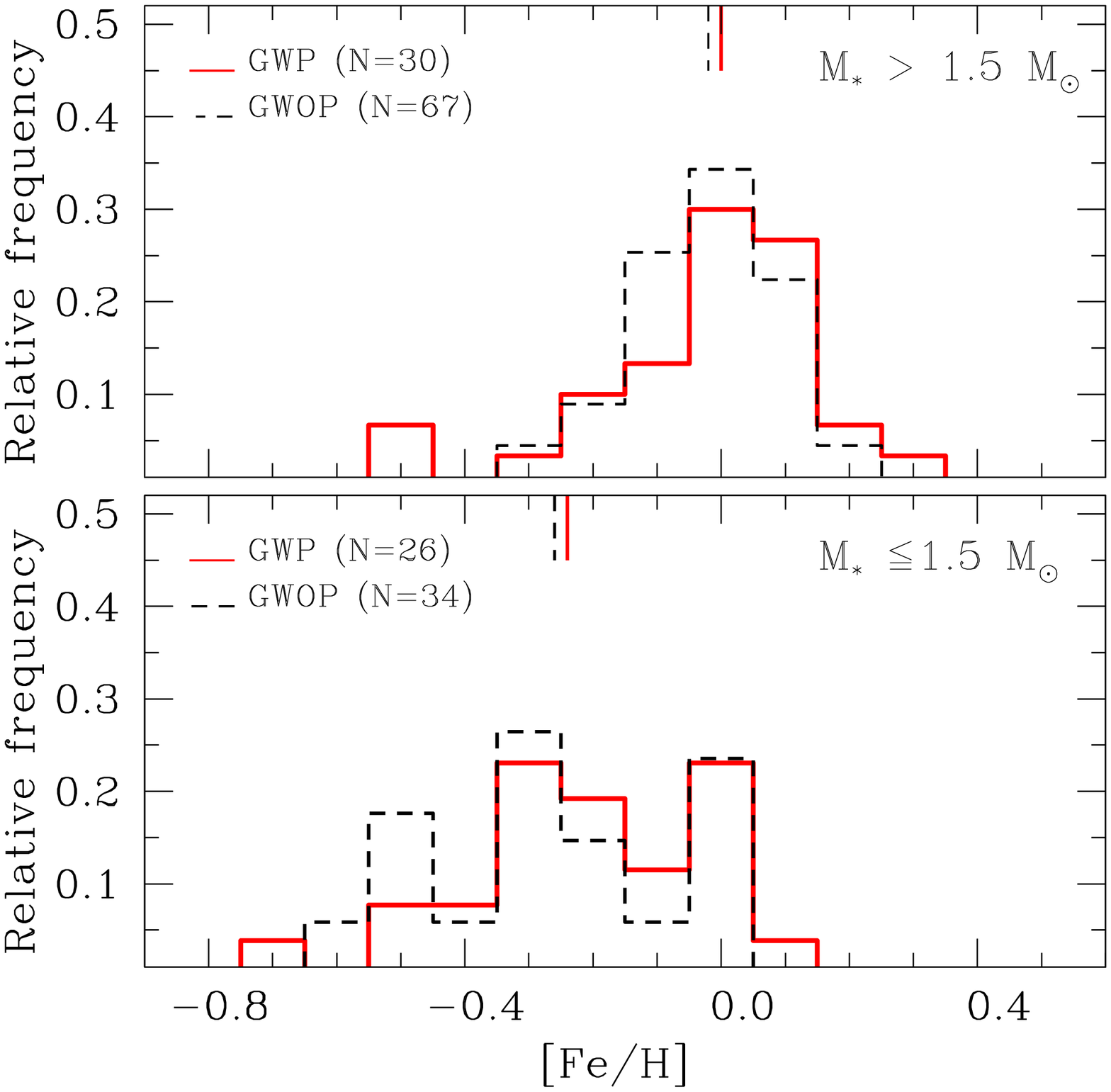}
      \includegraphics[width=.46\textwidth]{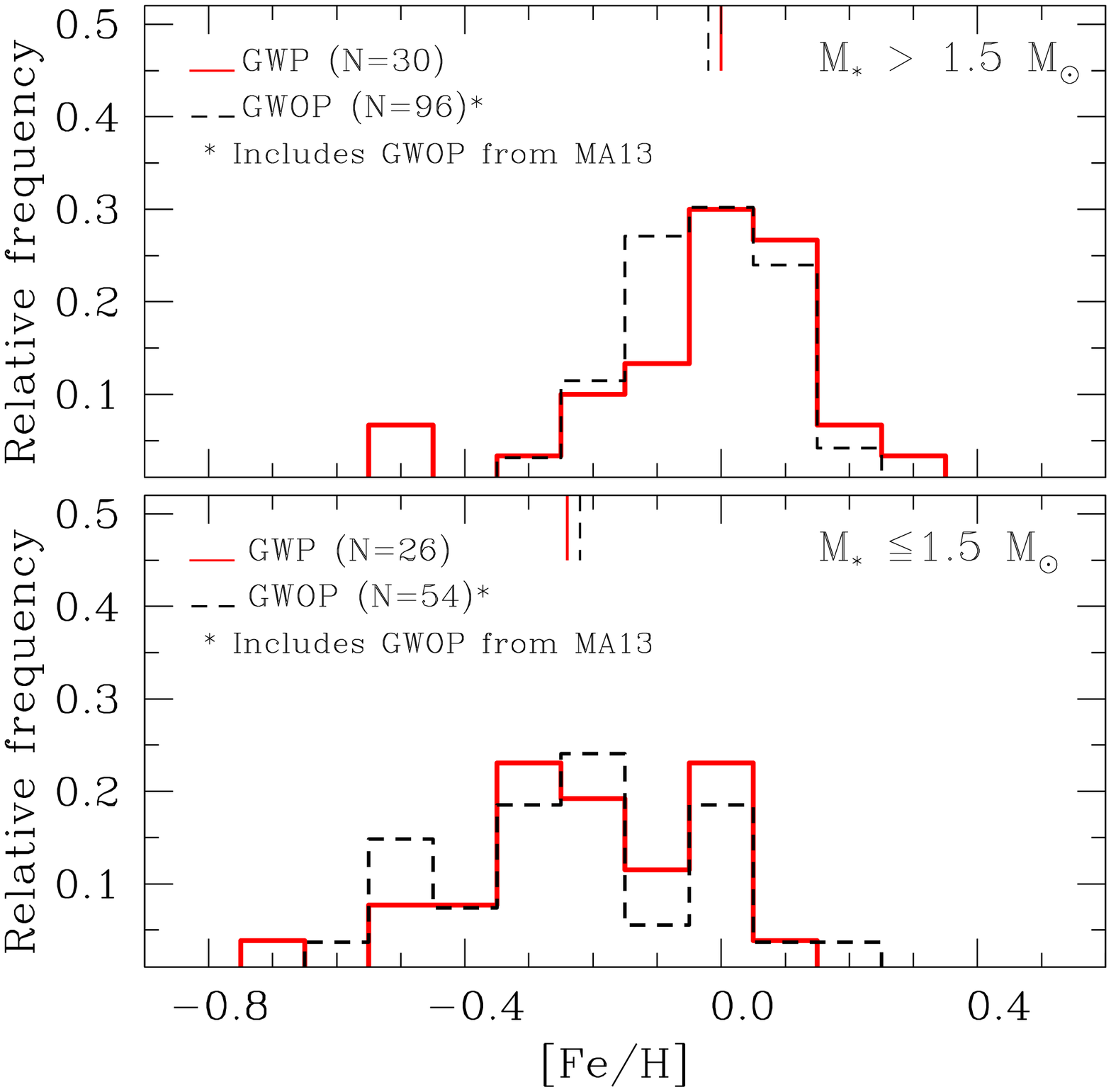}
   \caption{\textit{Left panels}: normalized metallicity distribution for the giant stars with $M_{\mathrm{\star}}$ > 1.5 $M_{\sun}$ (top panel) and $M_{\mathrm{\star}} < 1.5 M_{\sun}$ (bottom panel). \textit{Right panels}: same histograms as in the left panels but including the giant control sample from Maldonado et al. (2013). In all cases, the red solid line corresponds to the sample of giant stars with planets, whereas the black dashed line represents the giant control sample without planets. Median values of the distributions are indicated with vertical lines.}
              \label{FigGam}%
    \end{figure*}

In this section we use the results listed in Tables 2 and 5 to build the metallicity distributions for giant and subgiant stars and to search for differences between the stars with and without planets.      
 
\subsection{Giant stars}

In Figure 10 we show the normalized metallicity distribution, along with the histogram of the cumulative frequencies for the giant stars with planets (N = 56, red solid line) in comparison with the control sample (N = 101, black dashed line). Vertical lines at the top depict the median of each distribution. Both distributions are similar and centered at subsolar values. GWP have a median of [Fe/H] = -0.05 dex with a sigma of 0.20 dex, whereas the control sample has a  median of -0.05 dex with a sigma of 0.18 dex. The Kolmogorov-Smirnov (KS) test gives a probability of $\sim$85\% that both distributions are drawn from the same parent distribution and a maximum difference of 0.09 between the two samples in the cumulative frequencies. Therefore, in opposition to the metallicity offset found for stars on the main-sequence harboring giant planets (Santos et al. 2004, 2005; Fischer \& Valenti 2005; Ghezzi et al. 2010), giant stars with planets are not metal-rich when compared with giants without known planets. Both samples (giants with and without planets) are, on average, metal-poor. This result agrees with other studies with the exception of \citet{Hekker2007}.

\citet{Maldonado2013} found a hint of a possible dependency of metallicity on the stellar mass (see Fig. 10 in Maldonado et al.), where planet-hosting giants with $M_{\mathrm{\star}}$ > 1.5 $M _{\sun}$ are systematically more metal-rich. These authors re-analyzed the metallicity distribution separating the giant stars in two mass groups, $M_{\mathrm{\star}}$ $>$ 1.5 $M_{\sun}$ and $M_{\mathrm{\star}}$ $\leq$ 1.5 $M_{\sun}$. For the first group they found a clear separation between GWP and GWOP, with an average difference of $\sim$0.1 dex and a KS test probability of 5\% that both samples are drawn from the same parent population. On the other hand, they did not find any significant difference for the second group (i.e., giants with $M_{\mathrm{\star}}$ $\leq$ 1.5 $M_{\sun}$). Applying the same stellar mass division to our data as Maldonado et al. we constructed the [Fe/H] distributions for each mass group. These distributions are shown in the left panels of Figure 11. It can be seen that the [Fe/H] distribution for the higher mass group is slightly shifted towards higher metallicities with respect to the distribution of the lower mass group. Giants with $M_{\mathrm{\star}}$ $>$ 1.5 $M_{\sun}$ have, on average, higher metallicities than giants with $M_{\mathrm{\star}}$ $\leq$ 1.5 $M_{\sun}$. However, unlike \citet{Maldonado2013}, we find no significant difference between the metallicity distributions of giants with and without planets for stars with $M_{\mathrm{\star}}$ > 1.5 $M _{\sun}$ (left, upper panel). The KS test gives a probability of $\sim$91\% for both distributions being identical. In the case of stars with $M_{\mathrm{\star}}$ $\leq$ 1.5 $M_{\sun}$, both GWP and GWOP are centered at lower metallicities (left, bottom panel). The KS test gives a probability of $\sim$ 61\% that both distributions derive from the same parent distribution. The [Fe/H] statistics for each group are summarized in Table 10. 

  \begin{figure*}
   \centering
   \includegraphics[width=.45\textwidth]{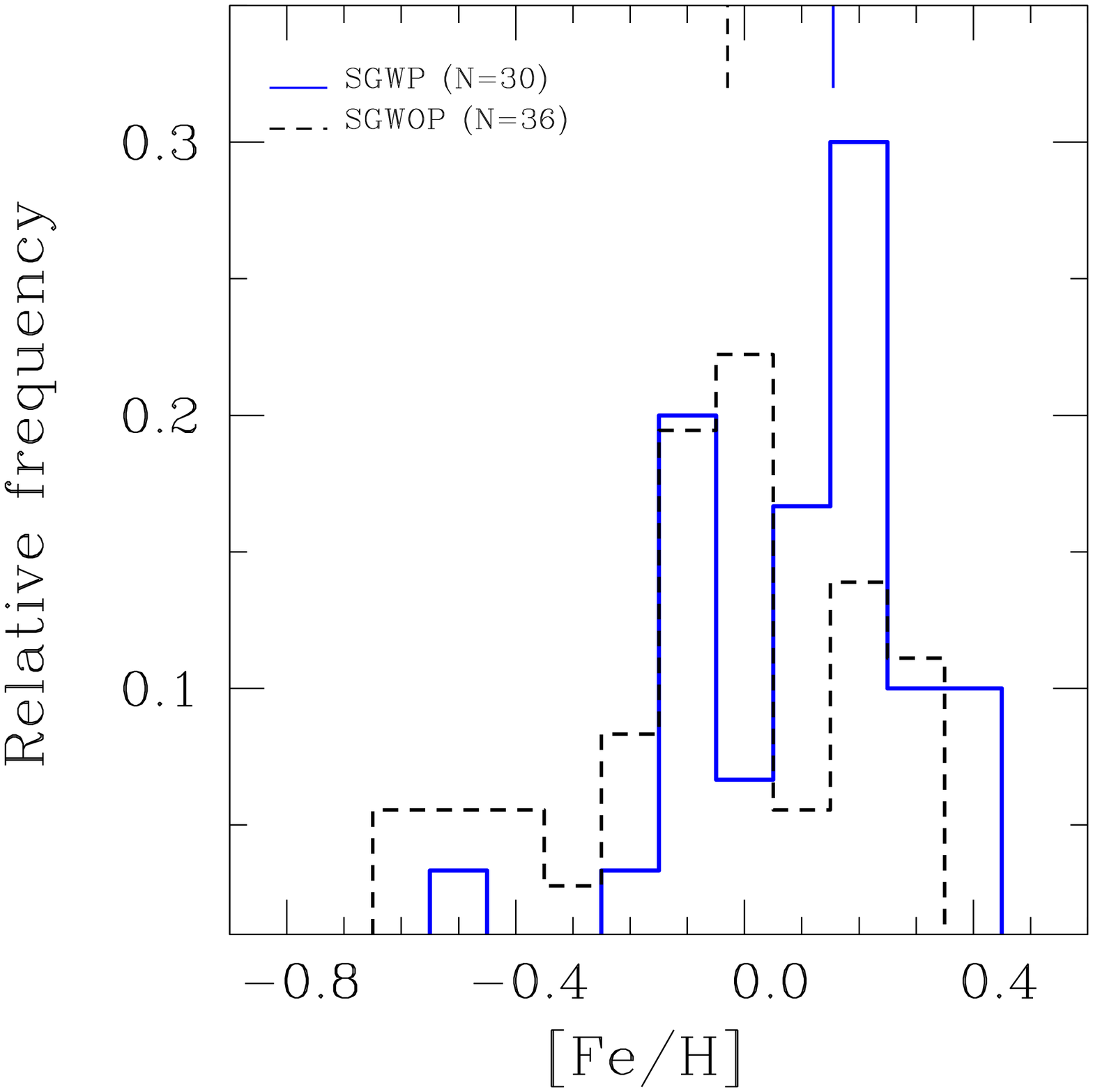}
   \includegraphics[width=.45\textwidth]{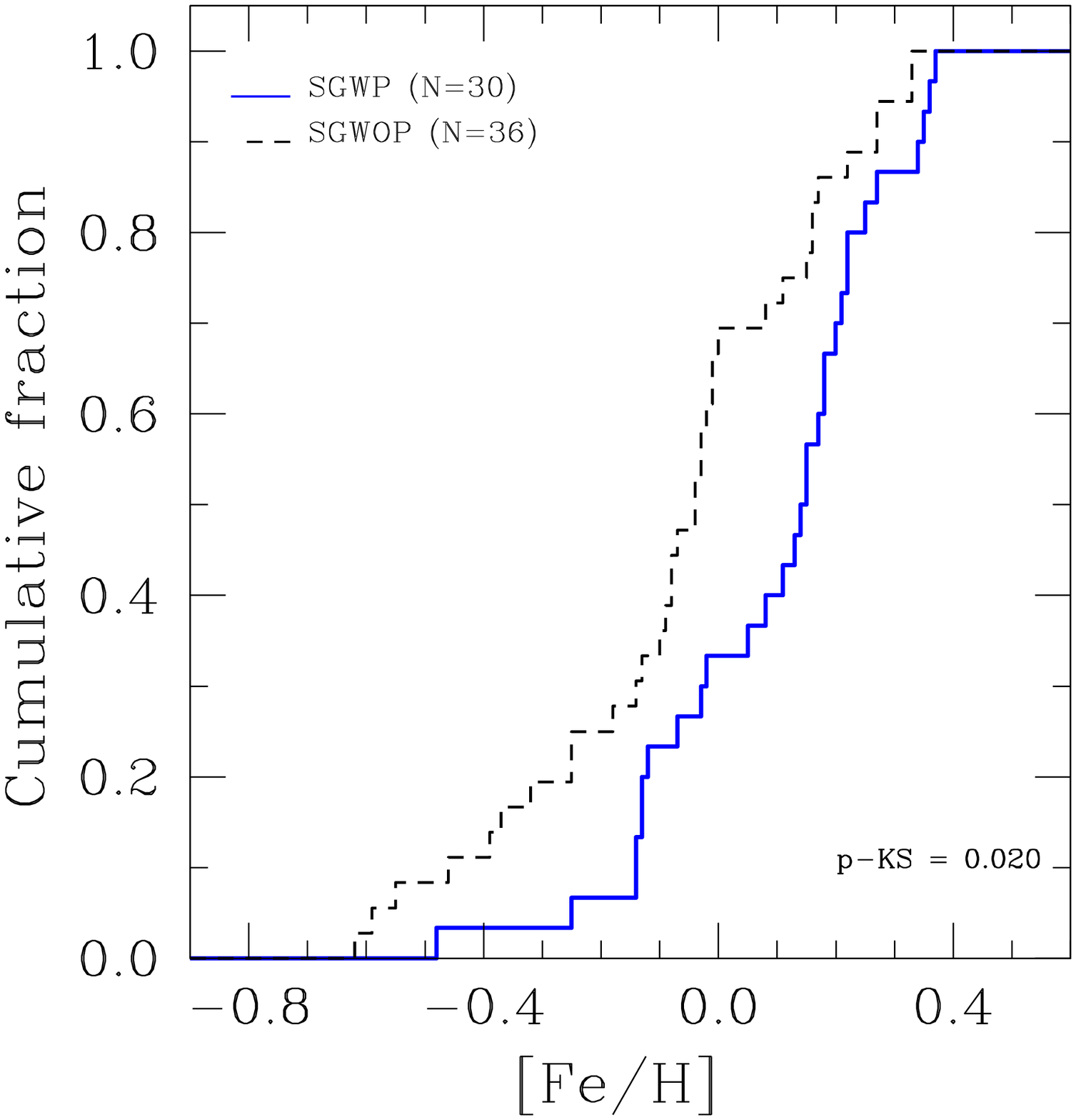}
\caption{Normalized metallicity distribution (\textit{left panel}) and histogram of cumulative frequencies (\textit{right panel}) for the SGWP sample (blue continuous line) in comparison with the SGWOP sample (black dashed line). Median values of the distributions are indicated with vertical lines. The Kolmogorov-Smirnov test shows that the probability that these samples derive from the same population is $\sim$ 0.02.}
              \label{FigGam}%
    \end{figure*} 

To investigate whether the difference between GWP and GWOP found by \citet{Maldonado2013} for stars with $M_{\mathrm{\star}}$ > $1.5$ $M _{\sun}$ might be related with the control sample, we combined our control stars with that of their study. We used 61 stars in common to define a relation to transform the [Fe/H] of Maldonado et al. to our scale\footnote{From a linear fit to the data we derived the following transformation: [Fe/H](this study) = (0.793$\pm$0.02) $\times$ [Fe/H](MA13) -- (0.032$\pm$0.04) , (rms= 0.08; $\chi_{r} ^{2}$ = 9.01).}. We also adopted the masses from these authors. The right panels of Figure 11 show the [Fe/H] distributions and the last row of Table 10 summarizes the statistics. For the group of stars with $M_{\mathrm{\star}}$ > $1.5$ $M _{\sun}$ the median is the same (-0.02 dex) and no clear difference between GWP and GWOP is found using the combined control sample. In addition, the KS test gives a probability of $\sim$ 90\% that both distributions derive from the same parent distribution. A similar outcome turned out for giants in the lower mass group. Thus, the discrepancy between our results and those of Maldonado et al. might be related to the different set of stars used to build the list of GWP with $M_{\mathrm{\star}}$ > $1.5$ $M _{\sun}$. Maldonado et al. analyzed 21 GWP, whereas we studied 30 GWP in this mass range. Adopting the same metallicity transformation used before, as an additional test, we extended our sample of giant stars with planets including all the GWP from the work of Maldonado et al. not in common with our original sample. However, even in this case no clear difference between GWP and GWOP was observed.

\subsection{Subgiant stars}
A comparison between the metallicity distribution of subgiants with planets (N = 30, blue solid line) and the control sample without planets (N = 36, black dashed line) is shown in Figure 12. Subgiants with planets are clearly shifted toward higher metallicities with respect to the control sample by $\sim$ 0.18 dex. The control sample has a median of -0.03 dex whereas the subgiants with planets have a median of +0.15 dex. The KS test shows a probability of only $\sim$ 2\% that both samples are drawn from the same parent distribution. The metal excess of the SGWP sample is very similar to the metallicity enhancement found in main-sequence stars with planets (Fischer \& Valenti 2005; Santos et al 2004, 2005; Ghezzi et al. 2010). These results for the subgiant stars agree with those previously found by other authors \citep{Ghezzi10b, Fischer2005, Johnson2010, Maldonado2013}.

      \begin{table}
      \tiny
      \caption[]{Metallicity statistics for the giant samples.}
         \label{table:1}
     \centering
         \begin{tabular}{c c c c c c}
         \hline
          
Sample	& Median &	Mean & Std deviation & N &	KS probability \\
\hline \hline
 \multicolumn{6}{c}{All masses} \\
 \hline \hline
GWP	&	-0.05	&	-0.08	&	0.20	&	56	&  0.850		\\
GWOP&	-0.05	&	-0.10	&	0.18	&	101	&		\\
          \hline
 \multicolumn{6}{c}{M > 1.5 $M_{\sun}$}        \\
      \hline \hline
GWP	&	0.00 &	-0.02	&	0.19	&	30	&	0.909	\\
GWOP&	-0.02 &	-0.02	&	0.11	&	67	&		\\
\hline
GWOP\tablefootmark{a} & -0.02 & -0.02 & 0.11 & 96 &0.900 \\
            \hline \hline
 \multicolumn{6}{c}{M $\leq$ 1.5 $M_{\sun}$}       \\
      \hline
GWP	&	-0.24	& -0.20	&	0.19	&	26	&	0.610	\\
GWOP&	-0.26	& -0.25	&	0.19	&	34	&		\\
          \hline
 GWOP\tablefootmark{a} &	-0.22 &	-0.22	&	0.20	&	54	&	0.929	\\         
          \hline
         
         \end{tabular}
         \tablefoottext{a}{Includes the giant control sample from Maldonado et al. (2013).}
        \end{table}

      \begin{table}
      \tiny
      \caption[]{Metallicity statistics for the subgiant samples.}
         \label{table:1}
     \centering
         \begin{tabular}{c c c c c c}
         \hline
          
Sample	& Median &	Mean & Std deviation & N &	KS probability \\
\hline \hline
SGWP	&	+0.15	&	+0.10	&	0.20	&	30	&  0.020		\\
SGWOP&	-0.03	&	-0.06	&	0.25	&	36	&		\\
          \hline
        
         \end{tabular}
        \end{table}

  \begin{figure}
   \centering
   \includegraphics[width=.47\textwidth]{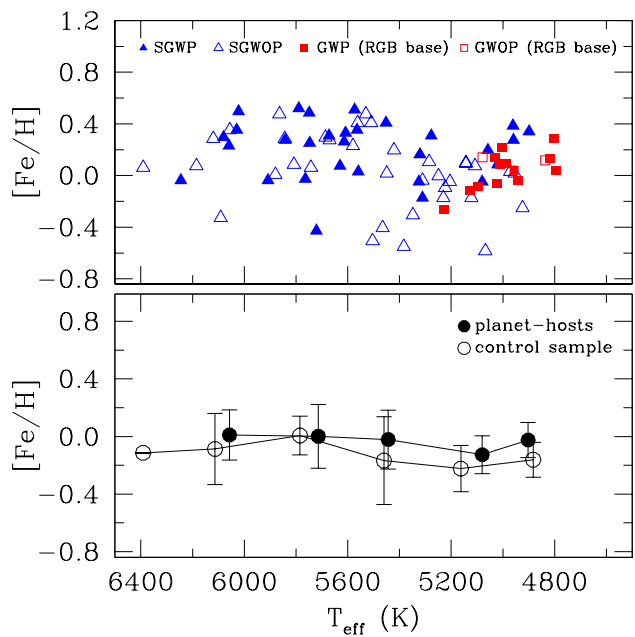}
   \caption{\textit{Upper panel}: [Fe/H] vs. $T_{\mathrm{eff}}$. Blue triangles and red squares represent subgiant and giant stars at the base of the RGB, respectively. Filled symbols correspond to stars with planets and the empty ones to stars without planets. \textit{Lower panel}: Same plot as the upper panel except that the data are averaged in bins of 300 K. Filled circles mark giants and subgiants with planets and empty circles indicate the stars without planets. }
              \label{FigGam}%
    \end{figure}

  \begin{figure*}
   \centering
   \includegraphics[width=.61\textwidth]{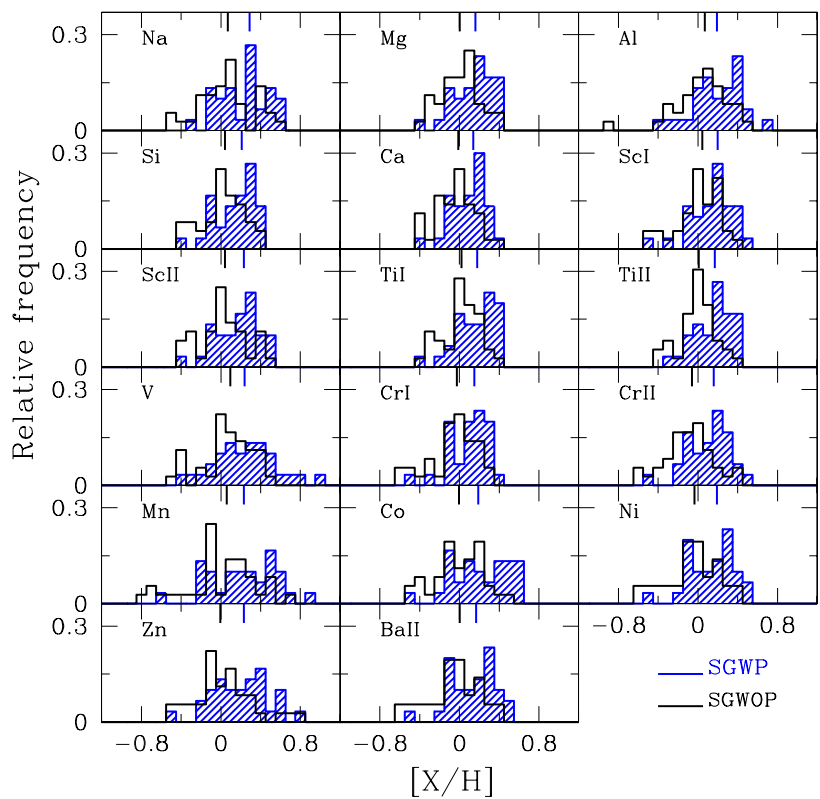}
  
   \caption{Normalized distributions of [X/H] (X=Na, Mg, Al, Si, Ca, \ion{Sc}{I}, \ion{Sc}{II}, \ion{Ti}{I}, \ion{Ti}{II}, V, \ion{Cr}{I}, \ion{Cr}{II}, Mn, Co, Ni, Zn, \ion{Ba}{II}) for subgiants with planets (shaded blue) and subgiants without planets (solid black lines).  Median values of the distributions of each element are indicated with vertical lines.}
              \label{FigGam}%
    \end{figure*}

\subsection{Metallicity as a function of $T_{\mathrm{eff}}$: evidence for dilution?}
The results of the metallicity distributions presented in Figures 10 and 12 are crucial to the understanding of the planet-metallicity correlation found for main-sequence stars with planets and, hence, essential in order to constrain giant planet formation models. As it was mentioned in the introduction, the lack of a metallicity enhancement in giant stars with planets has been related with the pollution hypothesis to explain the high metal content of the main-sequence stars with planets. If the excess in metal content of the planet-host stars is the result of the accretion of metal-rich material, this excess will only lie on the external layers of the atmosphere. However, as the star evolves this excess is expected to be diluted as the convective zones greatly deepens, reaching a mass of $\sim$0.7 $M_{\sun}$ for a 1 $M_{\sun}$ star along the red giant branch \citep{Pasquini2007}. Therefore, if pollution is the mechanism that operates to increase the metal contents, it should be expected that not only  giant stars with planets, but also subgiants show systematically lower metallicity than main-sequence stars with planets. 

     \begin{figure*}
   \centering
    \includegraphics[width=.61\textwidth]{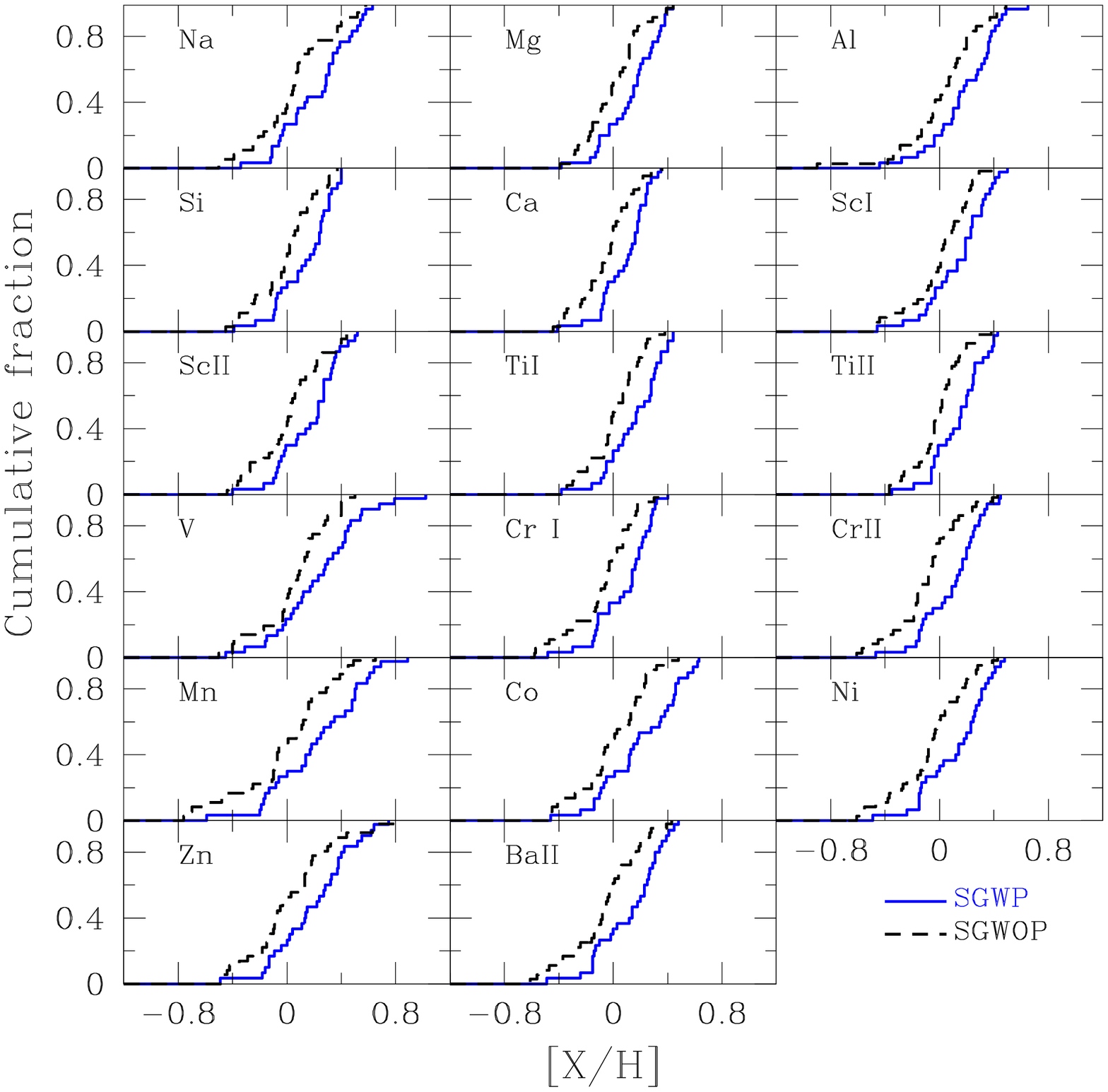}
   
   \caption{Cumulative functions of [X/H] (X=Na, Mg, Al, Si, Ca, \ion{Sc}{I}, \ion{Sc}{II}, \ion{Ti}{I}, \ion{Ti}{II}, V, \ion{Cr}{I}, \ion{Cr}{II}, Mn, Co, Ni, Zn, \ion{Ba}{II}) for subgiants with planets (solid blue lines) and subgiants without planets (dashed black lines).}
              \label{FigGam}%
    \end{figure*}

As another test to study signs of pollution, several authors have searched for a metallicity gradient as a function of $T_{\mathrm{eff}}$ (or stellar mass) among main-sequence samples \citep{Pinsonneault2001, Santos2001, Santos2003, Santos2004, Gonzalez2001, Fischer2005}. An increase in the upper boundary of the [Fe/H] distribution with increasing $T_{\mathrm{eff}}$ (which implies a decrease in the size of the convection zone), would support the idea of a stellar atmosphere pollution with metal-rich material. Most of these studies did not find any clear evidence of a metallicity gradient as a function of the $T_{\mathrm{eff}}$, and it has been suggested that such a trend on dwarfs might be hard to detect because of extra mixing zones \citep{Vauclair2004}. However, there should be a more evident trend between the hot and cool part of the subgiant branch, where the convection zone is expected to deepen by a factor of 10-100 \citep{Pasquini2007}.   

   \begin{figure*}
   \centering
  \includegraphics[width=.61\textwidth]{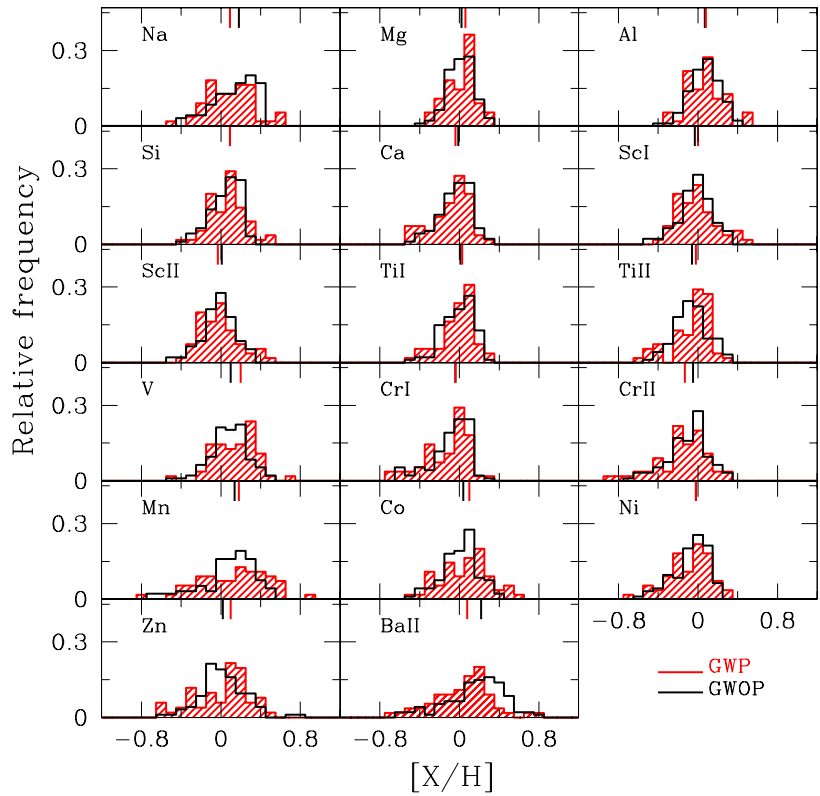}
   \caption{Normalized distributions of [X/H] (X=Na, Mg, Al, Si, Ca, \ion{Sc}{I}, \ion{Sc}{II}, \ion{Ti}{I}, \ion{Ti}{II}, V, \ion{Cr}{I}, \ion{Cr}{II}, Mn, Co, Ni, Zn, \ion{Ba}{II}) for giants with planets (shaded red) and giants without planets (solid black lines).  Median values of the distributions of each element are indicated with vertical lines.}
              \label{FigGam}%
    \end{figure*}

   \begin{figure*}
   \centering
   \includegraphics[width=.61\textwidth]{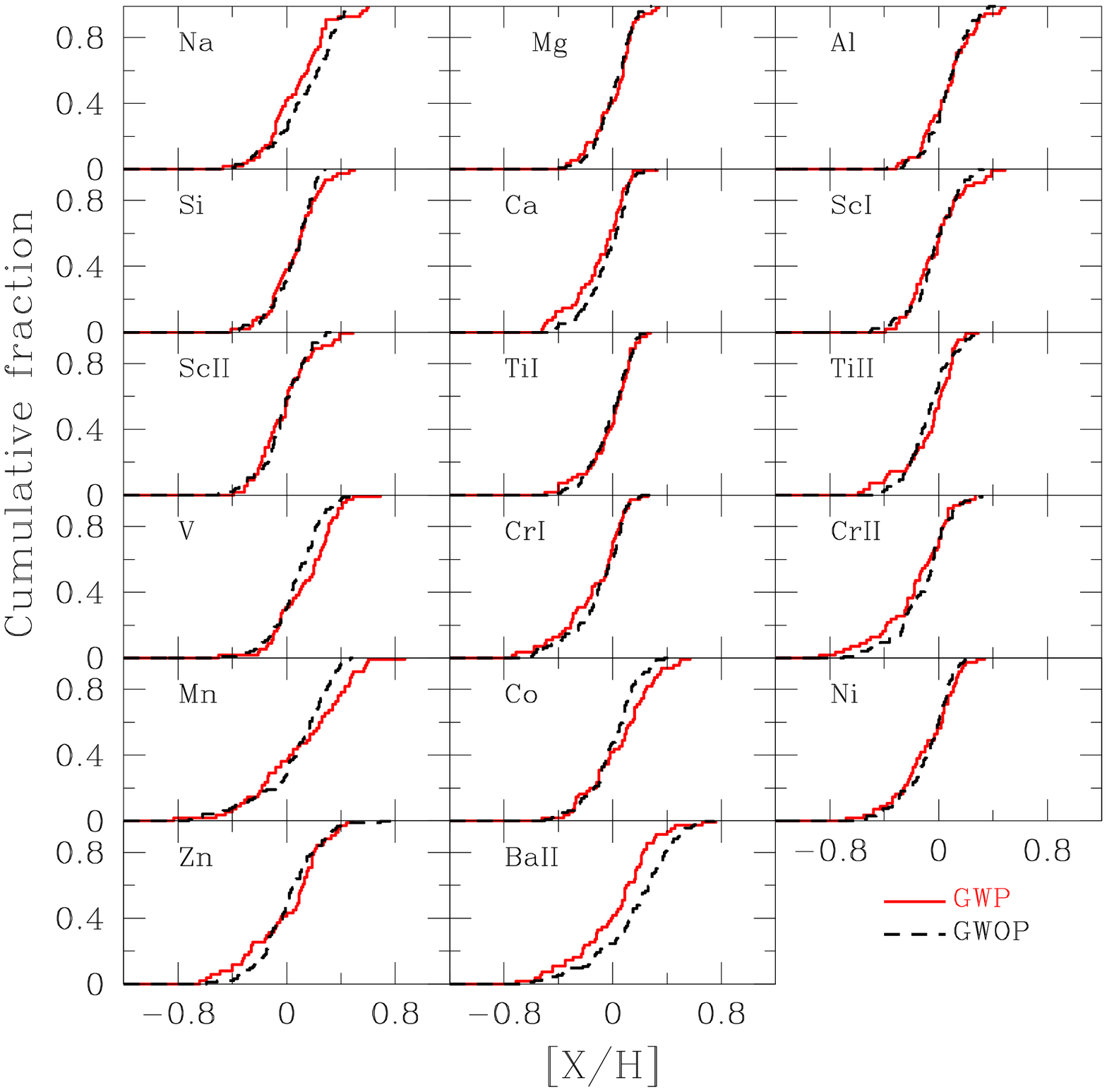}
   \caption{Cumulative functions of [X/H] (X=Na, Mg, Al, Si, Ca, \ion{Sc}{I}, \ion{Sc}{II}, \ion{Ti}{I}, \ion{Ti}{II}, V, \ion{Cr}{I}, \ion{Cr}{II}, Mn, Co, Ni, Zn, \ion{Ba}{II}) for giants with planets (solid red lines) and giants without planets (dashed black lines).}
              \label{FigGam}%
    \end{figure*}

Although the metal-poor distribution for the planet-host giants obtained here might agree with the dilution hypothesis, the relatively high average metallicity obtained for the subgiant stars with planets does not seem to fit with this scenario. However, if we plot our derived metallicities as a function of $T_{\mathrm{eff}}$ for 30 SGWP along with 12 GWP that lie at the base of the red giant branch (RGB) (Top panel, Figure 13), a suggestive drop in metallicity seems to happen at $T_{\mathrm{eff}} \sim$ 5400 K.  At cooler temperatures the upper boundary of the metallicity distribution seems to have a slight decrease of $\sim$ 0.10 dex. For comparison purpose, control samples are also plotted. In the bottom panel of Figure 13 we show the same graphic but binning $T_{\mathrm{eff}}$ in 300 K intervals. It is important to note that the apparent drop also occurs for the control sample stars. \citet{Murray2001} found evidence of lower average metallicity in 19 subgiant stars belonging to the Hertzsprung gap in relation to the dwarf sample, for which the enhance metal content is attributed to the accretion of up to 0.6 $M_{\mathrm{\oplus}}$ of iron onto the surfaces of dwarfs. In contrast, \citet{Fischer2005} found no evidence of a metallicity gradient with $T_{\mathrm{eff}}$ for 86 subgiant stars, of which only nine host planets. Similar results were reported by \citet{Ghezzi10b} for a sample of 14 subgiants and 6 giants at the bottom of the RGB. In view of previous works and the subtle decline in Figure 13, the evidence for dilution is very weak or even negligible. However it is intriguing enough to encourage researchers in the field to test this suggestion on larger samples of subgiants with planets.

\section{Chemical abundances: comparison of the samples}
The analysis of chemical abundances of other elements than iron may provide valuable information about the elements that might have an active role in the process of planetary formation, particularly for low metallicity stars, which seem to be the case of giant-hosts. Therefore, in this section we use the results listed in Tables 6 and 7 to search for possible differences in the chemical abundances of planet-host stars relative to the control samples, both for the giant and subgiant samples. Although the samples of stars with planets are relatively small (N = 56 for giants and N = 30 for subgiants), they represent significant fractions of the current total samples of evolved stars with planets ($\sim$ 67\% for giants and $\sim$ 70\% for subgiants).

\subsection{[X/H] distributions}

In Figures 14 and 16 we show the normalized distributions of [X/H] (X = Na, Mg, Al, Si, Ca, Sc, \ion{Sc}{II}, \ion{Ti}{I}, \ion{Ti}{II}, V, \ion{Cr}{I}, \ion{Cr}{II}, Mn, Co, Ni, Zn, \ion{Ba}{II}) for subgiant and giant stars, respectively. In these plots, stars with planets are indicated by shaded histograms and control samples are represented by black lines. Vertical lines on each histogram mark the median values. In addition, Figures 15 and 17 show the corresponding cumulative functions of [X/H] for subgiants and giants. Here, stars with planets are indicated by solid lines and the control samples with dashed black lines. Tables 12 and 13 summarize the statistics for each element and group of stars. SGWP show an evident metallicity excess with respect to the control sample without planets for all the analyzed species. This metallicity excess is about 0.17 dex, which agrees with the iron enhancement derived in the previous section. Similar results were obtained for main-sequence planet hosts \citep{Adibekyan2012b, Neves2009, Gilli2006, Beirao2005}. 

As it has been already noted in previous works studying solar-type stars \citep{Gilli2006, Bodaghee2003, Beirao2005, Neves2009}, we also find that the [X/H] distributions for SGWP are not symmetrical. Most of the elements show an increase in [X/H] until they reach a cut-off after which the distributions suffer a deep fall. This effect is clear for Mg, Si, Ca, \ion{Sc}{I}, \ion{Sc}{II}, \ion{Ti}{I}, \ion{Ti}{II}, and \ion{Cr}{I} for which the cut-off value is around [X/H] $\sim$ 0.5 dex. Interpretations for this behavior have been discussed in Santos et al. (2001, 2003, 2004c) and Neves et al. (2009). In addition to this cut-off, SGWP distributions for elements such as Al, Cr, Ni, and Ba might be bimodal. 

In the case of the giant stars, showed in Figures 16 and 17, as it was the case for [Fe/H], there is no metallicity excess in the [X/H] distributions of GWP for most of the elements. In addition, these distributions completely overlaps with the GWOP distributions. Only V shows a slight excess ($\sim$ 0.10 dex) for GWP, whereas elements like Ba, and Na are under abundant by $\sim$0.14 dex and $\sim$ 0.09 dex, respectively. These distributions appear to be more symmetrical, and only elements like  Mg, \ion{Ti}{I}, \ion{Ti}{II}, and \ion{Cr}{I} show abrupt falls around 0.25 dex.     

     \begin{table*}
      \tiny
      \caption[]{[X/H] statistics for the subgiant sample.}
         \label{table:1}
     \centering
         \begin{tabular}{l |c c c |c c c |c |c |c}
                   
\hline \hline  
	&		&	SGWP	&		&		&	SGWOP	&		&		&		&		\\
$[X/H]$	&	Average	&	Median	&	rms	&	Average	&	Median	&	rms	&	Diff. of averages 	&	Diff. of medians	&	KS probability	\\
\hline
Na	&	0.23	&	0.30	&	0.25	&	0.06	&	0.06	&	0.28	&	0.17	&	0.22	&	0.03	\\
Mg	&	0.15	&	0.18	&	0.20	&	0.01	&	0.00	&	0.20	&	0.13	&	0.15	&	0.02	\\
Al	&	0.19	&	0.20	&	0.24	&	0.05	&	0.07	&	0.23	&	0.14	&	0.12	&	0.06	\\
Si	&	0.15	&	0.22	&	0.20	&	0.01	&	0.03	&	0.22	&	0.14	&	0.18	&	0.02	\\
Ca	&	0.10	&	0.15	&	0.17	&	-0.04	&	-0.01	&	0.19	&	0.14	&	0.15	&	0.01	\\
\ion{Sc}{I}	&	0.15	&	0.20	&	0.22	&	0.02	&	0.03	&	0.22	&	0.13	&	0.16	&	0.06	\\
\ion{Sc}{II}	&	0.18	&	0.24	&	0.22	&	0.03	&	0.04	&	0.24	&	0.15	&	0.20	&	0.02	\\
\ion{Ti}{I}	&	0.17	&	0.19	&	0.20	&	0.02	&	0.01	&	0.19	&	0.15	&	0.16	&	0.01	\\
\ion{Ti}{II}	&	0.15	&	0.17	&	0.19	&	0.00	&	0.01	&	0.18	&	0.15	&	0.16	&	0.00	\\
V	&	0.25	&	0.26	&	0.32	&	0.07	&	0.09	&	0.25	&	0.18	&	0.14	&	0.05	\\
\ion{Cr}{I}	&	0.10	&	0.15	&	0.21	&	-0.07	&	-0.04	&	0.24	&	0.16	&	0.18	&	0.01	\\
\ion{Cr}{II}	&	0.11	&	0.15	&	0.22	&	-0.08	&	-0.06	&	0.25	&	0.19	&	0.20	&	0.00	\\
Mn 	&	0.24	&	0.25	&	0.33	&	0.00	&	0.05	&	0.35	&	0.24	&	0.18	&	0.05	\\
Co 	&	0.22	&	0.20	&	0.28	&	0.01	&	0.00	&	0.25	&	0.21	&	0.19	&	0.01	\\
Ni 	&	0.14	&	0.20	&	0.24	&	-0.05	&	-0.04	&	0.26	&	0.19	&	0.22	&	0.02	\\
Zn 	&	0.21	&	0.24	&	0.28	&	0.04	&	0.00	&	0.31	&	0.17	&	0.23	&	0.07	\\
\ion{Ba}{II}	&	0.14	&	0.17	&	0.17	&	-0.04	&	0.01	&	0.28	&	0.18	&	0.16	&	0.03	\\
      
     \hline
         
         \end{tabular}
        \end{table*}

      \begin{table*}
      \tiny
      \caption[]{ [X/H] statistics for the giant sample.}
         \label{table:1}
     \centering
         \begin{tabular}{l |c c c |c c c |c |c |c}
                   
\hline \hline
	&		&	GWP	&		&		&	GWOP	&		&		&		&		\\
$[X/H]$	&	Average	&	Median	&	rms	&	Average	&	Median	&	rms	&	Diff. of averages 	&	Diff. of medians	&	KS probability	\\
\hline
Na	&	0.09	&	0.09	&	0.23	&	0.14	&	0.18	&	0.22	&	-0.05	&	-0.09	&	0.064	\\
Mg	&	0.02	&	0.06	&	0.16	&	0.02	&	0.02	&	0.14	&	0.00	&	0.03	&	0.894	\\
Al	&	0.08	&	0.08	&	0.18	&	0.07	&	0.07	&	0.15	&	0.00	&	0.01	&	0.851	\\
Si	&	0.07	&	0.09	&	0.18	&	0.05	&	0.09	&	0.15	&	0.01	&	0.00	&	0.737	\\
Ca	&	-0.09	&	-0.04	&	0.20	&	-0.03	&	-0.01	&	0.16	&	-0.06	&	-0.03	&	0.460	\\
\ion{Sc}{I}	&	-0.01	&	0.00	&	0.20	&	-0.02	&	-0.03	&	0.17	&	0.01	&	0.03	&	0.828	\\
\ion{Sc}{II}	&	-0.05	&	-0.03	&	0.21	&	-0.02	&	0.01	&	0.16	&	-0.02	&	-0.04	&	0.765	\\
\ion{Ti}{I}	&	-0.01	&	0.03	&	0.17	&	-0.01	&	0.01	&	0.15	&	0.00	&	0.01	&	0.988	\\
\ion{Ti}{II}	&	-0.06	&	-0.02	&	0.20	&	-0.06	&	-0.06	&	0.17	&	0.00	&	0.04	&	0.236	\\
V	&	0.15	&	0.20	&	0.22	&	0.09	&	0.10	&	0.19	&	0.06	&	0.10	&	0.055	\\
\ion{Cr}{I}	&	-0.13	&	-0.04	&	0.23	&	-0.09	&	-0.04	&	0.20	&	-0.04	&	0.00	&	0.485	\\
\ion{Cr}{II}	&	-0.16	&	-0.13	&	0.25	&	-0.10	&	-0.05	&	0.21	&	-0.06	&	-0.08	&	0.397	\\
Mn	&	0.13	&	0.18	&	0.34	&	0.08	&	0.14	&	0.27	&	0.04	&	0.05	&	0.080	\\
Co	&	0.06	&	0.10	&	0.24	&	0.01	&	0.04	&	0.19	&	0.05	&	0.06	&	0.061	\\
Ni	&	-0.07	&	-0.02	&	0.22	&	-0.07	&	-0.02	&	0.18	&	-0.01	&	0.01	&	0.647	\\
Zn	&	0.01	&	0.10	&	0.28	&	0.02	&	0.02	&	0.23	&	-0.01	&	0.08	&	0.283	\\
\ion{Ba}{II}	&	0.03	&	0.08	&	0.30	&	0.17	&	0.22	&	0.29	&	-0.13	&	-0.14	&	0.004	\\

          \hline
         
         \end{tabular}
        \end{table*}

   \begin{figure*}
   \centering
   \includegraphics[width=.61\textwidth]{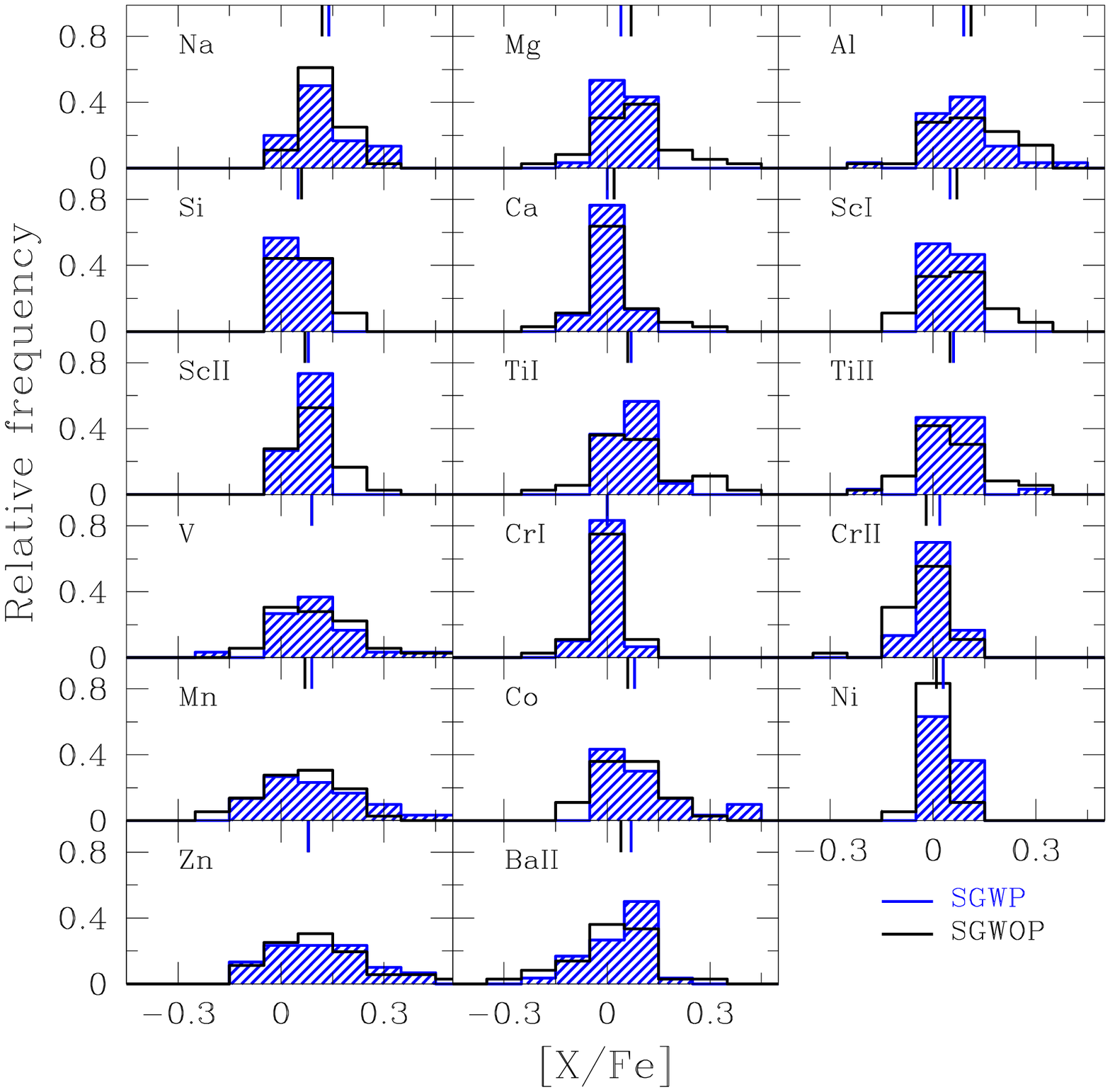}
   \caption{[X/Fe] normalized distributions for subgiants with planets (shaded blue) and subgiants without planets (solid black lines). Median values of the distributions of each element are indicated with vertical lines.}
              \label{FigGam}%
    \end{figure*}

   \begin{figure*}
   \centering
      \includegraphics[width=.61\textwidth]{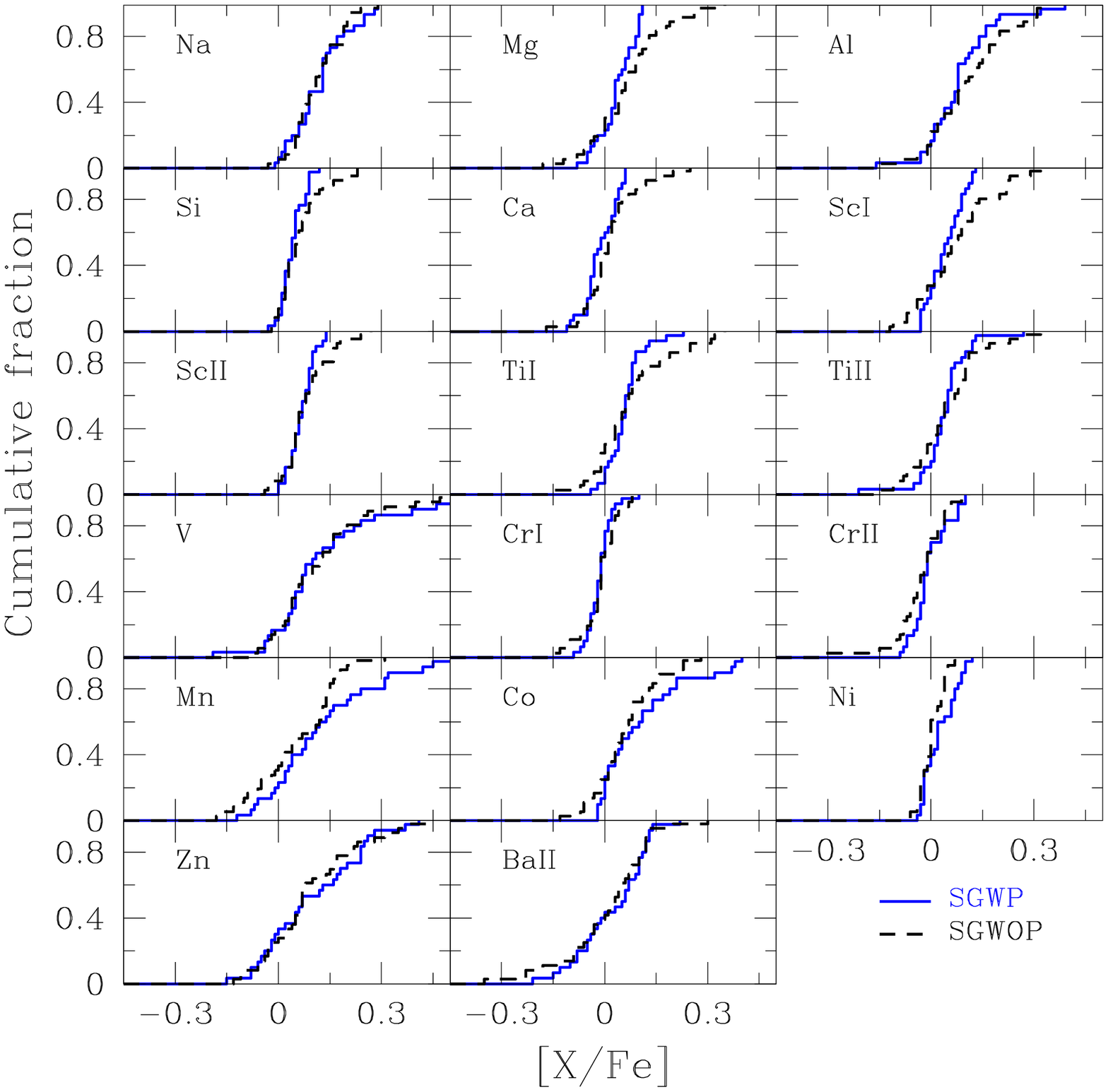}
   \caption{[X/Fe] cumulative functions for subgiants with planets (solid blue lines) and subgiants without planets (dashed black lines).}
              \label{FigGam}%
    \end{figure*}

    \begin{figure*}
   \centering
   \includegraphics[width=.61\textwidth]{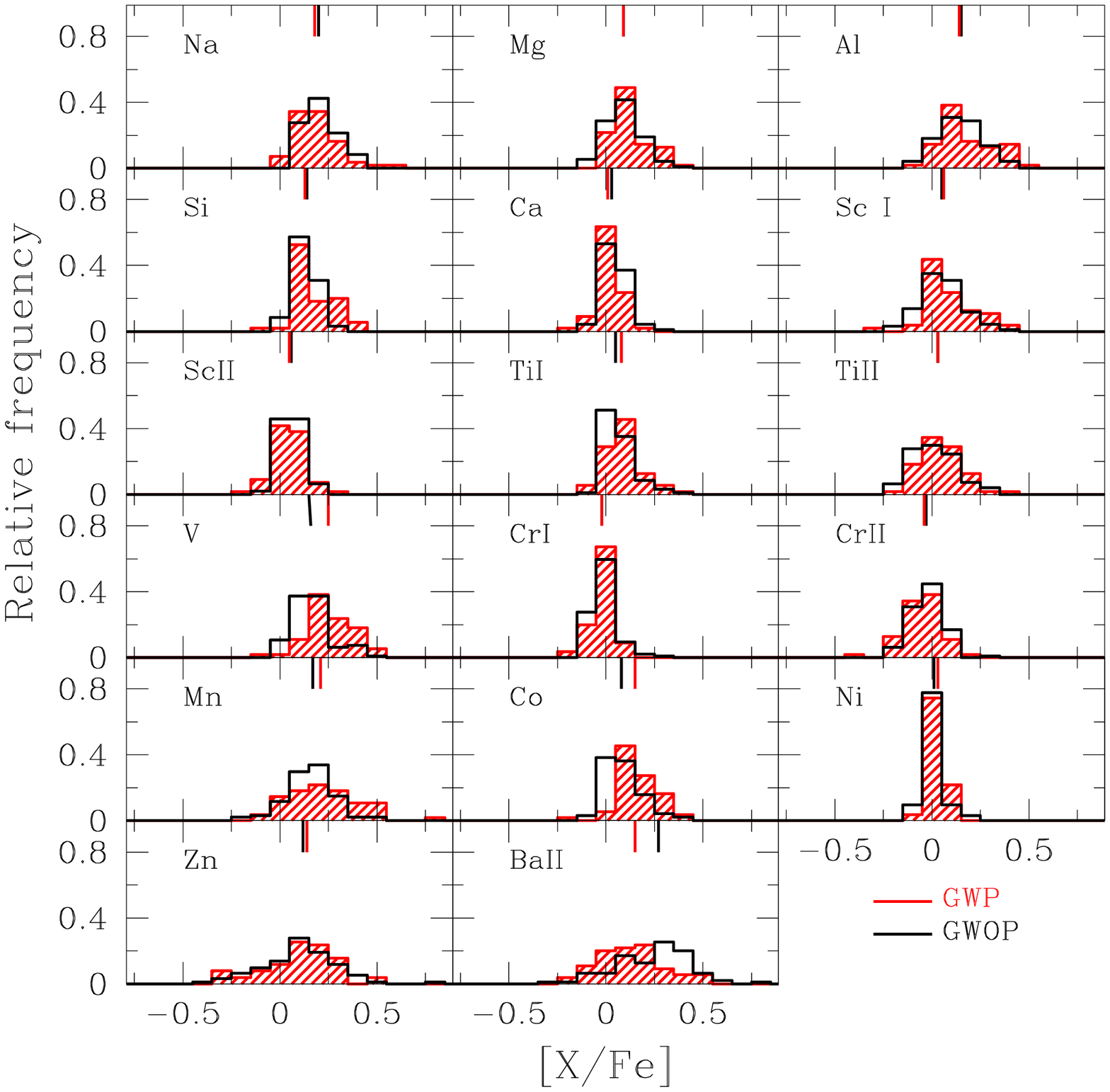}
   \caption{[X/Fe] normalized distributions for giants with planets (shaded red) and giants without planets (solid black lines). Median values of the distributions of each element are indicated with vertical lines.}
              \label{FigGam}%
    \end{figure*}         
    
      \begin{figure*}
   \centering
      \includegraphics[width=.61\textwidth]{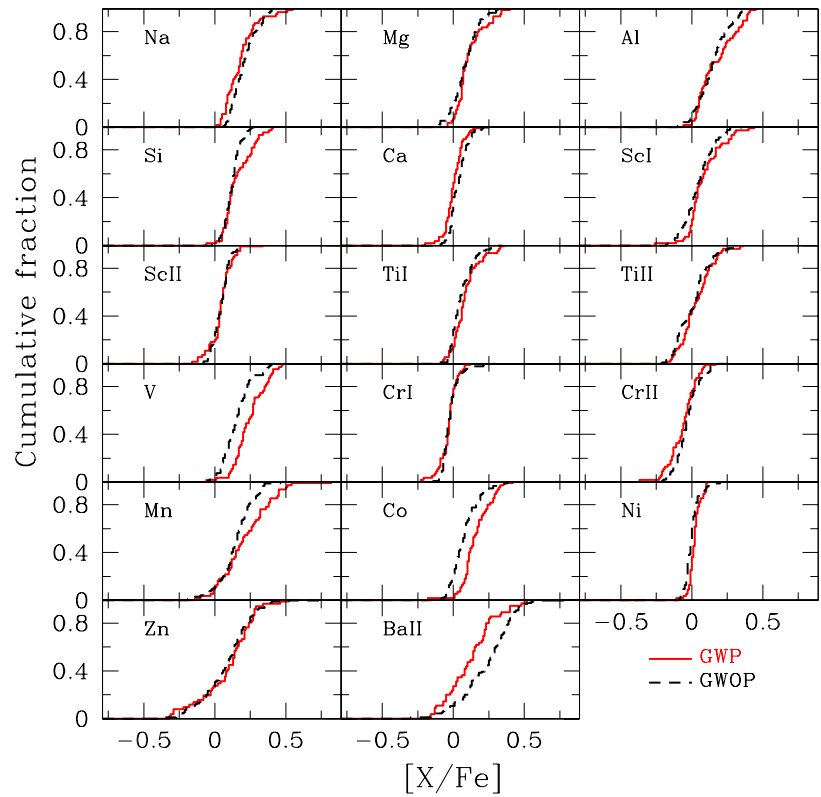}

   \caption{[X/Fe] cumulative functions for giants with planets (solid red lines) and giants without planets (dashed black lines).}
              \label{FigGam}%
    \end{figure*}           
       
\subsection{The [X/Fe] ratios}

The [X/Fe] distributions and the cumulative functions for subgiants are presented in Figures 18 and 19 and those for giants in Figures 20 and 21. Colors and line types are as in previous figures. Tables 14 and 15 list the statistics for each group, including the differences of the averages and medians, and the KS probabilities that samples with and without planets were drawn from the same parent population.

In the case of the subgiant sample, it can be seen that, in general, the distributions of planet-host stars and those of stars without planets have very similar behaviors. The differences between the average values of [X/Fe] of SGWP and SWOP are relatively small, with the largest corresponding to Mn (0.07 dex). This difference in Mn, although within the dispersion values, has been previously reported by other studies analyzing dwarf stars with planets \citep{Bodaghee2003, Kang2011, Adibekyan2012a}. Nonetheless, the KS probabilities turned out to be, in all cases, relatively high, ranging from 8\% for Ni to 100\% for V.

     \begin{figure*}[]
   \centering
   \includegraphics[width=.70\textwidth]{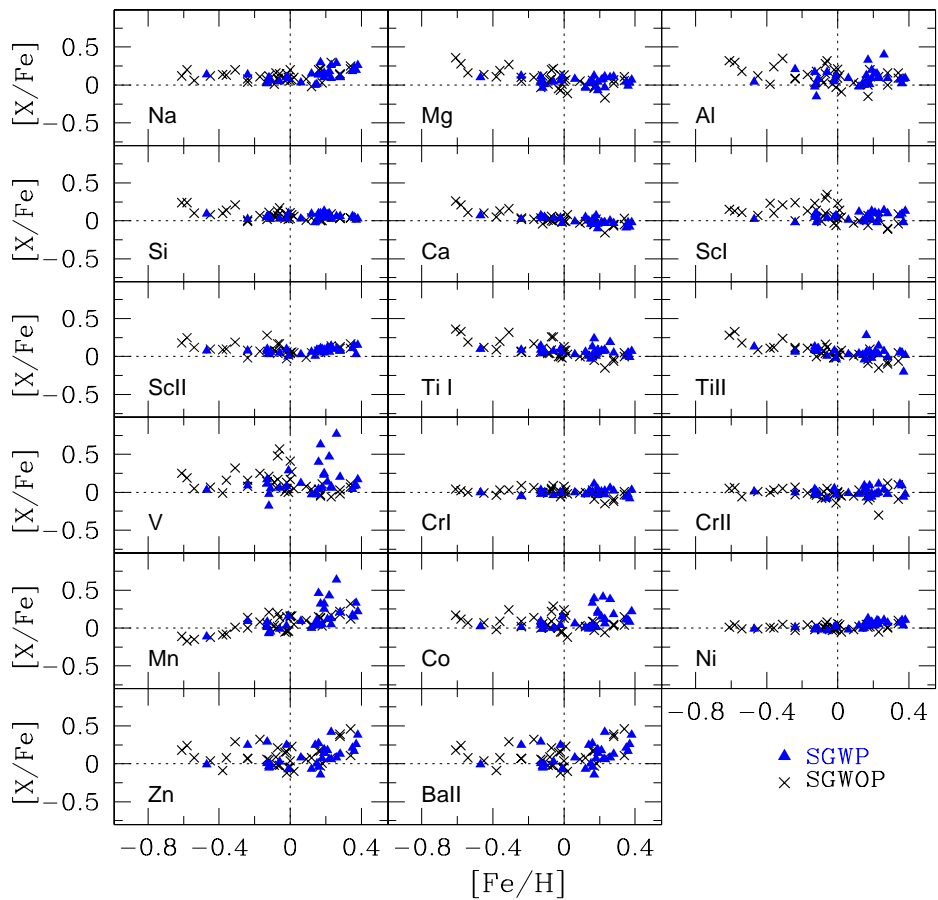}
   \caption{[X/H] vs. [Fe/H] for subgiant stars with planets (filled triangles) and the control sample (black crosses). Dashed lines represent the solar values.}
              \label{FigGam}%
    \end{figure*}

     \begin{table*}
      \tiny
      \caption[]{[X/Fe] statistics for the subgiant sample.}
         \label{table:1}
     \centering
         \begin{tabular}{l |c c c |c c c |c |c |c}               
\hline \hline  
	&		&	SGWP	&		&		&	SGWOP	&		&		&		&		\\
$[X/Fe]$	&	Average	&	Median	&	rms	&	Average	&	Median	&	rms	&	Diff. of averages 	&	Diff. of medians	&	KS probability	\\
\hline
Na	&	0.13	&	0.14	&	0.08	&	0.12	&	0.12	&	0.07	&	0.01	&	0.02	&	0.93	\\
Mg	&	0.05	&	0.04	&	0.05	&	0.08	&	0.07	&	0.11	&	-0.03	&	-0.03	&	0.13	\\
Al	&	0.09	&	0.09	&	0.10	&	0.12	&	0.11	&	0.12	&	-0.02	&	-0.01	&	0.63	\\
Si	&	0.05	&	0.05	&	0.04	&	0.07	&	0.06	&	0.06	&	-0.02	&	-0.01	&	0.54	\\
Ca	&	0.00	&	0.00	&	0.05	&	0.03	&	0.02	&	0.08	&	-0.02	&	-0.02	&	0.43	\\
\ion{Sc}{I}	&	0.05	&	0.05	&	0.05	&	0.08	&	0.07	&	0.11	&	-0.03	&	-0.02	&	0.21	\\
\ion{Sc}{II}	&	0.08	&	0.08	&	0.04	&	0.09	&	0.07	&	0.07	&	-0.01	&	0.00	&	0.51	\\
\ion{Ti}{I}	&	0.07	&	0.07	&	0.06	&	0.08	&	0.06	&	0.11	&	-0.01	&	0.01	&	0.57	\\
\ion{Ti}{II}	&	0.06	&	0.06	&	0.08	&	0.06	&	0.05	&	0.10	&	0.00	&	0.01	&	0.53	\\
V	&	0.15	&	0.09	&	0.20	&	0.13	&	0.09	&	0.14	&	0.02	&	-0.01	&	1.00	\\
\ion{Cr}{I}	&	0.00	&	0.00	&	0.04	&	0.00	&	0.00	&	0.05	&	0.00	&	0.00	&	0.80	\\
\ion{Cr}{II}	&	0.01	&	0.00	&	0.05	&	-0.02	&	-0.02	&	0.08	&	0.03	&	0.01	&	0.36	\\
Mn 	&	0.14	&	0.09	&	0.17	&	0.06	&	0.07	&	0.12	&	0.07	&	0.02	&	0.43	\\
Co 	&	0.12	&	0.08	&	0.12	&	0.07	&	0.06	&	0.09	&	0.05	&	0.02	&	0.65	\\
Ni 	&	0.04	&	0.03	&	0.05	&	0.01	&	0.01	&	0.04	&	0.02	&	0.02	&	0.08	\\
Zn 	&	0.11	&	0.08	&	0.14	&	0.10	&	0.08	&	0.14	&	0.01	&	0.00	&	0.95	\\
\ion{Ba}{II}	&	0.04	&	0.07	&	0.11	&	0.02	&	0.04	&	0.13	&	0.02	&	0.03	&	0.75	\\

          \hline
         
         \end{tabular}
        \end{table*}

     \begin{table*}
      \tiny
      \caption[]{ [X/Fe] statistics for the giant sample.}
         \label{table:1}
     \centering
         \begin{tabular}{l |c c c |c c c |c |c |c}
                   
\hline \hline  
	&		&	GWP	&		&		&	GWOP	&		&		&		&		\\
$[X/Fe]$	&	Average	&	Median	&	rms	&	Average	&	Median	&	rms	&	Diff. of averages 	&	Diff. of medians	&	KS probability	\\
\hline
Na	&	0.19	&	0.18	&	0.12	&	0.22	&	0.20	&	0.09	&	-0.03	&	-0.02	&	0.047	\\
Mg	&	0.12	&	0.09	&	0.10	&	0.09	&	0.09	&	0.09	&	0.03	&	0.00	&	0.415	\\
Al	&	0.18	&	0.14	&	0.13	&	0.15	&	0.15	&	0.11	&	0.03	&	-0.01	&	0.369	\\
Si	&	0.17	&	0.13	&	0.10	&	0.13	&	0.14	&	0.06	&	0.04	&	-0.01	&	0.045	\\
Ca	&	0.02	&	0.01	&	0.07	&	0.05	&	0.03	&	0.06	&	-0.03	&	-0.02	&	0.050	\\
\ion{Sc}{I}	&	0.10	&	0.06	&	0.13	&	0.06	&	0.05	&	0.11	&	0.04	&	0.01	&	0.150	\\
\ion{Sc}{II}	&	0.06	&	0.05	&	0.08	&	0.06	&	0.06	&	0.06	&	0.00	&	-0.01	&	0.852	\\
\ion{Ti}{I}	&	0.10	&	0.08	&	0.10	&	0.07	&	0.05	&	0.08	&	0.03	&	0.03	&	0.079	\\
\ion{Ti}{II}	&	0.05	&	0.03	&	0.12	&	0.02	&	0.03	&	0.12	&	0.02	&	0.00	&	0.268	\\
V	&	0.26	&	0.25	&	0.11	&	0.17	&	0.16	&	0.11	&	0.09	&	0.09	&	0.000	\\
\ion{Cr}{I}	&	-0.02	&	-0.02	&	0.06	&	-0.01	&	-0.02	&	0.07	&	-0.02	&	0.00	&	0.510	\\
\ion{Cr}{II}	&	-0.05	&	-0.04	&	0.10	&	-0.02	&	-0.03	&	0.09	&	-0.03	&	-0.01	&	0.198	\\
Mn	&	0.23	&	0.21	&	0.18	&	0.16	&	0.17	&	0.13	&	0.07	&	0.04	&	0.033	\\
Co	&	0.17	&	0.15	&	0.10	&	0.09	&	0.08	&	0.10	&	0.08	&	0.07	&	0.000	\\
Ni	&	0.03	&	0.03	&	0.04	&	0.01	&	0.01	&	0.05	&	0.02	&	0.02	&	0.013	\\
Zn	&	0.11	&	0.14	&	0.19	&	0.10	&	0.12	&	0.19	&	0.00	&	0.03	&	0.884	\\
\ion{Ba}{II}	&	0.14	&	0.15	&	0.17	&	0.25	&	0.27	&	0.19	&	-0.11	&	-0.12	&	0.000	\\
          \hline
         
         \end{tabular}
        \end{table*}

For giant stars, most of the species show no significant differences between GWP and GWOP, with exception of Ba, Na, Ca, V, Co, and Mn. GWP have, on average, lower Ba abundances in comparison with the control sample by $\sim$0.11 dex. The KS test indicates, with a high confidence level, that both distributions are different. A similar behavior is observed for Na and Ca, although to a lesser degree. The opposite trend is found for V and Co: GWP have, on average, higher abundances in comparison with GWOP by $\sim$ 0.09 dex. In both cases the KS test gives a null probability for both distributions being identical. Mn shows a similar trend, although much less evident. Recently, \citet{Maldonado2013} reported possible differences for Na and Co. However, as we stated in section 3.3, due to the small number of lines used to determine the abundances of Na and Ba, trends regarding these elements should be taken with caution.

       \begin{figure*}[]
   \centering
\includegraphics[width=.70\textwidth]{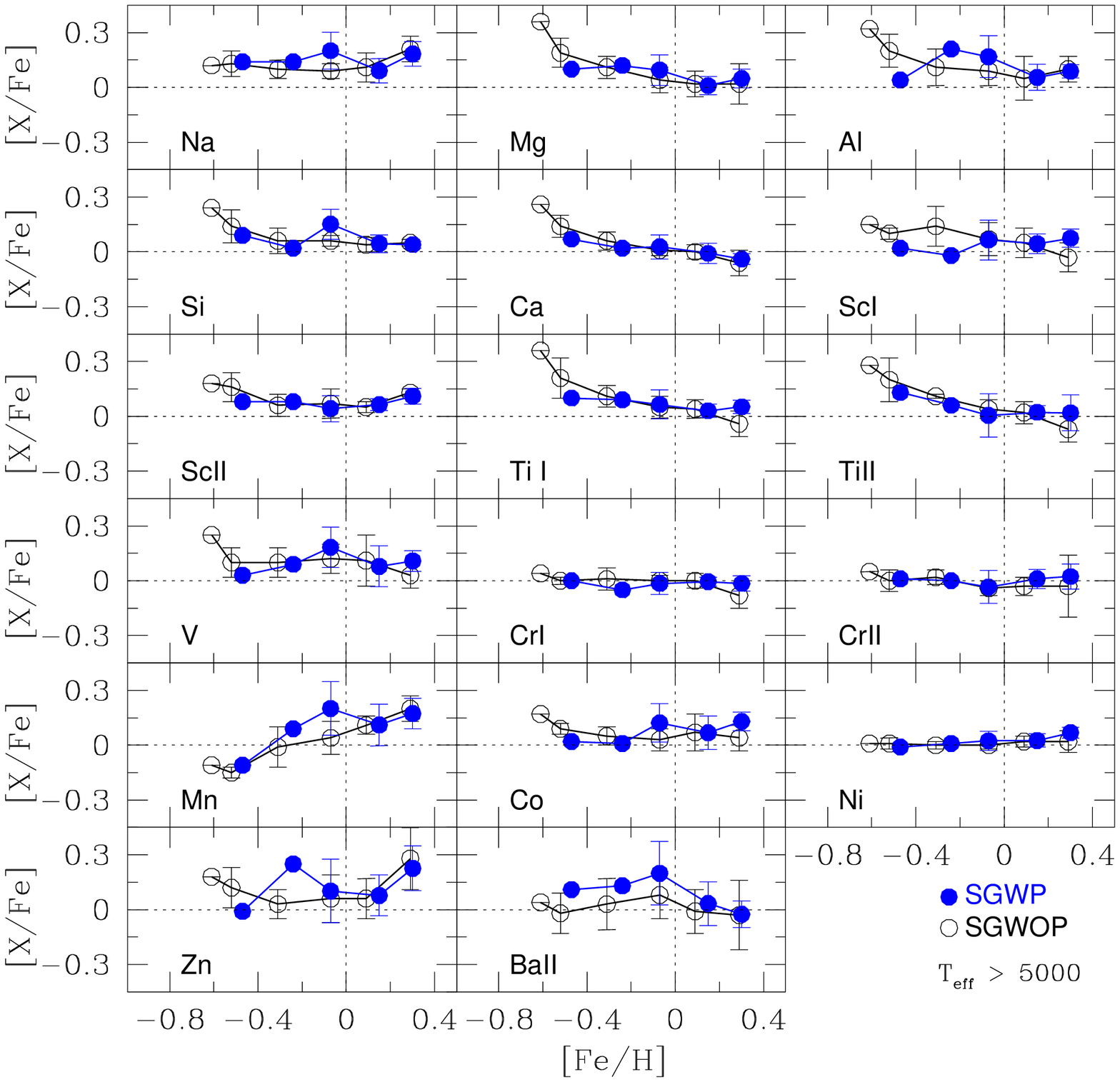}
   \caption{Same plot as Fig. 22, except that the data is averaged in [Fe/H] bins of 0.2 dex, centered at -0.6, -0.4, -0.2, 0.0, 0.2, and 0.4 dex. Dashed lines represent the solar values. Cool stars ($T_{\mathrm{eff}}$ < 5000 K) have been removed. The error bars represent the standard deviation about the mean value.}
              \label{FigGam}%
    \end{figure*}

    \begin{figure*}
   \centering
   \includegraphics[width=.70\textwidth]{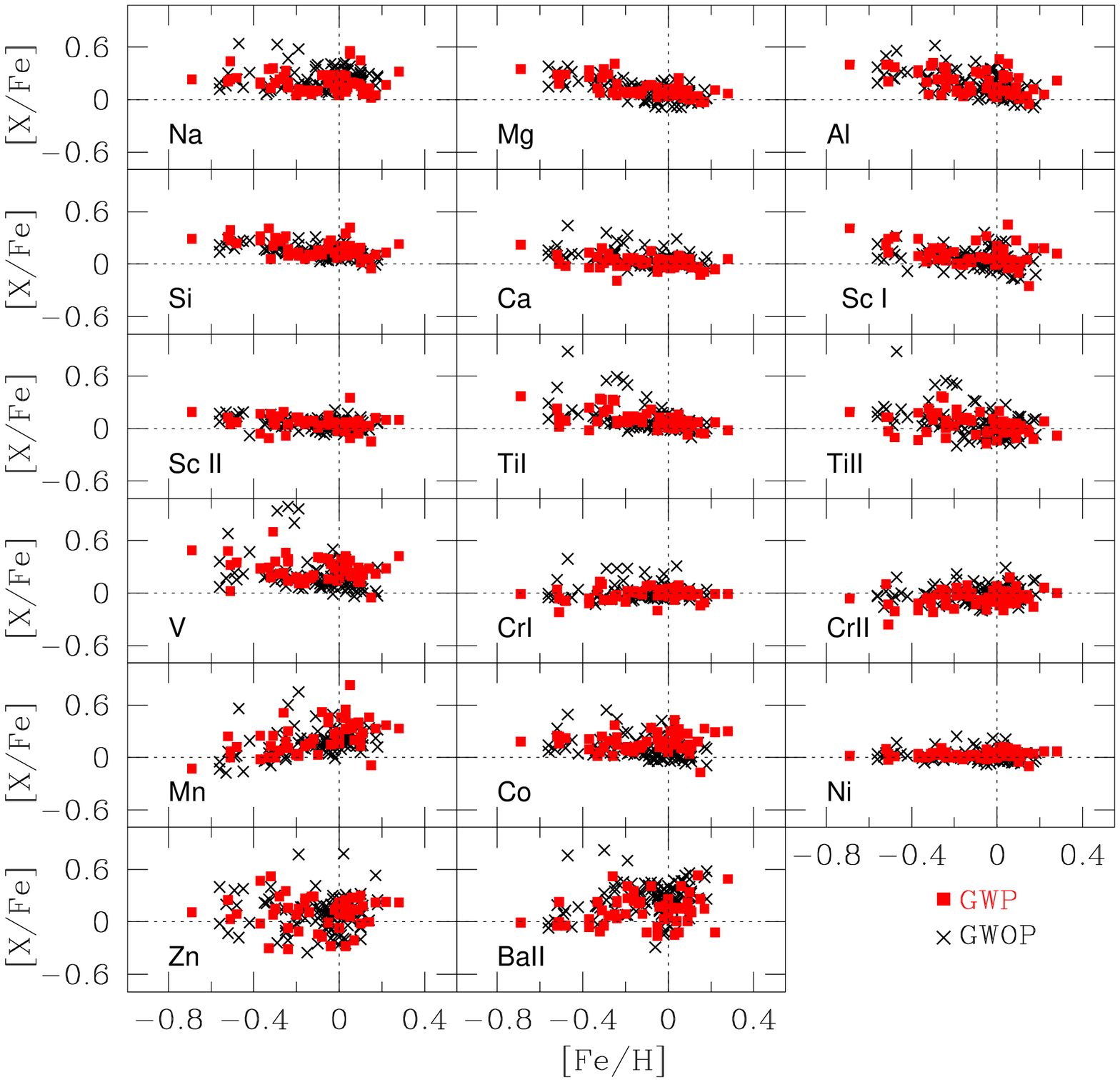}
   \caption{[X/H] vs. [Fe/H] for giant stars with planets (filled squares) and the control sample (black crosses). Dashed lines represent the solar values.}
              \label{FigGam}%
    \end{figure*} 

    \begin{figure*}
   \centering
   \includegraphics[width=.70\textwidth]{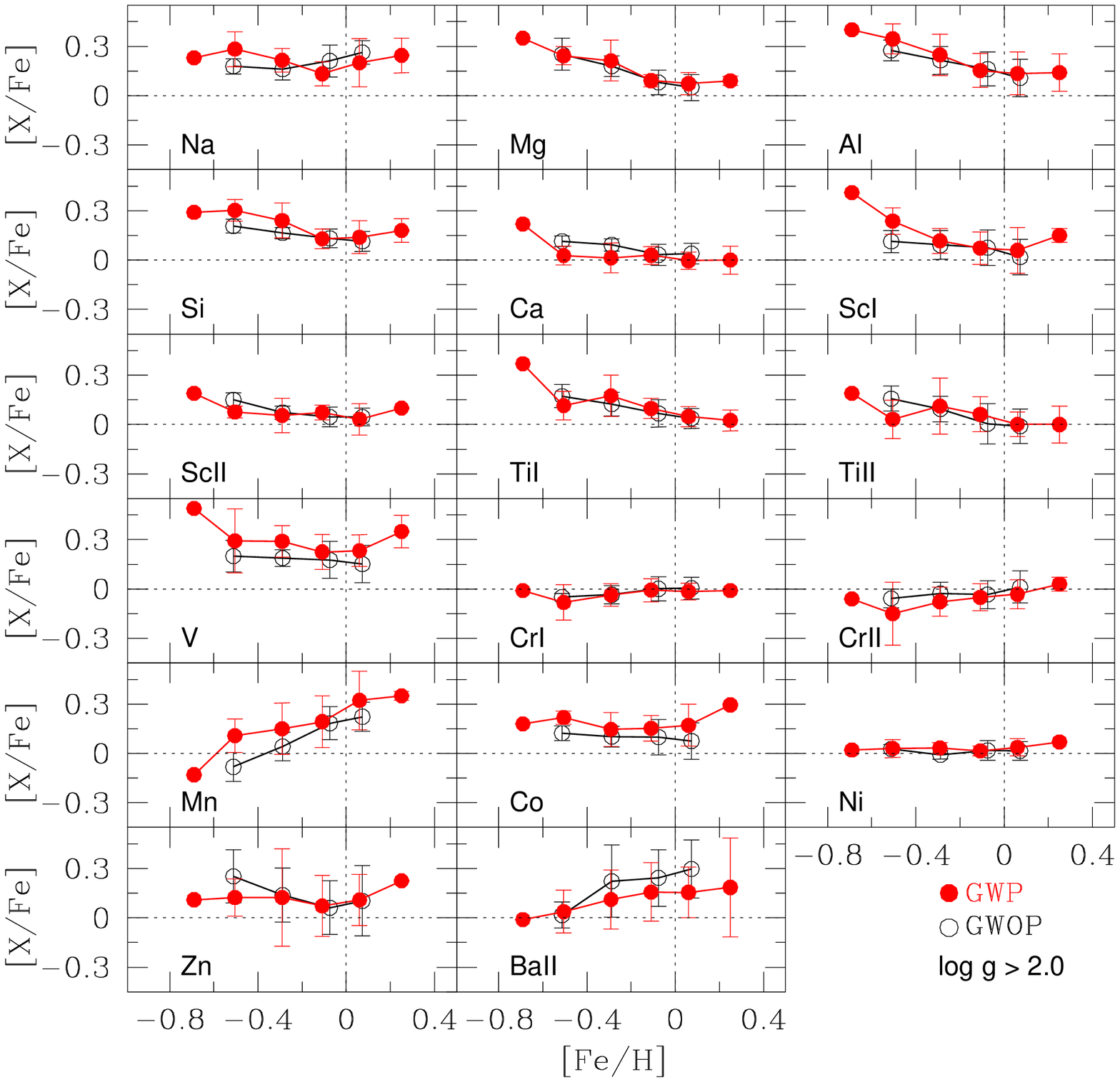}
   \caption{Same plot as Fig. 24, but data is averaged in [Fe/H] bins of 0.2 dex, centered at -0.6, -0.4, -0.2, 0.0, 0.2, and 0.4 dex. Low-gravity stars ($\log g $ < 2.0) have been removed. Dashed lines represent the solar values. The error bars represent the standard deviation about the mean value.}
              \label{FigGam}%
    \end{figure*}

\subsection{Search for differences in the [X/Fe] vs. [Fe/H] plane}       

The [X/Fe] vs. metallicity plots, generally used to study the chemical evolution trends of the Galaxy \citep{Edvardsson1993, Bensby2003, Fuhrmann2004}, have been recently used on main-sequence samples to analyze possible small differences between stars with and without planets for the same [Fe/H] \citep[e.g.,][]{Bodaghee2003, Neves2009, Kang2011, Adibekyan2012a}. In Figures 22 and 24 we show the plots [X/Fe] vs. [Fe/H] corresponding to 17 ions, for subgiant and giant stars in our sample. As before, stars with planets are marked with filled symbols (triangles for subgiants and squares for giants) and stars without planets are indicated with black crosses. Dashed lines mark the solar values. 

For the subgiant sample, in general, most of the species show no significant differences between stars with planets and the control sample. Both samples overlap in most of the [Fe/H] bins. However, for [Fe/H] > 0 we note a slight overabundance in V, Co, Mn and less apparent in Ti and Cr for the SGWP over the SGWOP. We note, however, that the four stars with the highest [X/Fe] values of these elements (\object{HD 158038}, \object{HD 73534}, \object{HD 27442}, and \object{HD 177830}) belong to the planet-host stars group with $T_{\mathrm{eff}}$ < 5000 K. A similar situation occurs for [Fe/H] < 0 where 3 stars from the control sample have $T_{\mathrm{eff}}$ < 5000 K. 

Gilli et al. (2006) and Neves et al. (2009) reported that the removal of cooler stars from the dataset reduces the dispersion, in particular for Ti, V, Co, Sc, and Al. An abundance overestimation in cooler stars could be related to blending effects, deviations from the excitation or ionization equilibrium conditions (Neves et al. 2009) and even NLTE effects \citep{Bodaghee2003}. In Figure 23 we present [X/Fe] vs. [Fe/H] trends using  binned average values for each element, and removing cool stars ($T_{\mathrm{eff}}$ < 5000). The [Fe/H] bins are 0.2 dex wide and centered at -0.6, -0.4, -0.2, 0.0, 0.2, and 0.4 dex. The differences for the elements discussed above exist only in the highest metallicity bins. For \ion{Ba}{II}, SGWP abundances seem to be systematically higher than those for SGWOP for the lower metallicity bins. However, we note that in all cases the discrepancies are within the scatter. Additionally, the two lowest metallicity bins for SGWP contain only one star. In summary, although there are some differences in the [X/Fe] vs. [Fe/H] plane, they are subtle and blurred by the high scatter. It would be very interesting, in the future, to re-make this analysis with a larger sample of subgiant stars with planets. 

In general, these results agree with previous works on main-sequence stars that find no significant differences between stars with and without planets \citep[see][]{Beirao2005, Bodaghee2003, Fischer2005, Gilli2006, Takeda2008, Neves2009}. However, these and other studies reported possible discrepancies for some elements, as for instance \citet{Sadakane2002} for V and Co, Bodaghee et al. for V, Mn, Ti, and Co; and Gilli et al. for V, Co, Mg, and Al. Other authors have found significant differences in other elements between dwarf stars with and without planets. For instance, \citet{Robinson2006} reported overabundance in Ni and Si whereas \citet{Gonzalez2007} found differences mainly for Al and Si. Recently, \citet{Kang2011} reported systematic overabundance of Mn in dwarf stars with planets while \citet{Adibekyan2012a} found an overabundance of $\alpha$-elements in main-sequence stars with planets at low metallicities.    
 
In the case of giant stars, the same general trends observed for the subgiant stars can be seen in Figure 24. Overall, the abundance differences for stars with planets, relative to the control sample, are not easily distinguished. This is in general agreement with studies that analyzed a sample of giants of similar size \citep[e.g.,][]{Luck2007, Takeda2008, Wang2011}. In comparison with the subgiant sample, the giant group shows a higher dispersion for species such as: V, Mn, Zn, and Ba. Furthermore, some outliers with abnormal high abundances can be easily identified for Na, \ion{Ti}{I}, \ion{Ti}{II}, V, Mn, Co, Zn, and Ba. These outliers correspond to one planet-host star (\object{HD 1690}) and to 7 stars from the control sample (\object{HD 32887}, \object{HD 50778}, \object{HD 107446}, \object{HD 124882}, \object{HD 131109}, \object{HD 151249}, and \object{HD 152980}), all of which have surface gravities below 2.0 dex. 

  \begin{figure}
   \centering
   \includegraphics[width=.40\textwidth]{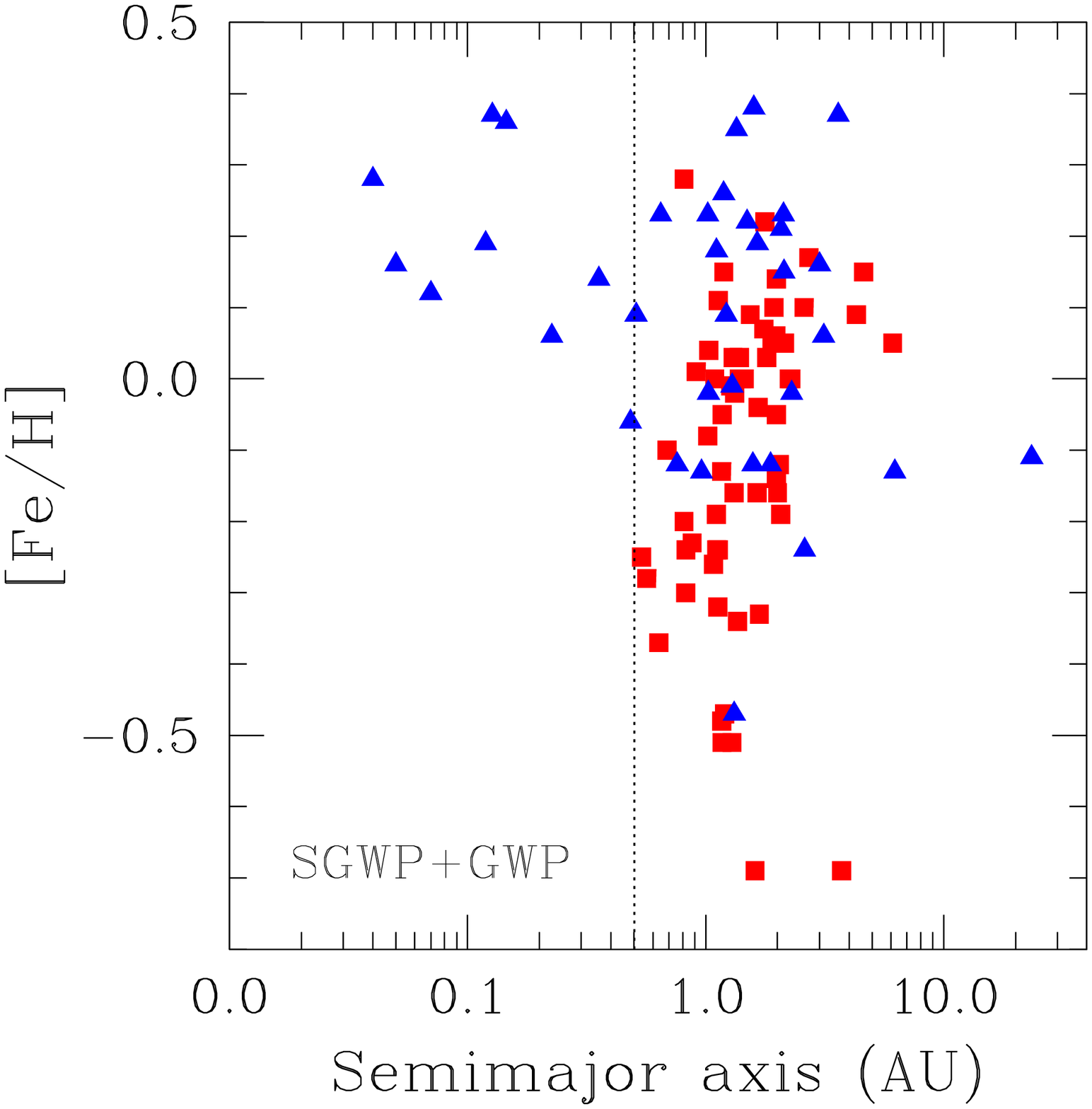}
   \includegraphics[width=.40\textwidth]{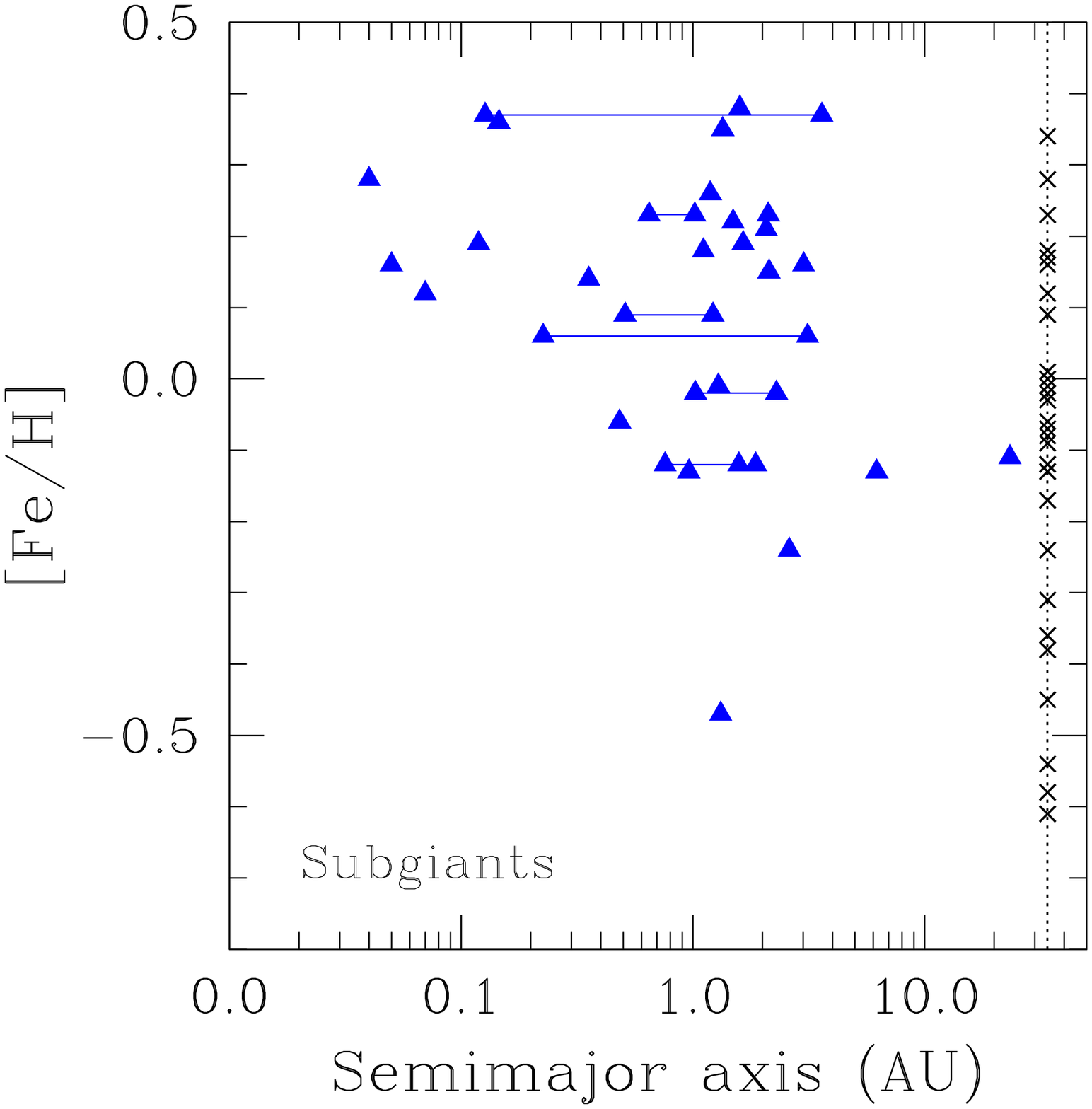}
   \includegraphics[width=.40\textwidth]{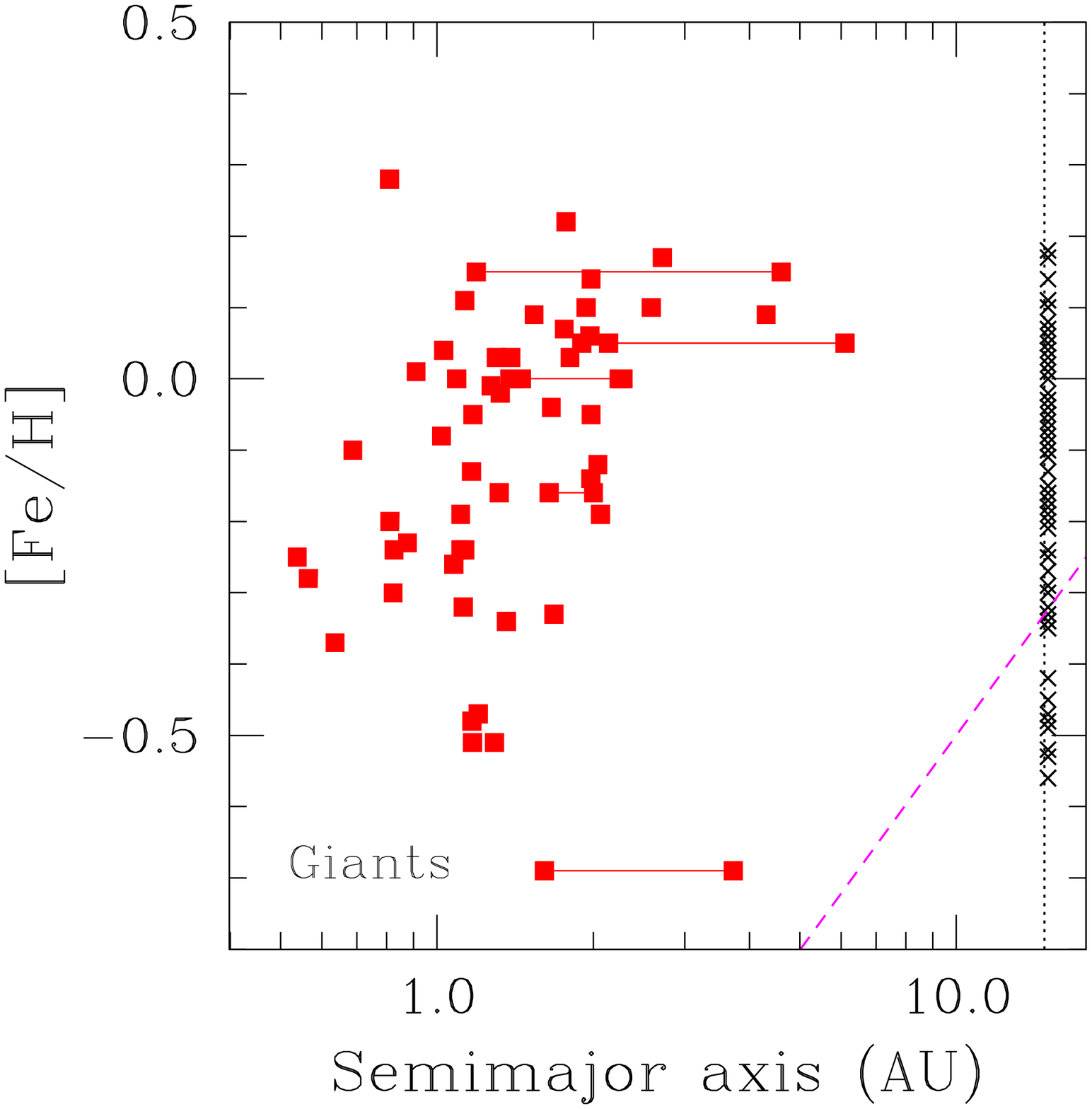}
   
   \caption{\textit{Top panel:} Stellar metallicity vs. semimajor axis of planets around evolved (giants and subgiants). Planets around giants are indicated by red squares and planets in subgiants by blue triangles. Black dashed line at 0.5 UA, indicates an observational limit for which none planet around giants has been detected. \textit{Middle and bottom panels:} [Fe/H] vs. semimajor axis   for planets around subgiants and giants, respectively. Here, the control samples are represented with black crosses. The magenta dashed line on the bottom panel represents the critical metalliticy for planet formation in the core accretion model, see Sec. 6.3 for details.}
              \label{FigGam}%
    \end{figure}
  
In Figure 25 we present the same plot as before but for binned average values, where the outliers have been eliminated. Small differences between the two samples for some of the elements show up in this figure. For example, Na presents a slope change: GWP, compared with GWOP, show higher abundances in the lower metallicity range. This situation is reversed from $\sim$ -0.10 dex for GWOP, having relatively higher abundances. A similiar situation, but without the reversal, seems to occur for Si at low metallicities, where GWP show a small excess in abundance compared with GWOP. On the other hand, planet-host stars have a systematic overabundance in V, Co, and Mn for almost the entire metallicity range, whereas the opposite seems to occur for Ba, for which GWP show systematically lower abundances than the control sample. As we stated before for subgiants, these results should be taken with caution because the differences are small and very close or within the dispersion.  

  \begin{figure}
   \centering
   \includegraphics[width=.50\textwidth]{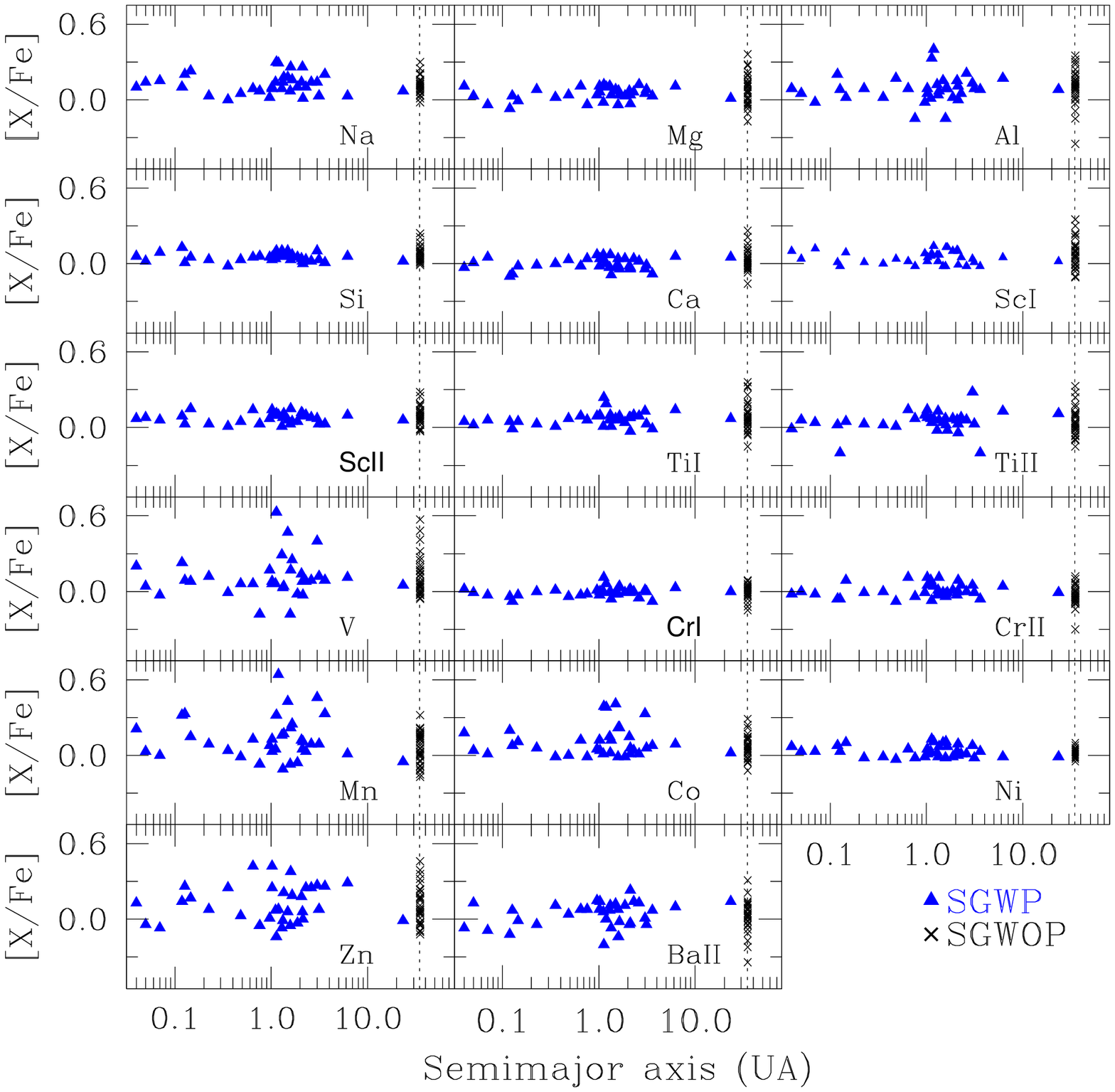}
      \includegraphics[width=.50\textwidth]{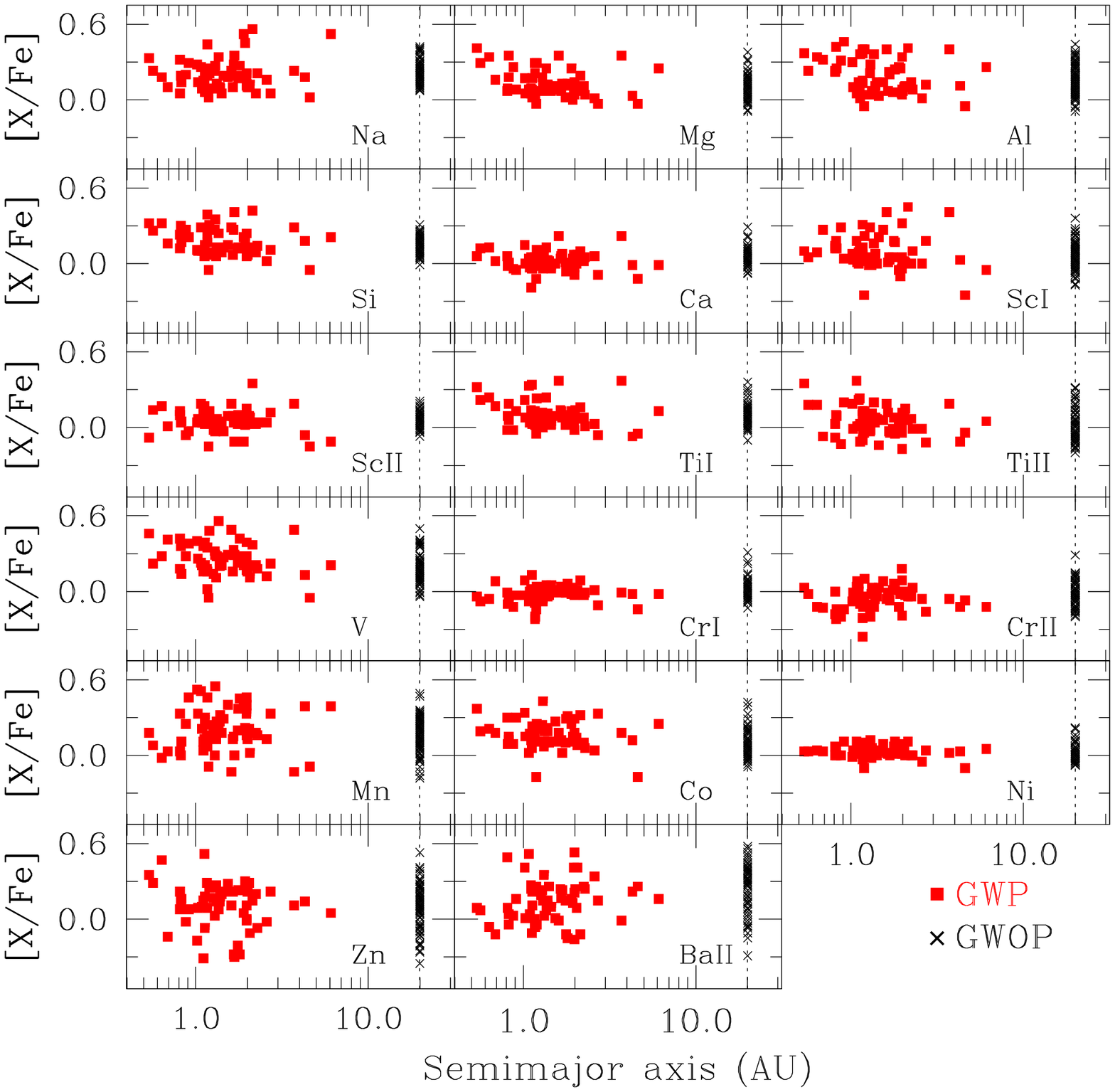}
   \caption{[X/Fe] ratios vs. orbital distance of planets around subgiants (\textit{top panel}) and giants (\textit{bottom panel}). Colors and symbols are as in Figure 26.}
              \label{FigGam}%
    \end{figure}

\section{Planetary properties}    
Many authors have searched for correlations between stellar and planetary properties. Most of these works have focused on solar-type \citep{Fischer2005, Beirao2005, Sousa2008, Kang2011, Adibekyan2013} and low-mass stars (Bonfils et al 2005, Johnson \& Apps 2009, Schlaufman \& Laughlin 2010, Buchhave et al. 2012). In this section we search for relations between the properties of planets (minimum mass, semimajor axis, orbital eccentricity, and multiplicity)\footnote{Planetary properties were compiled from the Extrasolar Planets Encyclopaedia and the Exoplanet Data Explorer.}  and the elemental abundances of their evolved-host stars. In addition, we explore potential differences between the planetary properties of planets around giants and subgiants. A positive identification of such correlations would provide insight into the formation and evolution scenarios of giant planets around more massive stars. 

\subsection{Multi-planet systems and metallicity}

\citet{Wright2009} studied 28 multi-planet systems and found that, on average, these systems are orbiting more metal-rich stars than single-planet systems by $\sim$0.1 dex, suggesting that metallicity traces not only planet occurrence, but also multiplicity among main-sequence stars with planets. 

To date, there are $\sim$16 multi-planet systems, with two planets, detected around evolved stars (exoplanet.eu), of which 11 are included in our sample (5 for giants and 6 for subgiants). This number is too small to make a thorough statistic analysis, however it is notable that 5 multi-planet systems around subgiants and 3 in giants orbit stars with [Fe/H] > 0. In addition, we find that single and multi-planet systems around subgiants have mean [Fe/H] values of +0.08 dex and +0.11 dex, respectively, and median values of +0.16 dex and +0.08 dex. On the other hand, single and multi-planet systems orbiting giants have mean values of -0.12 dex and -0.13 dex, respectively, but their median values are -0.08 dex and +0.0 dex. In the last case we note the inclusion of the very metal-poor star \object{HD 47536} with [Fe/H] = -0.69 dex. This giant is a peculiar object which stellar parameters depart from average values of the GWP and GWOP samples, being a very low-mass star (0.91 $M_{\mathrm{\sun}}$) from the thick disk, very luminous with a large radius (22.4 $R_{\mathrm{\sun}}$) and one of the oldest in the giant sample (10.15 Gyr). If this particular object is excluded, the mean and median for multi-planet systems around giants are +0.01 dex and +0.02 dex, respectively. In this case, multi-planet systems around giants result, on average, more metal-rich than single-planet systems by $\sim$0.13 dex. Even though a small metallicity excess seems to emerge for multi-planet systems around evolved stars, a larger sample is needed to test this initial trend.

\subsection{Orbital distance} 
Figure 26 shows [Fe/H] as a function of the semimajor axis $a$ for planets around giant and subgiant stars. In the top panel, the black dashed line at $\sim$ 0.5 AU represents the limit from which no planets have been detected orbiting giant stars \citep[e.g.,][]{Sato2008, Sato2010, Niedzielski2009}. It is possible that the occurrence of short-period planets around massive stars be scarce or null as a consequence of a different planet formation and evolution scenario. On the other hand, short period variations of giant stars \citep[e.g.,][]{Hatzes1993, Hatzes1994, Hekker2006} might hide the radial velocity signals of close-in planets. However, it has also been suggested that, as the star evolves to the giant branch, close-in planets might be swallowed by the expanded star \citep[e.g.,][]{Johnson2007a, Sato2008, Villaver2009, Nordhaus2010}. \citet{Siess1999} investigated the effects of the accretion of sub-stellar companions by red giant stars and disscussed several observational signatures of such events, including increased stellar rotation and modifications of the photospheric chemical composition, among others. 

A significant drop in the rotational velocities occurs as the stars leave the main-sequence and evolve to the giant branch. The reduction in the rotational velocity is the result of the stellar radius expansion as well an efficient process of magnetic braking \citep{Gray1989, doNascimento2000}. Consequently, giant stars are characterized by slow rotational velocities, typically v$\sin i \lesssim$ 2-3 km $ s^{-1}$ \citep{Gray1981, Gray1982, deMedeiros1996, Massarotti2008a, Carlberg2011}. However, a small percentage of red giants departs from this behavior and shows a v$\sin i$ excess of at least $\simeq$ 10 km $s^{-1}$ \citep[e.g.,,][]{deMedeiros1999, Massarotti2008b, Carlberg2011}. Several studies have suggested that viscous and tidal forces can cause a substellar companion to spiral-in toward the star, transfering orbital angular momentum into the stellar envelope, and producing a substantial increase in the rotational velocity of the star \citep[e.g.,][]{Siess1999, Carney2003, Massarotti2008b, Carlberg2011}. Adopting the cut-off v$\sin i$ of 8 km $ s^{-1}$ between rapid and slow rotators given by \citet{Carlberg2012}, we find no evidence of rapid rotation among our sample of giants with or without planets. All the giants listed in Table 2 have v$\sin i \lesssim$ 4.67 km $ s^{-1}$ and errors below 1.5 km $ s^{-1}$. In a forthcoming paper (Jofr\'e et al., in prep) we will analyze other possible signals of planet engulfment by red giant stars. 

In the middle panel of Figure 26 the semimajor axis vs. [Fe/H] for subgiants is shown. Multi-planet systems are connected by solid lines and stars without planets are represented with black crosses. As it can be seen, planets with $a$ $\gtrsim$ 0.5 AU orbit around stars with a wide range of metallicities, including subsolar values, whereas planets closer than $\sim$ 0.5 AU are found around subgiants with [Fe/H] > 0. Recently, \citet{Adibekyan2013}, analyzing a large sample of FGK dwarf hosts, suggested that planets orbiting metal-poor stars have longer periods than those in metal-rich systems. On the other hand, for giant stars (bottom panel), planets with $a$  $\lesssim$ 1 AU are hosted by giants with subsolar metallicities, whereas planets with $a$ > 1 AU are also found orbiting giants with [Fe/H] $\gtrsim$ 0. 

Finally, in Figure 27 we show the [X/Fe] ratios vs. the orbital distance for subgiants (\textit{top panel}) and giants (\textit{bottom panel}). No correlation between abundances and orbital distance seems evident in these figures.

\subsection{Planetary mass} 

\begin{figure}
   \centering
    \includegraphics[width=.40\textwidth]{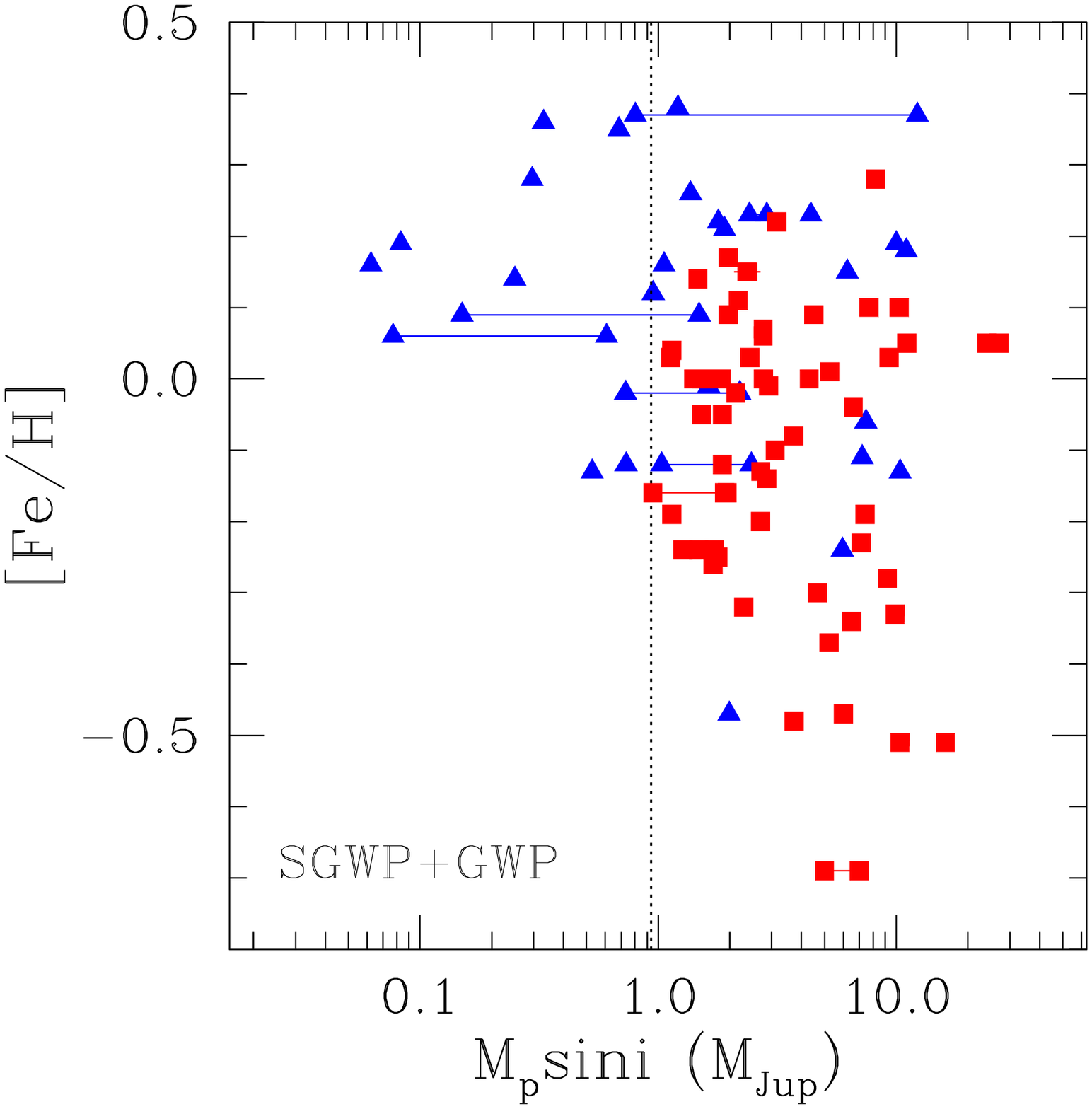}
   \includegraphics[width=.40\textwidth]{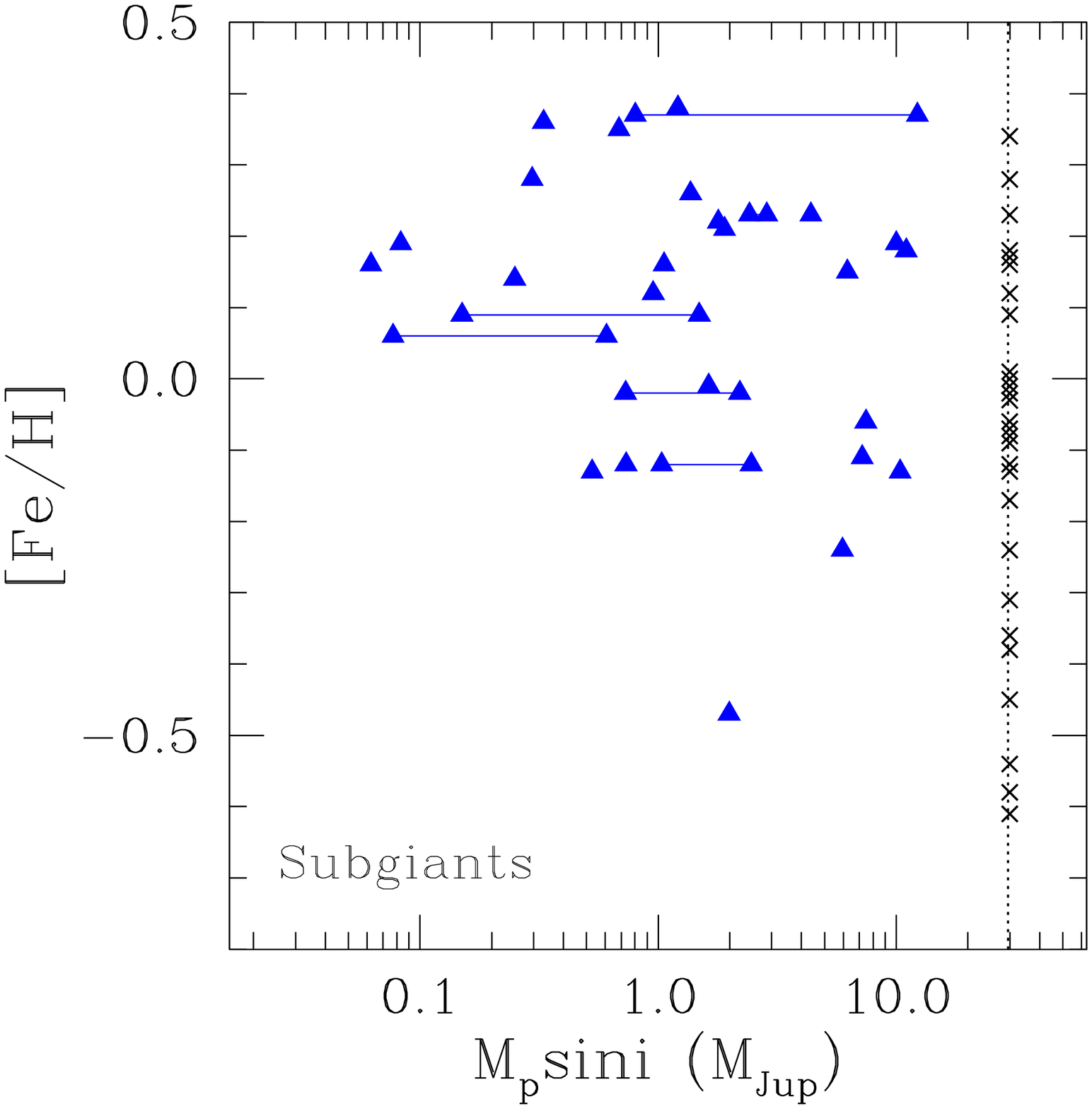}
   \includegraphics[width=.40\textwidth]{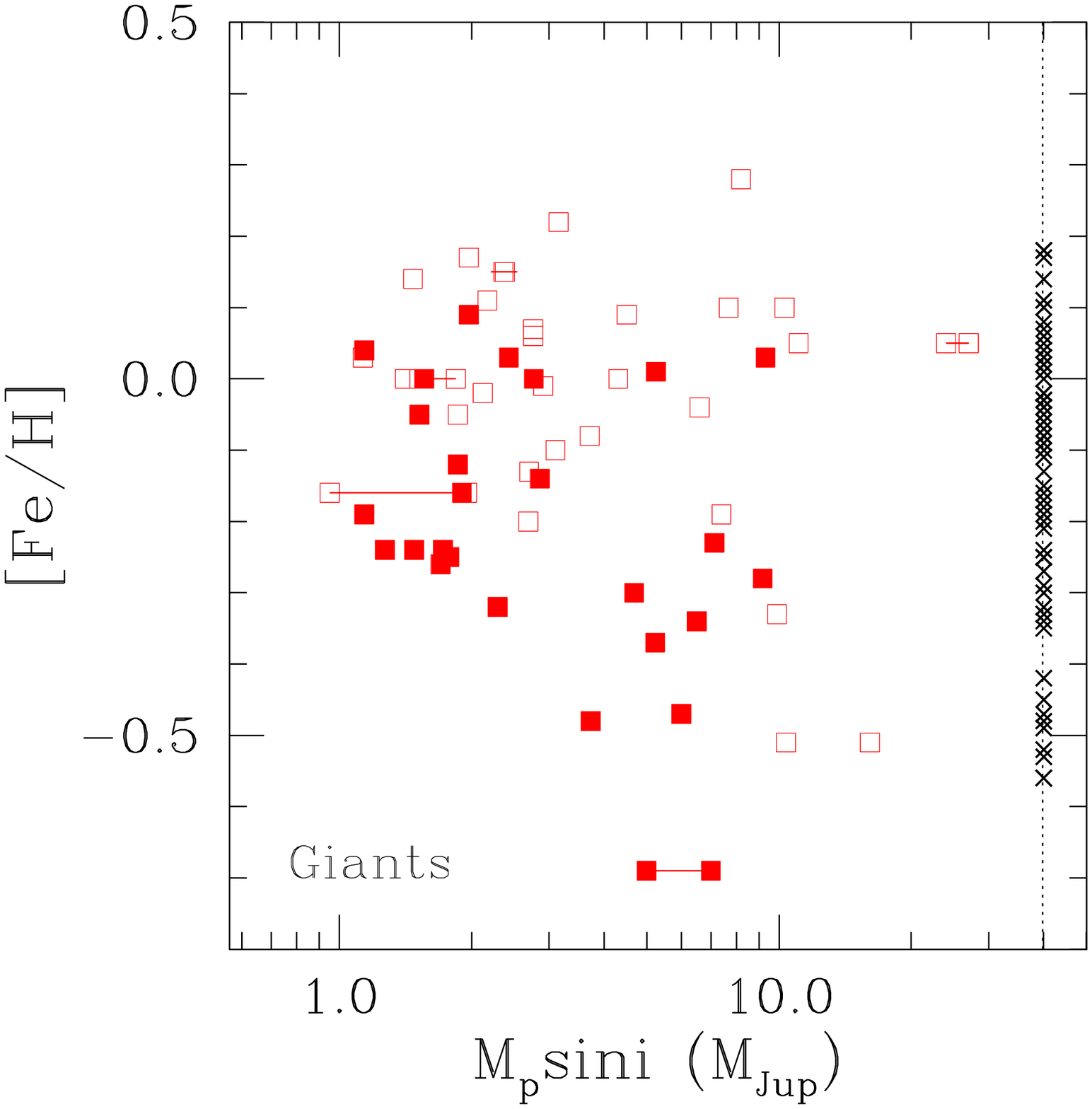}
   \caption{\textit{Top panel:} Stellar metallicity vs. the minimum planetary mass of planets around evolved stars (giants and subgiants). Red squares and blue triangles represent giant and subgiant stars, respectively. The dotted black line represents 0.95 $M_{\mathrm{Jup}}$. Multi-planet systems are connected by solid lines. \textit{Middle panel:} [Fe/H] vs. $M_{\mathrm{p}} \sin i$ for planets around subgiants. \textit{Bottom panel:} [Fe/H] vs. $M_{\mathrm{p}} \sin i$ for planets around giants. Here, empty red squares indicate giants with $M_{\star}$ > 1.5 $M_{\mathrm{\sun}}$ and filled squares indicate giants with $M_{\star} \leq$ 1.5 $M_{\mathrm{\sun}}$. In all cases, control samples are marked with black crosses.}
              \label{FigGam}%
    \end{figure}

\begin{figure}
   \centering
   \includegraphics[width=.50\textwidth]{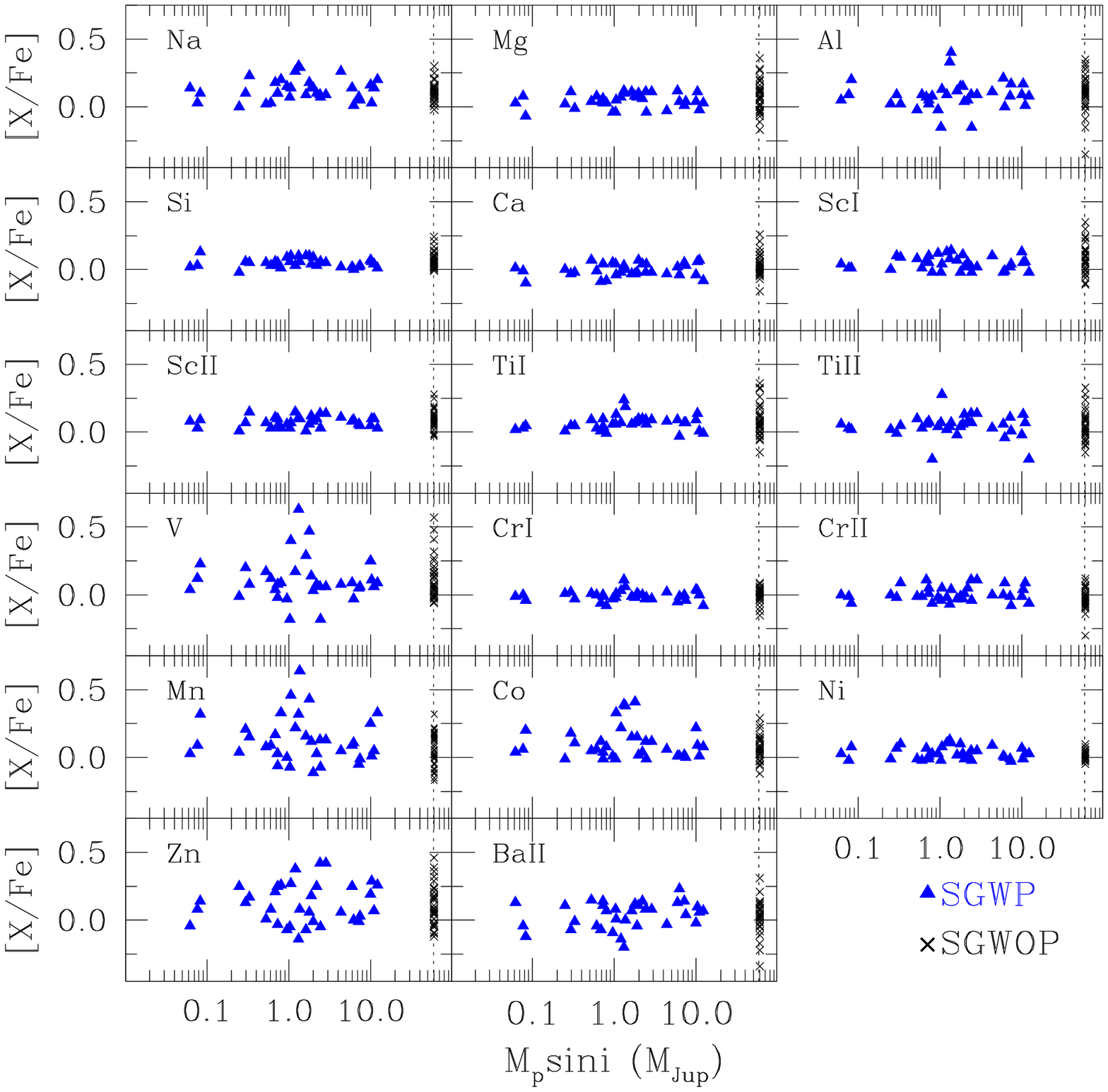}
      \includegraphics[width=.50\textwidth]{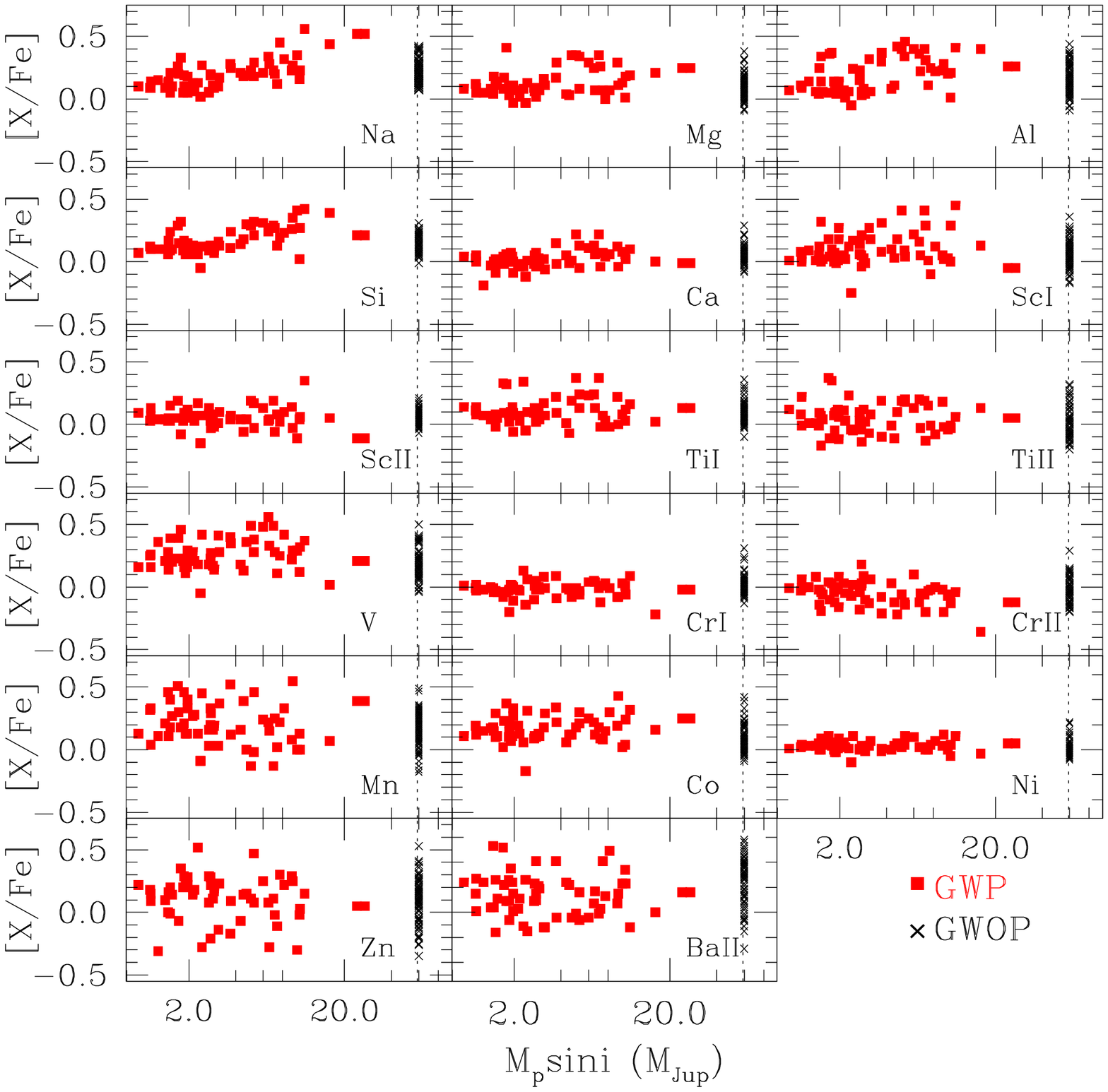}

   \caption{[X/Fe] ratios vs. $M_{\mathrm{p}} \sin i$  of planets orbiting subgiants (\textit{top panel}) and giants (\textit{bottom panel}). Colors and symbols are as in Fig. 26.}
              \label{FigGam}%
    \end{figure}

Figure 28 shows the stellar metallicity as a function of the planetary minimum mass for planets around giants and subgiants. From the top panel, it can be seen that the mass of planets around subgiants include the Neptune-class, but only planets with  $M_{\mathrm{p}} \sin i \gtrsim$ 0.95 $M_{\mathrm{Jup}}$ (dotted line) have been detected around giant stars. This probably represents an observational bias of the Doppler technique due to the larger intrinsic variability of giants that, combined with their relatively higher masses, restrict the detection only to gas giant planets. However, it has been suggested that, at least in dwarf stars, the detected planets might represent the bulk of planetary mass in the inner few AU of the host stars \citep{Fischer2005}. 

In addition, several authors have shown that giant planets around evolved stars are, on average, more massive than planets around solar-type stars \citep{Lovis2007, Pasquini2007, Dollinger2009}. However, from the top panel of Figure 28, it can be seen that, neglecting planet around subgiants with masses below 0.95 $M_{\mathrm{Jup}}$, no significant difference seems to be present between the planets around giants and subgiants. If the planets around subgiants with $M_{\mathrm{p}} \sin i \lesssim$ 0.95 $M_{\mathrm{Jup}}$ are excluded, given the possible observational bias of these type of planets around giants,  we find that planets around subgiants and giants have median $M_{\mathrm{p}} \sin i$ values of 2.31 $M_{\mathrm{Jup}}$ and 2.76 $M_{\mathrm{Jup}}$, respectively. Moreover, 55\% and 43\% of planets around giants and subgiants, respectively, have $M_{\mathrm{p}} \sin i >$ 2.5 $M_{\mathrm{Jup}}$. The KS test gives a $\sim$ 90\% probability that both planetary mass distributions are drawn from the same population. Recently, \citet{Jones2014} arrived to a different conclusion, suggesting that the mass distributions of planets around giants and subgiants are different. These authors considered all the planets around subgiants, including those with $M_{\mathrm{p}} \sin i \lesssim$ 0.95 $M_{\mathrm{Jup}}$, which we have excluded in this analysis. Jones et al. applied, however, a cut-off in stellar mass, considering stars within the 0.9 -- 2.0 $M_{\mathrm{\odot}}$ range. Furthermore, Jones et al. found no relationship between planetary mass and stellar mass, which might support our results.    

Several studies, on solar-type stars with planets, have found evidence that the planet-metallicity correlation might be weaker for low-mass planets, suggesting that stars hosting less massive planets have, on average, lower metallicity than stars with higher-mass planets \citep{Udry2007, Sousa2008, Sousa2011, Johnson2009, Bouchy2009, Ghezzi10a, Mayor2011, Buchhave2012, Neves2013}. In particular, \citet{Mayor2011} found that the metallicity distribution of dwarfs hosting planets with masses less than 30-40 $M_{\mathrm{\oplus}}$ (Super-Earth and Neptune-type planets) is clearly shifted to lower [Fe/H] values compared with that of stars hosting gas giant planets with masses above 50 $M_{\mathrm{\oplus}}$. Interestingly, for planets around subgiants (Figure 28, middle panel), it can be seen a similar planetary-mass limit to that found  by \citet[Fig.17]{Mayor2011}: subgiants hosting planets less massive than $\sim$ 0.11 $M_{\mathrm{Jup}}$ (35 $M_{\mathrm{\oplus}}$) have [Fe/H] $\lesssim$ +0.20, whereas subgiants hosting more massive planets have a higher scattering in metallicities, including [Fe/H] > +0.2. However, because of the difference in size between our sample and that of Mayor et al., specially for stars with low-mass planets, this result represents only an emerging trend and should be taken with caution.   

On the other hand, for planets around giants (Figure 28, bottom panel), although not clear, metallicity seems to decrease for higher planetary masses. However, as first noted by \citet{Maldonado2013}, this trend is mainly due to stars less massive than 1.5  $M_{\mathrm{\sun}}$. These giants are indicated by filled red squares in the bottom panel of Figure 28, whereas giant stars with masses above 1.5 $M_{\mathrm{\sun}}$, represented by empty squares, seem to follow the general trend of subgiants. 

The lack of an overabundance of metals in main-sequence stars hosting low-mass planets in comparison with those hosting gas giant planets has been explained within the core accretion theory for planet formation \citep{Mayor2011, Buchhave2012}. A high metallicity enviroment (high dust-to-gas ratio) allows the rapid formation of dust grains which coagulate to form the planetesimals. These planetesimals merge to form planetary cores and when they reach $\sim$ 10 $M_{\mathrm{\oplus}}$ start to accrete gas from the surrounding disk to form gas giant planets \citep{Pollack1996}. The accretion of gas continues until the dissipation of the disk, usually in a few million years. However, low-metallicity enviroments might not form the planetary cores with the critic masses rapidly enough to accrete a considerable amount of gas before the disk is photo-evaporated. 

The low-metallicity of giant stars with planets, including 11 objects with [Fe/H] $\lesssim$ -0.30, raises the question about the giant planet formation within the, metallicity-dependent, core accretion model. On the bottom panel of Figure 26 we plot, with a magenta dashed line, the minimum metallicity required for planet formation in the core accretion model found by \citet{Johnson2012}. This lower critical metallicity function depends on the semimajor axis as: $[Fe/H]_{\mathrm{critic}}$ $\simeq$ --1.5 + $\log(a/1 AU)$. If a planetary system lies at the right side of this line (forbidden zone),  it would be a challenge to the core accretion model \citep{Johnson2012}. The closest object to the forbidden zone corresponds to the multi-planet sytem around HD 47536, but no planet is beyond the critic limit. Furthermore, it has been suggested that the more massive proto-planetary disks around higher-mass stars \citep{Natta2000}, such as giants, might compensate or counterbalance their lower metallicities allowing giant-planet formation \citep{Takeda2008, Ghezzi10a, Maldonado2013}. Several numerical modeling in the framework of the core acretion theory, suggest that the planet-formation efficiency depends on the mass of protoplanetary disks \citep[e.g.,][]{Kennedy2008}. These results seem to be supported by the observational findings of \citet{Johnson2010}, that suggest a planet ocurrence increase from 3\% for red dwarfs up-to 14\% for A-type stars, at Solar metallicity. Another possibility is that gas giant-planet formation occurs by the disk instability model \citep{Boss1997, Boss1998, Boss2010}. In this scenario, if the protoplanetary disk is massive enough ($\simeq$ 0.1$ M_{\mathrm{\sun}}$) it can become gravitationally unstable and fragments into dense, self-gravitating clumps of gas and dust which would contract to form giant protoplanets. Furthermore, in this model, giant-planet formation would occur independently of the metallicity enviroment \citep{Boss2002}. In this context, planets around metal-poor giants might be naturally explained. Moreover, \citet{Cai2006} and \citet{Meru2010} suggest that low-metallicity enviroments enhance planet-formation, which might explain the trend observed for giants less massive than 1.5 $M_{\mathrm{\sun}}$, where the more massive planets seem to orbit the more metal-poor stars (filled squares on the bottom panel of Figure 28).

\begin{figure}
   \centering
   \includegraphics[width=.47\textwidth]{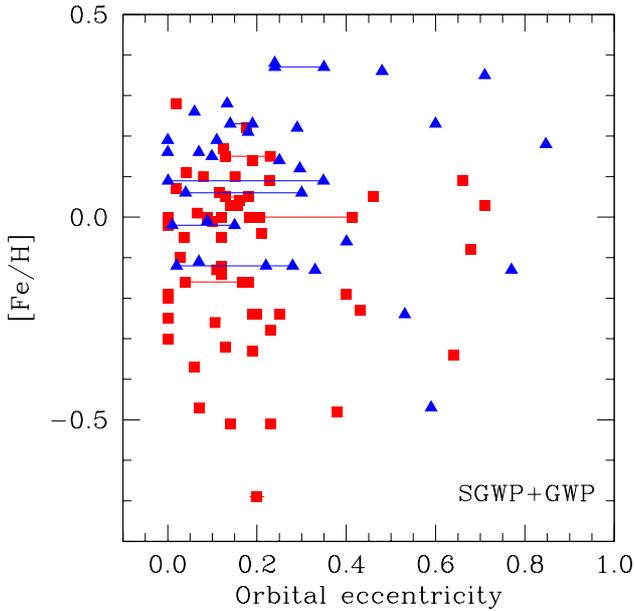}
   \caption{Stellar metallicity vs. orbital eccentricity of planets around evolved stars (giants and subgiants). Colors and symbols are as in Fig. 26.}
              \label{FigGam}%
    \end{figure}

      \begin{figure}
   \centering
   \includegraphics[width=.50\textwidth]{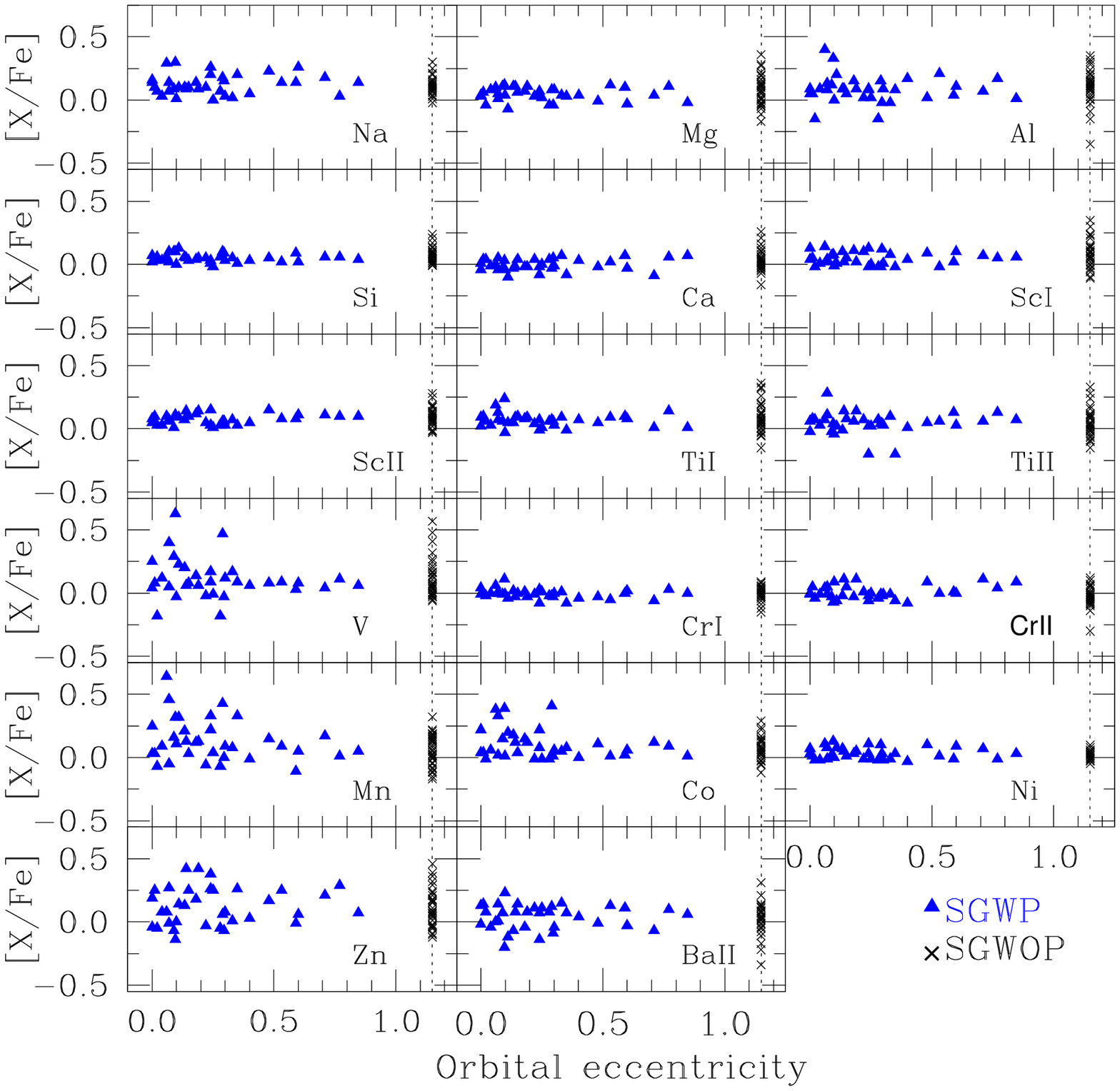}
      \includegraphics[width=.50\textwidth]{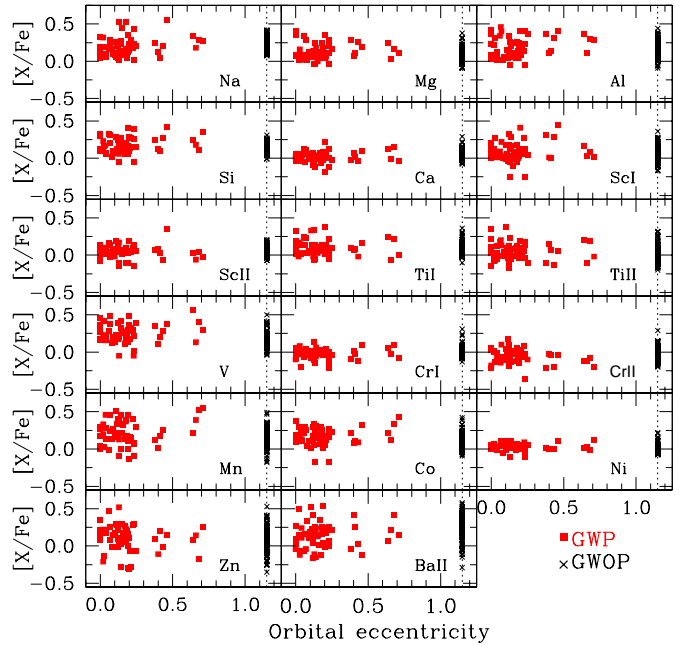}

   \caption{[X/Fe] ratios vs. orbital eccentricity of planets around subgiant (\textit{top panel}) and giant stars (\textit{bottom panel}). Colors and symbols are as in Fig. 26.}
              \label{FigGam}%
    \end{figure}

Finally, Figure 29 (top panel) shows  [X/Fe] vs. planetary mass for subgiant stars. No clear relation between the abundances studied here and planetary mass was found. This result is similar to that obtained for the main-sequence stars \citep{Beirao2005, Kang2011}. However, for giant stars with planets (bottom panel), there appears to be a slight correlation between Na and Si abundances and the planetary mass. For these two elements the [Na/Fe] and [Si/Fe] ratios seem to increase with the planetary mass. The same might be also true for Mg, Al and \ion{Sc}{I}, although the dispersion is larger. We have no explanation for these observational trends and, at this point, it is worth to raise a word of caution regarding the correlations for Na, Al, and Mg due to the limited number of lines (2-3) employed in their abundances computation.

\subsection{Eccentricity}

In Figure 30 we plotted metallicity vs. orbital eccentricity for the planets around subgiants and giants\footnote{The planet HD 47536 c is not included in the analysis of this subsection because, so far, the eccentricity has not been determined.}. No clear trends are evident, which seems to agree with the lack of correlations found for solar-type stars \citep{Fischer2005, Wright2009}. In addition, Figure 31 shows the plots for [X/Fe] vs. eccentricity for subgiants (top) and giants (bottom). As in the case of iron, there are no evident correlations neither for subgiants nor for giants.

On the other hand, several studies have suggested that planets around evolved stars display, on average, lower eccentricities than planets orbiting solar-type stars \citep{Maldonado2013, Jones2014}. We confirm this observational result, comparing the median eccentricity of planets around main-sequence stars, $\sim$ 0.25 \citep{Maldonado2013}, with the median eccentricity of planets orbiting our sample of evolved stars, which is 0.15. However, if evolved stars are divided in giants and subgiants, we obtain a median eccentricity of 0.14 for the planets around giants and of 0.23 for planets around subgiants (and of 0.20 if we only consider planets with $M_{\mathrm{p}} \sin i$ > 0.95 $M_{\mathrm{Jup}}$). Furthermore, 74\% of the planets around giants have eccentricities below 0.2, while this fraction is 56\% for planets orbiting subgiants. The KS test yields a probability of $\sim$ 1\% that eccentricities of planets around giant and subgiant stars derive from the same distribution. This probability is about 5\% if planets around subgiants with masses above 0.95 $M_{\mathrm{Jup}}$ are excluded. This result might suggest a possible difference between the eccentricity distribution of planets around giants and subgiants. Hence, the eccentricity distribution of planets orbiting subgiants would be similar to that of planets hosted by solar-type stars.   

\citet{Jones2014}, as a first interpretation, suggested that the different eccentricity distributions for planets around evolved stars and those orbiting solar-type stars might be the result of tidal circularization due to the larger radii of evolved stars. According to \citet{Jackson2008}, the circularization timescale might be reduced if the influence of tides in the star, which depends on the stellar radius, are taken into account. In other words, planets around larger radii giants (<$R_{\mathrm{\star}}$> $\sim$ 9.9 $R_{\mathrm{\sun}}$ for our sample) should show lower eccentricities than those planets planets orbiting smaller radii stars, such as subgiants (<$R_{\mathrm{\star}}$> $\sim$ 2.1 $R_{\mathrm{\sun}}$ for our sample). The median eccentricities values that we find for the planets around giants and subgiants along with the relatively low KS probability would support this scenario. However, it is worth to mention that \citet{Jones2014} do not report differences between the eccentricities of planets around giants and those orbiting subgiants. Since the list of stars used as well the criterion employed to classify stars as subgiants or giants are not included in the work of Jones et al., it is not possible for us to identify the cause of disagreement with our result.

Another possibility for the low eccentricities for planets around evolved stars, might be the result of a different formation and evolution mechanism as a consequence of the higher-mass of the evolved stars in comparison with solar-type stars \citep{Johnson2008, Johnson2007a, Jones2014}. On the other hand, it would be possible that the scarcity of planets around giants with eccentricities above 0.2 may represent additional evidence of the engulfment of close-in planets. The presence of several planets around dwarf stars with eccentricities above 0.2 in the range $\sim$ 0.09 - 0.6 AU (See Fig. 8 in Jones et al. 2014), which curiously is the same interval where no planets are observed around giants, suggests that these planets might be destroyed when stars evolve to the red giant branch. Intriguingly, the first results of a survey to search for planets around earlier A-F type dwarfs \citep[e.g.,][]{Galland2005, Lagrange2009} have revealed planets with orbital eccentricities above 0.2 \citep{Galland2005b, Desort2008, Borgniet2014}. It will be interesting to see more results from this ongoing survey.

\section{Summary and conclusions}

In this work we have determined atmospheric stellar parameters ($T_{\mathrm{eff}}$, $\log g$, [Fe/H], $\xi_{t}$) and chemical abundances for a sample of 86 evolved stars with planets (56 giants and 30 subgiants) and for a control sample of 137 stars (101 giants and 36 subgiants) without reported planetary companions to date. Stars from the control sample belong to three RV surveys for planets around evolved stars. They were carefully chosen to have enough RV measurements to rule out, within a high confidence level, the existence of planets similar to those already reported. The stellar parameters were derived homogeneously based on a classical spectroscopic analysis using the EWs automatically measured from both high resolution and signal-to-noise spectra. We found a good agreement with other determinations from the literature, and also with other techniques, which ensures the reliability and consistency of our analysis.

We calculated space-velocity components of the stars in our samples following the method of \citet{Johnson1987}. In addition, adopting the classification criteria of \citet{Reddy2006}, we found that 93\% of the stars belong the thin disk, 2\% are thick disk stars, and 4\% are transition stars. No difference between stars with and without planets was detected.

We also derived projected stellar rotational velocities following the procedure of \citet{Fekel1997}. Our determinations are in good agreement with the literature values. We found no evidence of rapid rotation neither among giants with planets nor among giants without planets. High rotational velocity has been suggested as an observational signal of planet engulfment by red giants.

The analysis of the metallicity of subgiants with and without planets indicates that the metallicity distribution of subgiants with planets is centered at $\sim$ +0.15 dex and more importantly,   subgiants with planets are, on average, $\sim$ 0.16 dex more metal-rich than subgiants without planets, confirming the results of previous studies. Thus, subgiants with planets follow the planet-metallicity correlation found for planet-host main-sequence stars.

The metallicity distribution of giants with planets is centered at slightly subsolar values ($\sim$ -0.05 dex) and we found no metallicity offset between giants with and without planets. Furthermore, as first suggested by \citet{Maldonado2013}, we found that the metallicity distribution of massive giant stars ($M_{\mathrm{\star}} > 1.5 M_{\sun}$) is slightly shifted towards higher metallicities compared to the distribution of giants with $M_{\mathrm{\star}} \leq 1.5 M_{\sun}$. However, contrary to Maldonado et al., for the more massive giants we found no metallicity difference between stars with and without planets.

The [Fe/H] vs. $T_{\mathrm{eff}}$ plot for the subgiants, including the giants at the base of the RGB, shows a slight decrease in the upper boundary of the metallicity distribution towards cooler temperatures. However, this drop is small ($\sim$ 0.10 dex) and needs to be confirmed with a larger sample of subgiants on the blue and red parts of the subgiant branch.

The chemical abundances of 14 elements (Na, Mg, Al, Si, Ca, Sc, Ti, V, Cr, Mn, Co, Ni, Zn, and Ba) relative to the Sun for giants and subgiants were obtained from the measurement of EWs. Our results agree reasonably well with those presented by other works. Analyzing the [X/H] ratios, we found that subgiants with planets show a clear overabundance compared with the control sample without planets, which agrees with the results of planet-host dwarfs. However, as in the case of Fe, we find no significant differences between the [X/H] distributions of giant stars with and without planets for most of the elements. However, the abundances of some elements present likely differences: giants with planets have an overabundance of V of about 0.10 dex when compared to the control sample, whereas for Ba, and Na, giants with planets show lower abundances than the control sample by at least $\sim$ 0.10 dex. In general, the [X/Fe] distributions display no clear differences between subgiants with and without planets. The same behavior occurs for most of the elements in the giant sample. However, a significant difference is observed for Ba, where stars with planets show, on average, lower abundances than stars without planets by 0.11 dex. A similar result is observed for Na and Ca, although to a lesser degree. The opposite trend is observed for V and Co, for which giants with planets have an average excess of $\sim$ 0.09 dex compared with giants without planets. A similar behavior is observed for Mn, although much less evident. We stress that the conclusions found for Na and Ba should be considered as preliminary results, due to the reduced number of lines used in their determinations.   

The subgiants with and without planets with the same [Fe/H] show a similar behavior in the [X/Fe] vs. [Fe/H] plane for most of the elements. Only for V, Co, and Ni, in the higher metallicity range, SGWP might have a very subtle overabundance compared with the control sample. The opposite trend seems to appear for Ba, where the abundances of stars with planets are systematically higher than those of the control sample in the lower metallicity bins. In the case of giants, we found a hint of systematic overabundances in V, Co, and Mn in stars with planets, whereas the opposite behavior occurs for Ba. A slope change is also observed for Na. However, differences are small and need to be taken with care.

Finally, we compared the properties of planet around giants and subgiants and searched for possible correlations between these planetary properties and the chemical abundances of their host stars. Interestingly, although the sample is not large enough to make a thorough statistic analysis, we found some preliminary emerging trends that are summarized as follows:

\begin{itemize}
\item Multi-planet systems around evolved stars, both in giants and subgiants, show a slight metallicity enhancement compared with single-planet systems, following the trends observed on main-sequence stars.

\item Planets with orbital distances larger than $\sim$ 0.5 AU are orbiting subgiants with a wide range of metallicities, but those planets with shorter semimajor axes, are only around subgiants with [Fe/H] > 0.  On the other hand, giants hosting planets with $a$ $\lesssim$ 1 AU, have subsolar metallicities. However, planets with larger semimajor axes orbit not only subsolar metallicity giants but also those with [Fe/H] > 0.

\item  The mass distribution of planets around subgiants includes, besides massive giant planets, the Neptune-type, but only planets with $M_{\mathrm{p}} \sin i$ $\gtrsim$ 0.95 $M_{\mathrm{Jup}}$ have been detected around giant stars. Considering this might be an observational bias of the Doppler technique and taking into account only those planets around subgiants with $M_{\mathrm{p}} \sin i$ $\gtrsim$ 0.95 $M_{\mathrm{Jup}}$, there is no significant difference between the mass distribution of planets around giants and subgiants. Gas giant planets around giants and subgiants are, on average, more massive than those around solar-type stars.

\item Subgiants hosting planets with $M_{\mathrm{p}} \sin i$ $\lesssim$ 0.11 $M_{\mathrm{Jup}}$, have [Fe/H] $\lesssim$ +0.20 dex, while those hosting more massive planets have a higher dispersion in metallicities, including hosts stars with [Fe/H] > +0.20 dex. This result might follow the planet-mass metallicity trend observed on dwarf hosts. On the other hand, as previously found by Maldonado et al. (2013), [Fe/H] seems to decrease with increasing planetary mass for giants with $M_{\mathrm{\star}}$ $\leq$ 1.5 $M_{\odot}$.

\item The [Na/Fe] and [Si/Fe] ratios seem to increase with the mass of planets around giants.

\item Planets around giants show, on average, lower orbital eccentricities than those orbiting subgiants and dwarfs. As previously suggested by \citet{Jones2014}, the orbits of planets around giants might experience a more efficient circularization by interactions with the host stars, as a result of their larger radii. On the other hand, it is possible that the lack of high eccentricity planets around giants be another signal of the engulfment of close-in planets.
\end{itemize}

Despite the low-metallicities of giants with planets, we found no planet orbiting stars with metallicities below the critical limit for planet formation wihtin the core accretion model \citep{Johnson2012}. Furthermore, it is possible that the strong metallicity dependence of this model might be counterbalanced by the relatively higher mass of giants. On the other hand, the negative correlation between planetary mass and the metallicity of the giants with $M_{\mathrm{\star}}$ < 1.5 $M_{\mathrm{\sun}}$, suggests that another planet formation mechanism might be taking place. According to \citet{Cai2006} and \citet{Meru2010}, a low-metallicity enviroment would favor planet-formation within the disk instability scenario.

\begin{acknowledgements}
      We thank Olga Pintado for obtaining the spectra of HD 4732 and HD 141680 and the ELODIE database team for providing us the calibration spectra. We are grateful to Rodrigo D\'iaz for sending us the calibration lamps obtained with SOPHIE and his useful comments and suggestions about control samples. We also express our gratitude to Guillermo Torres for his helpful comments on the detectability of giant planets around massive stars by radial velocities. We also thank Luca Pasquini for useful comments. In addition, we thank the anonymous referee for his/her useful suggestions which helped us to improve significantly the manuscript. We would also like to acknowledge S. Sousa, L. Girardi, R. L. Kurucz, C. Sneden, L. Sbordone, P. Bonifacio, F. Castelli, and F. Arenou for making their codes publicly accessible. E. J, and R. P. gratefully acknowledge the financial support from CONICET in the forms of PhD fellowships. This study has made extensive use of The Extrasolar Planets Encyclopaedia; The Exoplanet Data Explorer; SIMBAD database operated at CDS, Strasbourg (France); and the NASA ADS database.
\end{acknowledgements}




\bibliographystyle{aa}
\bibliography{ref} 

\listofobjects

\end{document}